\documentclass[aps,prx,article,twocolumn,preprintnumbers,amsmath,amssymb,superscriptaddress]{revtex4-2}

\usepackage{graphicx}  
\usepackage{dcolumn}   
\usepackage{bm}        
\usepackage{amssymb}   
\usepackage{amsmath}
\usepackage{mathrsfs}
\usepackage{epigraph}
\usepackage{braket}
\usepackage{gensymb}
\usepackage{lmodern}
\usepackage{lipsum}
\usepackage{marvosym}
\usepackage{ tipa }
\usepackage{bbold}
\usepackage{dsfont}
\usepackage{esint}
\usepackage{mathdots}
\usepackage{xcolor}
\usepackage{appendix}
\usepackage{natbib}
\usepackage{multirow}
\usepackage{float}
\usepackage{bbm}
\usepackage{comment}
\usepackage[colorlinks=true, linkcolor=blue, urlcolor=blue, citecolor=blue]{hyperref}

\begin{document}
\newcommand{\Ham}{\mathcal{H}}
\newcommand{\kbf}{\mathbf{k}}
\newcommand{\qbf}{q}
\newcommand{\Qbf}      {\textbf{Q}}
\newcommand{\lbf}      {\textbf{l}}
\newcommand{\ibf}      {{i}}
\newcommand{\jbf}      {{j}}
\newcommand{\rbf}      {{r}}
\newcommand{\Rbf}      {\textbf{R}}
\newcommand{\Schrdg} {{Schr\"{o}dinger}}
\newcommand{\eps} {{\bm{\varepsilon}}}
\newcommand{\probA}      {{\mathsf{A}}}
\newcommand{\HamPump}      {{\Ham_{\rm pump}}}
\newcommand{\HamPr}      {{\Ham_{\rm probe}}}
\newcommand{\timeMax} {{t_{\rm m}}}
\newcommand{\qin} {{q_{\rm i}}}
\newcommand{\qout} {{q_{\rm s}}}
\newcommand{\epsin} {{\eps_{\rm i}}}
\newcommand{\epsout} {{\eps_{\rm s}}}
\newcommand{\win} {{\omega_{\rm i}}}
\newcommand{\wout} {{\omega_{\rm s}}}

\title{Witnessing Light-Driven Entanglement using Time-Resolved\\ Resonant Inelastic X-Ray Scattering} 	
\author{Jordyn Hales}\thanks{J.H. and U.B. contributed equally to this work.}
\affiliation{Department of Physics and Astronomy, Clemson University, Clemson, SC 29634, USA}
\author{Utkarsh Bajpai}\thanks{J.H. and U.B. contributed equally to this work.}
\affiliation{Department of Physics and Astronomy, Clemson University, Clemson, SC 29634, USA}
\author{Tongtong Liu}
\affiliation{Department of Physics, Massachusetts Institute of Technology, Cambridge, Massachusetts 02139, USA}
\author{Denitsa R. Baykusheva}
\affiliation{Department of Physics, Harvard University, Cambridge, Massachusetts 02138, USA}
\author{Mingda Li}
\affiliation{Department of Nuclear Science and Engineering, Massachusetts Institute of Technology, Cambridge, Massachusetts 02139, USA}
\author{Matteo Mitrano}
\email[\href{mailto:mmitrano@g.harvard.edu}{mmitrano@g.harvard.edu}]{}
\affiliation{Department of Physics, Harvard University, Cambridge, Massachusetts 02138, USA}
\author{Yao Wang}
\email[\href{mailto:yaowang@g.clemson.edu}{yaowang@g.clemson.edu}]{}
\affiliation{Department of Physics and Astronomy, Clemson University, Clemson, SC 29634, USA}	
\date{\today}
\begin{abstract}
\textbf{Abstract:}
Characterizing and controlling entanglement in quantum materials is crucial for the development of next-generation quantum technologies. However, defining a quantifiable figure of merit for entanglement in macroscopic solids is theoretically and experimentally challenging. At equilibrium the presence of entanglement can be diagnosed by extracting entanglement witnesses from spectroscopic observables and a nonequilibrium extension of this method could lead to the discovery of novel dynamical phenomena. Here, we propose a systematic approach to quantify the time-dependent quantum Fisher information and entanglement depth of transient states of quantum materials with time-resolved resonant inelastic x-ray scattering. Using a quarter-filled extended Hubbard model as an example, we benchmark the efficiency of this approach and predict a light-enhanced many-body entanglement due to the proximity to a phase boundary. Our work sets the stage for experimentally witnessing and controlling entanglement in light-driven quantum materials via ultrafast spectroscopic measurements.
\end{abstract}
\maketitle

\section{Introduction}
Quantum materials are systems featuring collective electronic behavior\,\cite{keimer2017physics} with broad technological applications, such as superconductivity\,\cite{matthias1963superconductivity}, topological order\,\cite{wen2017colloquium}, and quantum spin liquidity\,\cite{balents2010spin,broholm2020quantum}. Underlying these emergent phenomena is the presence of entanglement among subparts of the electronic wavefunctions\,\cite{amico2008entanglement, horodecki2009quantum}, which has important fundamental and applied consequences. On one hand, quantum fluctuations caused by entanglement play an important role in the appearance of quantum phase transitions characterized by unconventional behavior\,\cite{imada1998metal,kondo2015point,faeth2020incoherent, xu2020spectroscopic,he2021superconducting,chen2022lattice}. On the other hand, entanglement constitutes a precious resource for material-based quantum computing, where information is encoded and manipulated via arbitrary entangled or even multipartite entangled states\,\cite{vedral2008quantifying,guhne2009entanglement}. Therefore, accurately characterizing and controlling entanglement in the solid state is a key step towards the realization of future quantum technologies\,\cite{de2021materials}.

The many-body wavefunction of a highly entangled quantum system cannot be expressed as the direct product of multiple single-particle states in any basis. Entangled wavefunctions in synthetic few-body quantum simulators can be experimentally characterized \,\cite{haffner2005scalable,esteve2008squeezing,guhne2009entanglement,gross2010nonlinear,van2012measuring} through the R\'{e}nyi entropy\,\cite{islam2015measuring,kaufman2016quantum,tubman2016measuring,linke2018measuring, brydges2019probing} and multi-point correlations\,\cite{schweigler2017experimental, hilker2017revealing, salomon2019direct, koepsell2019imaging, vijayan2020time, prufer2020experimental, zache2020extracting, koepsell2020microscopic}. However, the measurement complexity increases with the Hilbert-space dimension and scales exponentially with the system size. As solid-state measurements are restricted to a limited number of macroscopic observables, a tomography of electronic wavefunctions in quantum materials becomes impractical\,\cite{wootters1981statistical,carvalho2004decoherence,mintert2005concurrence, aolita2006measuring}.

A more efficient approach for investigating entanglement in real materials relies on determining the ``entanglement depth''\,\cite{sorensen2001entanglement, guhne2005multipartite, acin2001classification,friis2019entanglement}, defined as the minimum number of entangled modes required to construct a specific many-body state. The bounds on entanglement depth of a quantum system can be quantified through expectation values of specific operators called entanglement witnesses\,\cite{terhal2000bell, coffman2000distributed, amico2004dynamics, roscilde2004studying,brukner2006crucial}. Recent developments along this direction have been enabled by use of the quantum Fisher information (QFI)\,\cite{pezze2009entanglement,hyllus2012fisher,toth2012multipartite}, which can be quantified from equilibrium spectroscopy\,\cite{hauke2016measuring}. By extracting the QFI from the imaginary part of the dynamical susceptibility, it is possible to witness a lower bound for the entanglement depth without relying on a full tomography of the wavefunction\,\cite{helstrom1969quantum,braunstein1994statistical,braunstein1996generalized}. This approach has been successfully applied to the study of magnetic materials \cite{mathew2020experimental,laurell2021quantifying,scheie2021witnessing}, demonstrating its feasibility in solid-state experiments at equilibrium.

Diagnosing entanglement in quantum materials out of equilibrium would be particularly impactful, as ultrafast lasers have led to the synthesis of nontrivial states of matter without equilibrium analogues\,\cite{zhang2014dynamics, basov2017towards,delatorre2021nonthermal}. Dynamical entanglement has been theoretically demonstrated in the wavefunctions of solvable toy models\,\cite{amico2004dynamics,alba2017entanglement,nahum2017quantum}, but so far it has been unclear how to experimentally characterize this phenomenon in nonequilibrium spectroscopic experiments. Inspired by quantum metrology, entanglement can be quantified by the fluctuations of local probes at a given distance or a finite momentum, and the recently developed time-resolved resonant inelastic x-ray scattering (trRIXS) technique\,\cite{dean2016ultrafast, cao2019ultrafast, mitrano2020probing} --- sensitive to collective charge, orbital, spin, and lattice excitations\,\cite{dean2016ultrafast, mitrano2019ultrafast, mazzone2021laser, mitrano2019evidence,parchenko2020orbital} --- opens new opportunities for diagnosing entanglement in light-driven systems.

\begin{figure}
	\includegraphics[width=8.5cm]{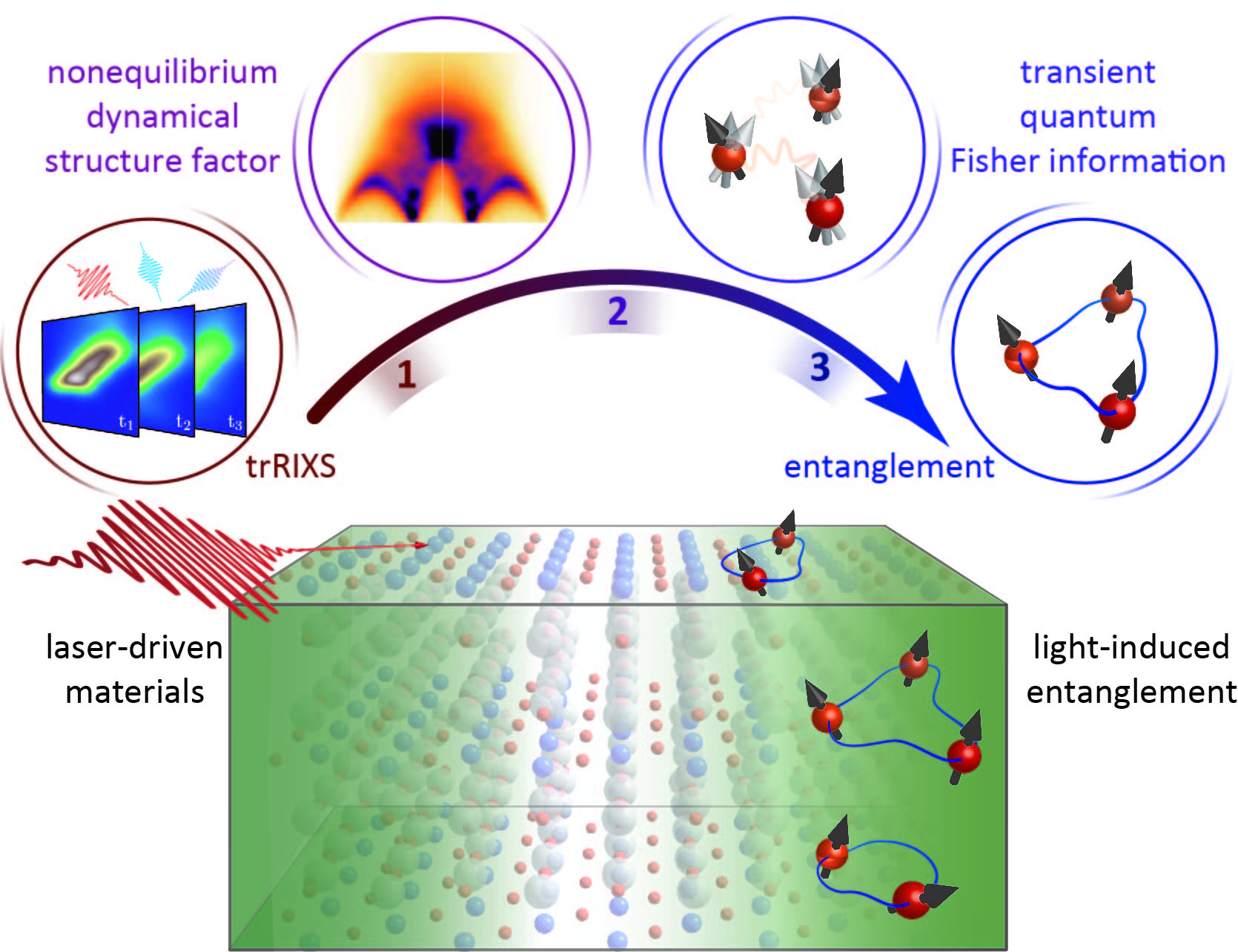}
	\caption{\textbf{Probing light-driven entanglement in quantum materials.} An intense pump laser drives a material out of equilibrium and its time-dependent collective excitations are probed by time-resolved Resonant Inelastic X-ray Scattering (trRIXS). One witnesses many-body entanglement by first extracting the nonequilibrium dynamical structure factor from the trRIXS response function, then calculating the quantum Fisher information associated with a specific operator using the transient dynamical structure factor, and, finally, comparing the transient quantum Fisher information with operator-specific quantum bounds. }
	\label{fig:cartoon}
\end{figure}

Witnessing entanglement with trRIXS requires measuring the scattering cross-section of a light-driven material, evaluating the transient dynamical structure factors, calculating the QFI, and determining relevant quantum bounds signaling the presence of entanglement for a certain observable (see Fig.~\ref{fig:cartoon}). Each step encapsulates a distinct challenge, and we illustrate them by focusing on the spin degrees of freedom. First, estimating accurate spin fluctuations using trRIXS requires separating the contributions of high-order excitations mediated by the intermediate state of the scattering process. Second, competing time and energy resolution effects imply that the QFI for the instantaneous wavefunctions\,\cite{pang2017optimal} cannot be directly extracted from the spectrum at a single time delay and must be deconvolved on the time axis. Third, assessing the entanglement depth requires calculating suitable theoretical bounds for the QFI operator. While the connection between QFI and multiparticle entanglement is well established for spin operators in magnetic materials and gapped fermionic lattice modes\,\cite{de2021entanglement}, this connection is unclear for interacting fermions with dopant carriers, which is particularly relevant to the study of light-driven materials. 

Here, we address these challenges and develop a procedure to witness entanglement using trRIXS. As a brief summary of the main results, we quantify the impact of the finite core-hole lifetime on the evaluation of dynamical structure factor, develop a self-consistent protocol to calculate the instantaneous QFI, and assess the multipartite entanglement dynamics using the QFI and appropriate theoretical bounds. Through this procedure, we focus on the study of light-induced entanglement when a system is close to a phase boundary and choose a one-dimensional (1D) extended Hubbard model (EHM). The EHM under consideration, with repulsive on-site interaction and attractive nearest-neighbor interaction, has been recently identified as the underlying microscopic description of 1D cuprate chains (e.g., Ba$_{2-x}$Sr$_x$CuO$_{3+\delta}$)\,\cite{chen2021anomalously}. By comparing its dynamics with that of a pure Hubbard model, we find that the nonlocal interaction plays a crucial role on the nonequilibrium enhancement of many-body entanglement.

\section{Results}

\subsection{Quantifying the Time-dependent Quantum Fisher Information from trRIXS}\label{sec:qfi}

For a time-dependent wavefunction $|\psi(t)\rangle$ and in the \Schrdg\ picture, the instantaneous QFI density associated with a local spin operator is defined as\,\cite{pezze2009entanglement,hyllus2012fisher,toth2012multipartite}
\begin{eqnarray}\label{eq:qfiDefinition}
f_\mathrm{Q}(q, t) &=& \frac{4}{N} \sum_{ij}e^{iq(r_i-r_j)}\left[ \left\langle\psi(t)\left| {S}_i^z{S}_j^z\right|\psi(t)\right\rangle  \right.\nonumber\\
&&\left.- \left\langle\psi(t)\left|{S}_i^z\right|\psi(t)\right\rangle \left\langle\psi(t)\left|{S}_j^z\right|\psi(t)\right\rangle \, \right]\,.
\end{eqnarray}
where $r_i$ is the real-space position vector of site $i$, $S_j^z$ is the localized spin operator ($z$ component), and $N$ is the total number of sites. The transient QFI $f_\mathrm{Q}(q, t)$ is completely determined by the instantaneous wavefunction $|\psi(t)\rangle$, or an ensemble represented by the density matrix $\rho(t)$. However, different from the integral relation for equilibrium states\,\cite{hauke2016measuring}, one cannot obtain the $f_\mathrm{Q}(q, t)$ by directly integrating an instantaneous pump-probe spectrum at a certain time delay $t$, due to the convolution with finite-width probe pulses. Such a finite width contributes to the energy resolution of time-resolved spectroscopy experiments and is usually comparable to the natural timescales of the investigated systems. We describe here a strategy to extract the time-dependent QFI from a sequence of nonequilibrium spectra.

Let us first assume that the dynamical structure factor $S({q}, \omega, t)$ is experimentally accessible, leaving the discussion of its extraction from x-ray measurements to Sec.~\ref{sec:RIXS}. The $S({q}, \omega, t)$ is defined as\,\cite{wang2017producing} 
\begin{eqnarray}\label{eq:Sqw}
S(q, \omega, t) 
&= & 
 \frac{1}{2\pi N}\sum_{ij}e^{-iq(r_i-r_j)} \iint\displaylimits_{-\infty}^{+\infty} dt_1 dt_2\, g\left(t_1; t\right) g\left(t_2; t\right) \nonumber\\
&& \left\langle\hat{S}_i^z\left(t_1\right)
\hat{S}_j^z\left(t_2\right) \right\rangle e^{i\omega (t_1-t_2)}, 
\end{eqnarray}
where $g(\tau; t)$ denotes the temporal probe envelope, usually approximated with a Gaussian profile\,\cite{freericks2009theoretical} 
\begin{eqnarray}
g(\tau; t) = \frac1{\sigma_{\mathrm{pr}}\sqrt{2\pi}} e^{-(\tau - t)^2/2\sigma_{\mathrm{pr}}^2}\,,
\end{eqnarray}
with pulse duration $\sigma_{\mathrm{pr}}$. This duration is physical, primarily set by the spectral content of the laser pulses but also renormalized by the instruments and the material's self-energy. We express the pump-probe spectra in the interaction picture (same below), where the operator $\hat{\mathcal{O}}(t) = \mathcal{U}(-\infty, t) \mathcal{O}\mathcal{U}(t, -\infty)$ evolves via the unitary operator $\mathcal{ U}(t,t_0)= \mathcal{\hat T}_{t} \left[e^{-i\int_{t_0}^t  \Ham(t')dt'}\right]$. Here, the time-dependent Hamiltonian $\Ham(t)$ only includes the pump (and not the probe) field.

Due to the effects of the probe pulse profile on time and energy resolution, the spectrum $S({q}, \omega, t)$ at time $t$ is determined not only by the instantaneous wavefunction $|\psi(t)\rangle$, but also by earlier or later wavefunctions in a finite time window. Therefore, the QFI density $f_\mathrm{Q}(q, t)$, which diagnoses the entanglement of the instantaneous wavefunction $|\psi(t)\rangle$, cannot be evaluated by simply integrating the $S({q}, \omega, t)$ along the energy axis. As we show in Supplementary Note 1, the relationship between the time-dependent QFI $f_\mathrm{Q}(q, t)$ and the transient structure factor $S({q}, \omega, t)$ becomes an implicit integral equation
\begin{eqnarray}\label{eq:qfi_gauss}
\int\displaylimits_{-\infty}^{+\infty} d\tau g(\tau; t)^2f_\mathrm{Q}( q, \tau) = 4 \int\displaylimits_{-\infty}^{+\infty} d\omega\, S( q, \omega, t)\,.
\end{eqnarray}
Note that we have assumed the absence of long-range magnetic order at the specific momentum $q$, which is the case for the simulations in this paper. If a long-range order is present, one should further subtract the elastic peak from the structure factor, whose intensity corresponds to the disconnected part (second term) of Eq.~\eqref{eq:qfiDefinition}. In the limit of ultrashort probe pulses, i.e., $\sigma_{\rm pr}$ smaller than any nonequilibrium physical timescale of the system, the envelope $g(\tau;t)^2$ can be approximated by a delta function $\delta(\tau-t)$, leading to an explicit solution consistent with the equilibrium sum-rule integral\,\cite{hauke2016measuring}. However, when the probe pulse has a finite time duration (as in spectroscopy experiments with high energy resolution), this approximation breaks down. In order to extract the instantaneous QFI from this implicit  Eq.~\eqref{eq:qfi_gauss}, we expand its left-hand side and convert the equation into a self-consistent integro-differential problem. As detailed in Supplementary Note 1, this leads to
\begin{equation}\label{eq:qfi_tay}
f_\mathrm{Q}(q, t) = 8\sigma_\mathrm{pr} \sqrt{\pi} \int\displaylimits_{-\infty}^{+\infty} d\omega\, S( q, \omega, t) + \sum_{m=1}^\infty \frac{\mathcal{C}_m}{(2m)!} \frac{\partial^{2m}f_\mathrm{Q}}{\partial t^{2m}},
 \end{equation}
where ${\mathcal{C}_m=-(1/\sigma_\mathrm{pr}\sqrt{\pi}) \int_{-\infty}^{\infty}e^{-x^2/\sigma_\mathrm{pr}^2}}x^{2m} dx = {-(\sigma_\mathrm{pr}^{2m}/\sqrt{\pi}) \Gamma(m + 1/2)}$. In the presence of time-translational invariance, the infinite series on the right-hand side of Eq.~\eqref{eq:qfi_tay} vanishes and we reproduce the equilibrium relation $f_\mathrm{Q}^{\rm eq}( q, \tau) = 8\sigma_\mathrm{pr} \sqrt{\pi} \int\displaylimits_{-\infty}^{+\infty} d\omega\, S^{\rm eq}( q, \omega, t)$\,\cite{hauke2016measuring}. However, high-order derivative terms on the right-hand side can play a significant role far from equilibrium, as discussed in Sec.~\ref{sec:EHM}.

The self-consistently calculated $f_\mathrm{Q}(q, t)$ serves the purpose of witnessing entanglement in a transient $k$-partite quantum state when exceeding its operator-specific boundary\,\cite{hauke2016measuring}. While we use the pure-state notation in the derivation of the QFI sum rule and choose a pure initial state in our simulations, this approach applies to both pure and mixed initial states (see Supplementary Note 1 for further details). This generalization relies on considering $\langle\cdots\rangle$ as an ensemble average and on the linearity of Eqs.~\eqref{eq:qfi_gauss} and \eqref{eq:qfi_tay}. We note that, as specified by Hauke {et al.} in Ref.\,\cite{hauke2016measuring}, this protocol requires a careful normalization of the scattering cross-section into absolute values by properly accounting for the trRIXS matrix elements.

\begin{figure}[!t]
	\includegraphics[width=8.5cm]{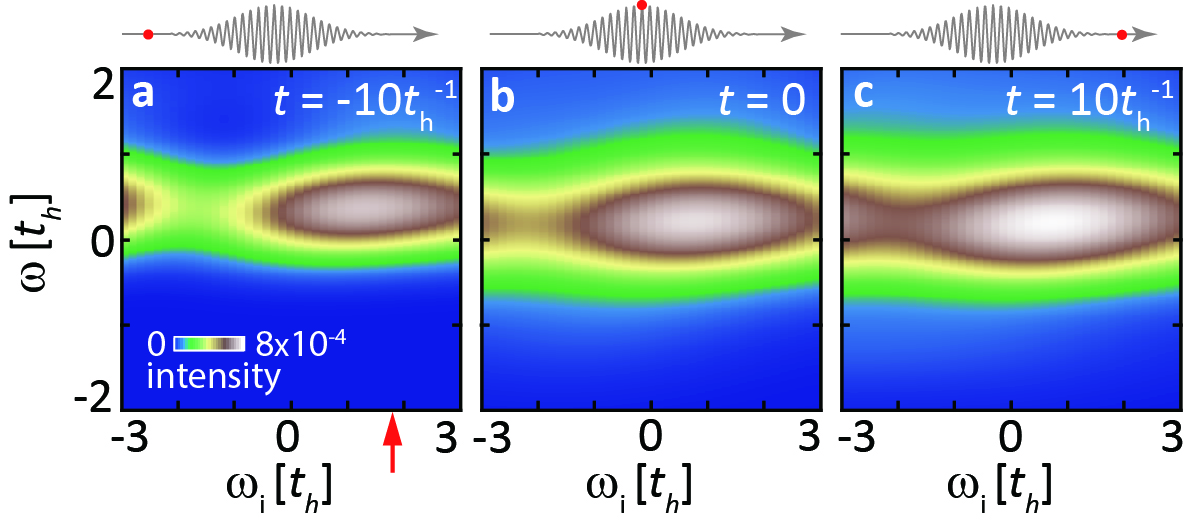}\vspace{-2mm}
	\caption{\textbf{Overview of time-resolved Resonant Inelastic X-ray Scattering (trRIXS) spectra.} \textbf{a}-\textbf{c} Snapshots of the trRIXS spectra with $q=\pi/6$ and for selected pump-probe time delays. The red arrow indicates the resonance incident energy $\win$ used to analyze the structure factors. The upper insets show the time relative to the pump pulse, whose amplitude is $A_0=1$ and frequency is $\Omega=10t_\mathrm{h}$. The core-hole lifetime is set as $\tau_\mathrm{core} = 1/3\,t_\mathrm{h}$.}
	\label{fig:fullRIXS}
\end{figure}

\subsection{Time-Resolved Resonant Inelastic X-Ray Scattering}\label{sec:RIXS}
Recent experimental work on diagnosing entanglement in the solid state focused on inelastic neutron scattering of low-dimensional spin systems \cite{laurell2021quantifying,scheie2021witnessing,mathew2020experimental}, as the neutron scattering cross section is directly proportional to the dynamical spin structure factor\,\cite{Lovesey1984theory}. While there is not yet an ultrafast incarnation of inelastic neutron scattering, the recent development of trRIXS provides an alternative pathway to access nonequilibrium dynamical structure factors of spin and charge degrees of freedom \,\cite{dean2016ultrafast, cao2019ultrafast, mitrano2019ultrafast,mitrano2020probing}. As detailed in Methods, trRIXS is a photon-in-photon-out x-ray scattering process involving an intermediate state with a finite lifetime. Due to the spin-orbit coupling at the core level (e.g.,~the $2p$ orbitals for the transition-metal $L$-edge RIXS), this intermediate state can involve spin flip events and couple to magnetic excitations of the valence band\,\cite{ament2009theoretical}. Therefore, trRIXS is sensitive to spin excitations and can be used to probe the nonequilibrium spin dynamics of light-driven materials\,\cite{wang2021x}. 

The trRIXS cross-section, denoted as $\mathcal{I}(q,\win-\omega,\win,t)$ in Methods, depends on energy, momentum, and polarization of both the incident and scattered photons. In our simulation, we select a scattering geometry with $\pi$-polarized (parallel to the scattering plane) incident photons and $\sigma$-polarized (perpendicular to the scattering plane) scattered photons, which maximizes the spin-flip contribution to the trRIXS cross-section\,\cite{ament2009theoretical, wang2021x}. Due to our focus on spin entanglement, we keep this polarization configuration fixed throughout the paper and omit the polarization subscripts in $\mathcal{I}$. The trRIXS spectrum comes with two energy axes for the incident photon energy $\win$ and the energy loss $\omega$ (difference between incident and scattering photon energies). Figure \ref{fig:fullRIXS} shows sample trRIXS spectra for a driven extended Hubbard model (see Sec.~\ref{sec:EHM}) and for a range of incident energies $\win$.  We select the resonance $\win$ by maximizing the trRIXS intensity. At fixed $\win$, the nonequilibrium dynamical structure factor $S( q, \omega, t)$ can be estimated by
\begin{equation}\label{eq:sqw_Iqw}
S( q, \omega, t) \approx \frac {\mathcal{I}(q,\win-\omega,\win,t)}{\tau_{\rm core}^2|{M}^{\rm (in)}_{\qin\epsin}{M}^{\rm (out)}_{\qout\epsout}|^2}\,,
\end{equation}
which implies replacing the excitation operator $\hat{\mathcal{D}}^\dagger_{{q}_s{\epsilon}_s}(t_1')\hat{\mathcal{D}}_{{q}_i{\epsilon}_i}(t_1) $ with a spin-flip operator $\hat{S}^+(t_1) $. Here, $\hat{\mathcal{D}}_{{q}_i{\epsilon}_i}$ and $\hat{\mathcal{D}}^\dagger_{{q}_s{\epsilon}_s}$ denote the dipole-transition operators for the incident and scattering processes, respectively; ${M}^{\rm (in)}_{\qin\epsin}$ and ${M}^{\rm (out)}_{\qout\epsout}$ indicate the corresponding matrix elements and $\tau_{\rm core}$ is the core-hole lifetime. The definition of these factors and the full expression of the trRIXS cross-section $\mathcal{I}(q,\win-\omega,\win,t)$ are explained in the Methods section. The estimation of Eq.~\eqref{eq:sqw_Iqw} is analogous to the ultrashort core-hole lifetime (UCL) approximation in equilibrium\,\cite{ament2009theoretical,jia2016using}, and it can be proven that Eq.~\eqref{eq:sqw_Iqw} becomes exact in the $\tau_{\rm core} \rightarrow 0$ limit.

\subsection{TrRIXS and QFI in a Driven Extended Hubbard Model}\label{sec:EHM}
In this work, we aim to witness entanglement with trRIXS in a prototypical correlated electron system. Given prior equilibrium RIXS experiments\,\cite{Schlappa2012spin}, we consider 1D cuprate chains as an ideal platform for the experimental verification of our results. Recent experiments in Ba$_{2-x}$Sr$_x$CuO$_{3+\delta}$ have identified the EHM with mixed-sign interactions\,\cite{chen2021anomalously} as their underlying model Hamiltonian (see Methods) and here we investigate its light-driven dynamics. Throughout this paper, we set the on-site ($U$) and the nearest-neighbor ($V$) interactions to $U=8t_{\rm h}$ and $V=-t_{\rm h}$, respectively, corresponding to the characteristic values for cuprate chains\,\cite{chen2021anomalously, wang2021phonon}. The existence of this nearest-neighbor term $V$ is crucial for the presence of many-body entanglement, as we discuss in Sec.~\ref{sec:entanglement}. 

We introduce the pump excitation in our second-quantized electrons through the standard Peierls substitution. The pump laser pulses are described by a vector potential in the form of an oscillatory Gaussian  ${A}(t)=A_0\, e^{-t^2/2\sigma_\mathrm{pump}^2} \cos(\Omega t)$ with fixed width $\sigma_\mathrm{pump} = 3 t_\mathrm{h}^{-1}$, variable amplitude $A_0$, and frequency $\Omega$. The ground state of the EHM is calculated by the parallel Arnoldi method with Paradeisos acceleration\,\cite{lehoucq1998arpack,  jia2017paradeisos} and the time evolution is evaluated by the Krylov subspace technique\,\cite{manmana2007strongly, balzer2011krylov}. We adopt a 12-site chain with periodic boundary conditions and quarter filling throughout this paper, due to its proximity to the triplet-pairing phase\,\cite{lin1995phase, qu2021spin}. We employ the ground state at zero temperature as the initial state, due to the computational complexity of simulating the trRIXS cross-section of an ensemble. The generality of this method is further discussed in Sec.~\ref{sec:qfi} and the SI.

\begin{figure}[!t]\hspace*{-3mm}
	\includegraphics[width=9cm]{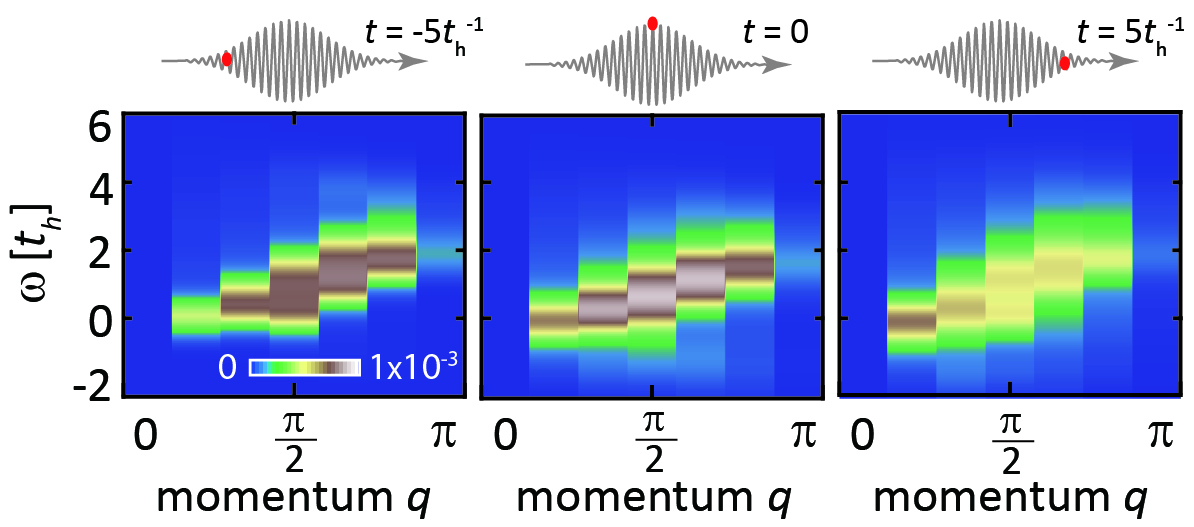}\vspace{-2mm}
	\caption{\textbf{Momentum distribution for time-resolved Resonant Inelastic X-ray Scattering (trRIXS) spectra.} From left to right: trRIXS spectra for the 1D momentum $q$ ranging from 0 to $\pi$, at $t = -5 t_\mathrm{h}^{-1}$, $t = 0$, and $t = 5 t_\mathrm{h}^{-1}$, respectively. The incident energy $\win$ is set as 1.8$t_h$ and the upper insets show the time relative to the pump pulse, whose condition is the same as Fig.~\ref{fig:fullRIXS}. }
	\label{fig:RIXSMomentum}
\end{figure}

\begin{figure}\hspace*{-3mm}
	\includegraphics[width=8.9cm]{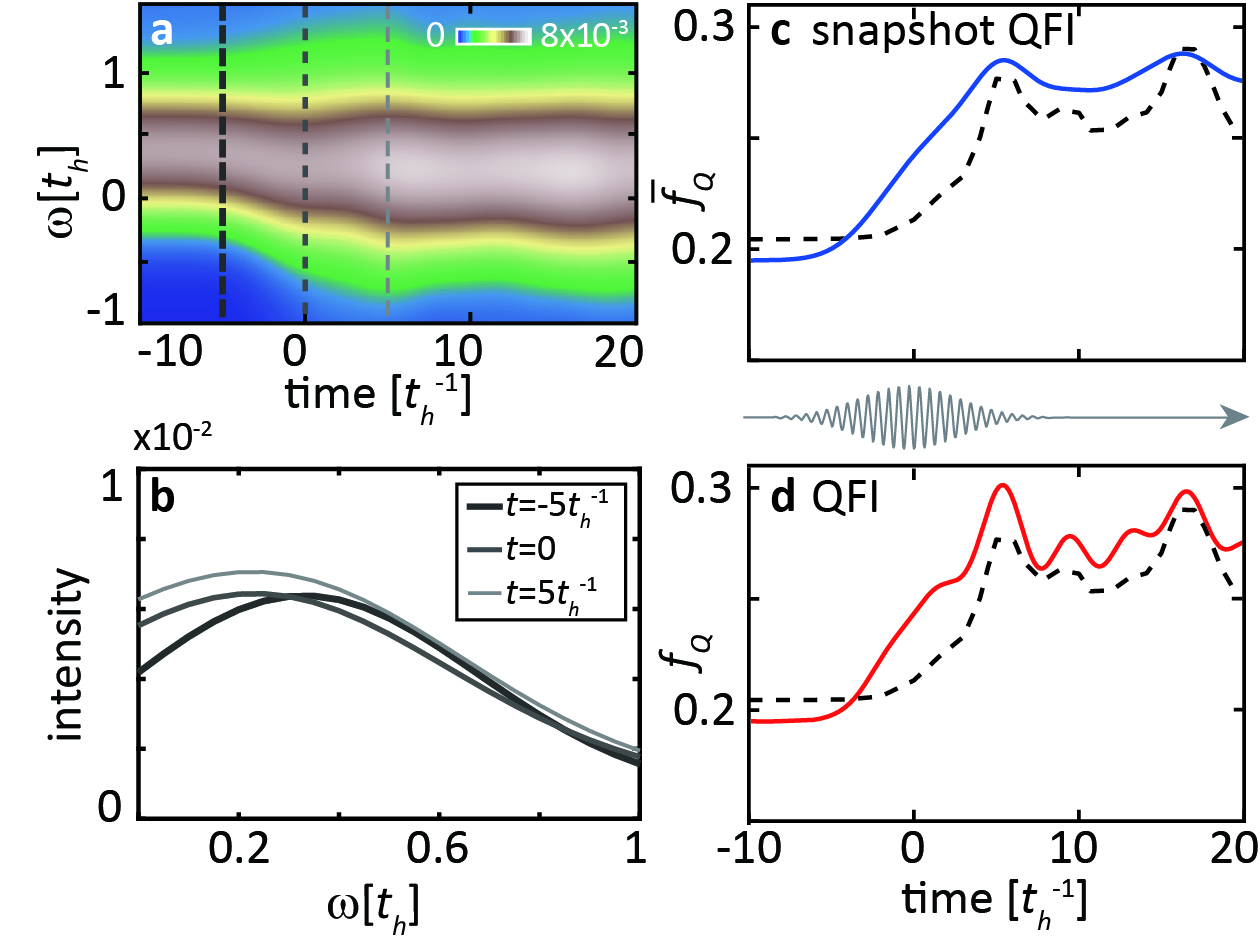}\vspace{-3mm}
	\caption{\textbf{Nonequilibrium dynamical structure factor and transient quantum Fisher information (QFI).} \textbf{a} Evolution of the dynamical spin structure factor $S(q,\omega,t)$ estimated from $\mathcal{I}(q, \win - \omega,\win, t)$ by fixing the incident energy at $\win=1.8t_h$, as in Fig. \ref{fig:fullRIXS}. \textbf{b} Spectral distribution for $t=-5t_h^{-1}$, 0, and $5t_h^{-1}$, respectively. \textbf{c} Time-dependence of the snapshot quantum Fisher information (QFI) density $\bar{f}_\mathrm{Q}(q, t)$ (blue curve) evaluated by integrating the instantaneous spectral cuts [e.g., each curve in panel (\textbf{b})] and of the exact QFI (black dashed line). \textbf{d} Time dependence of the QFI density ${f}_\mathrm{Q}(q, t)$ (red curve) evaluated by the self-consistent iteration in Eq.~\eqref{eq:qfi_tay} and comparison with the exact QFI (black dashed line).
	The core-hole lifetime for these data is $\tau_\mathrm{core} = 1/3\,t_\mathrm{h}$, while the momentum transfer is fixed to $q=\pi/6$.
	}
	\label{fig:QFI}
\end{figure}

	Figure~\ref{fig:fullRIXS}{a} shows selected theoretical $\pi$-$\sigma$-polarized trRIXS spectra. The incident energy $\win=1.8 t_\mathrm{h}$ (defined relative to the absorption edge) is determined according to the equilibrium resonance profile for the singly-occupied initial state. By varying the momentum transfer $q$, the equilibrium RIXS spectrum of a quarter-filled EHM displays a two-spinon continuum ($t\!=\!-5t_h^{-1}\!<\!-\sigma_{\rm pump}$), as shown in Fig.~\ref{fig:RIXSMomentum}. Different from undoped antiferromagnets\,\cite{scheie2021witnessing}, low-energy spin excitations of a doped extended Hubbard model mainly lie near the nesting vector $q=2k_F$ of spinon Fermi surface\,\cite{parschke2019numerical}. At large momenta ($q\geq\pi/2$), the trRIXS spectral weights are gradually suppressed by the x-ray scattering matrix elements\,\cite{jia2014persistent, jia2016using}, which we divide out as shown in Eq.~\eqref{eq:sqw_Iqw} when evaluating the spin QFIs. When calculating the time-dependent QFI, we focus on a specific momentum, $q = \pi/6$. Distinct from undoped antiferromagnets previously employed to witness equilibrium entanglement\,\cite{scheie2021witnessing}, this small-momentum wavevector exhibits the most evident spectral changes and captures the longest-range correlations supported by our system size. The 1D system simulated in this paper does not have a spontaneous symmetry breaking, which simplifies the entanglement analysis of the spin fluctuations.	
At the selected resonance and momentum transfer, we calculate the time-dependent $S(q,\omega,t)$ from the trRIXS intensity following Eq.~\eqref{eq:sqw_Iqw}, as shown in Fig.~\ref{fig:QFI}{a}. At the center of pump pulse, i.e., $t = 0$, the excitation spectrum experiences an overall softening [also see Fig.~\ref{fig:QFI}{b}] due to a Floquet renormalization of the spin-exchange energy, and a slight broadening of the spectral peak\,\cite{mentink2015ultrafast,wang2021x}. These spectral changes persist long after the pump pulse as a result of strong correlation effects\,\cite{wang2017producing}. However, different from the case of light-driven spin spectra at the top of a magnon band\,\cite{wang2021x}, the softening here is accompanied by an increase of the spectral intensity. To analyze the nonequilibrium entanglement, the evolution of the intensity is more relevant than that of the peak position.	

To test our framework for evaluating the transient QFI of light-driven spin degrees of freedom, we first calculate the exact values of ${f}_\mathrm{Q}(q, t)$ [dashed curves in Figs.~\ref{fig:QFI}{c} and {d}] by plugging the simulated time-dependent wavefunctions $|\psi(t)\rangle$ into Eq.~\eqref{eq:qfiDefinition}. Note that this direct evaluation is theoretically possible due to the access to the instantaneous wavefunctions, which cannot be measured in experiments. We then proceed to evaluate the QFI as it would be done in real trRIXS experiments, i.e., by only assessing the sequence of trRIXS snapshots without additional theoretical knowledge about the state of the driven system. A direct extension of the equilibrium formula in Ref.~\onlinecite{hauke2016measuring} entails treating each single time delay as an equilibrium spectrum. To distinguish it from the nonequilibrium definition in Eq.~\eqref{eq:qfi_tay}, we rename it as ``snapshot QFI'' 
\begin{eqnarray}\label{eq:snapshotQFI}
\bar{f}_\mathrm{Q}(q, t) = 8\sigma_\mathrm{pr} \sqrt{\pi} \int d\omega\, S( q, \omega, t)\,.
\end{eqnarray}
Compared with the exact QFI densities, the $\bar{f}_\mathrm{Q}(q, t)$ (blue curve) overestimates the transient increase near $t\sim 0$ and does not capture the oscillations after the pump ends [see Fig.\ref{fig:QFI}c]. Such a deviation at ultrafast timescales reflects the presence of convolution effects caused by the finite probe width.

\begin{figure}[!t]
	\includegraphics[width=8.5cm]{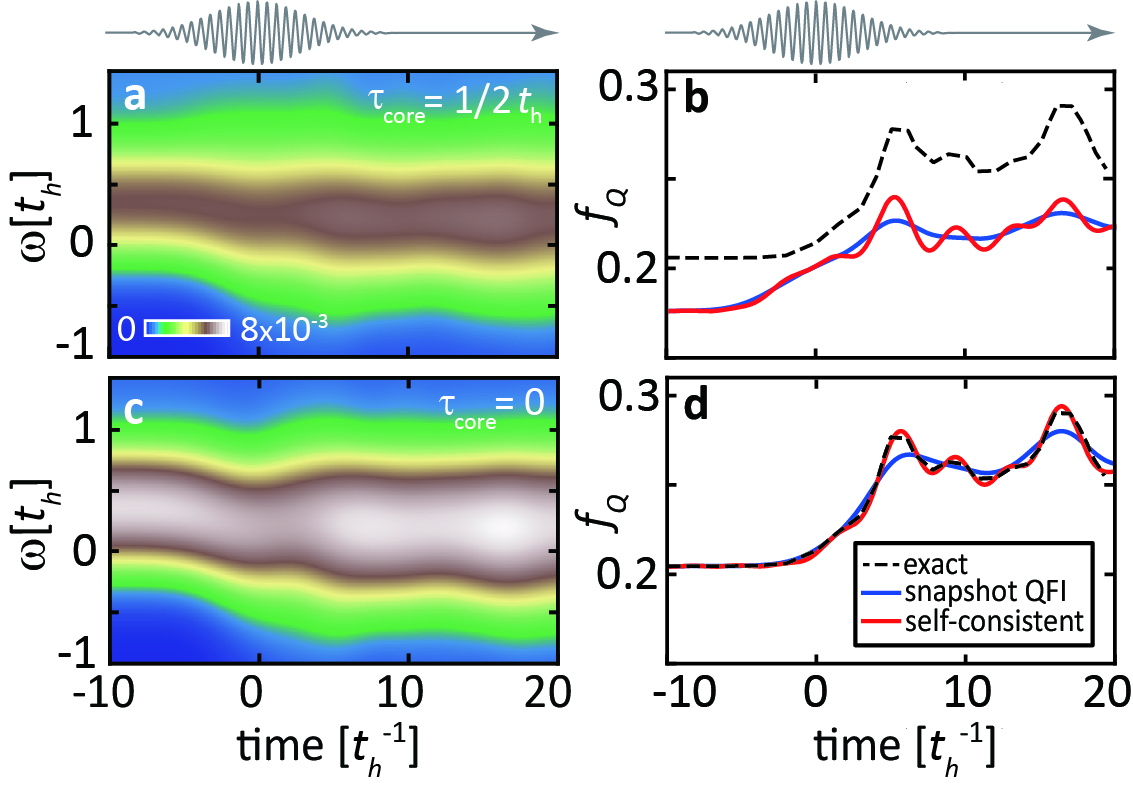}\vspace{-3mm}
	\caption{\textbf{Core-hole lifetime contribution to the quantum Fisher information (QFI) evaluation.} \textbf{a} Evolution of $S(q,\omega,t)$ extracted from the trRIXS intensity for core-hole lifetime set as $\tau=1/2t_h$. \textbf{b} Comparison of the exact results of instantaneous QFI (black dashed line), the snapshot QFI density $\bar{f}_\mathrm{Q}(q, t)$ (blue curve) evaluated the direct integration, and the QFI density ${f}_\mathrm{Q}(q, t)$ (red curve) evaluated by the self-consistent iteration. \textbf{c},\textbf{d} Same as \textbf{a},\textbf{b} but for the zero core-hole lifetime situation, where trRIXS is identical to $S(q,\omega,t)$. The upper insets show the evolution of the pump-field vector potential. }
	\label{fig:lifetime}
\end{figure}

The failure of the snapshot QFI in capturing the exact QFI evolution requires the introduction of the full self-consistent iteration Eq.~\eqref{eq:qfi_tay}. Compared to the snapshot QFI, the latter contains high-order time-derivatives, which are non-negligible when the spectrum varies rapidly in time. We evaluate these high-order derivatives using finite difference methods, starting from the time sequence of trRIXS spectra, and then solving the self-consistent equations. This procedure leads us to the ${f}_\mathrm{Q}(q, t)$ indicated by the red curve in Fig.~\ref{fig:QFI}{d}.  The full self-consistent QFI is closely aligned with the exact QFI behavior and captures the time-dependent oscillations induced by the pump for $t>5t_h^{-1}$ [see Fig.~\ref{fig:QFI}{d}]. This implies that the self-consistent calculation of the QFI is essential for capturing fast coherent dynamics and it is not sufficient to approximate each time delay as a quasi-equilibrium spectrum. 	
The remaining deviation with respect to the exact evolution of QFI, which occurs both at equilibrium and at the center of the pump, can be attributed to the fact that the trRIXS spectrum is not identical to the spin structure factor $S(q,\omega,t)$. Such a discrepancy is known in equilibrium RIXS, which captures the poles of $S(q,\omega,t)$ but is less accurate in yielding its spectral weight\,\cite{jia2016using}. Physically, the intermediate state in trRIXS has a finite lifetime and contains additional dynamics, such as multi-magnon or spin-charge excitations, besides instantaneous spin-flip events in the UCL limit. Due to the admixture with these excitations, Eq.~\ref{eq:sqw_Iqw} is only a good approximation but not an identity. The trRIXS spectrum underestimates the QFI at equilibrium, where the local moment is maximal and other processes are irrelevant (see Supplementary Figure 2), while overestimating the QFI at the center of the pump pulse where there are more charge carriers induced by the laser.

To quantify the effect of a finite core-hole lifetime, we compare simulations with $\tau_{\rm core}=1/2 t_h$ and $\tau_{\rm core}=0$. As shown in Fig.~\ref{fig:lifetime}, the QFIs extracted from the trRIXS spectra converge towards the exact calculations with decreasing core-hole lifetime. In the limit of an infinitesimal $\tau_{\rm core}$ [see Figs.~\ref{fig:lifetime}d], the self-consistently calculated $f_\mathrm{Q}(q, t)$ precisely matches the exact QFI density obtained from the instantaneous wavefunctions. In contrast, the ``snapshot QFI'' $\bar{f}_\mathrm{Q}(q, t)$ (blue curve) still deviates from them, reflecting the intrinsic error caused by the finite time resolution due to the probe pulse.  When the lifetime is not negligible, the conversion of the equilibrium RIXS intensity into the $S(q,\omega)$ requires a more systematic approach. One method consists of calculating the four-particle response function as the lowest-order perturbative expansion of the lifetime $\tau_{\rm core}$\,\cite{jia2016using}. For trRIXS, one can also correct for the finite lifetime effects by using an overall scaling factor determined by the equilibrium RIXS intensity and $S(q,\omega)$, both of which can be independently measured. As discussed in Sec.~II of the SI, this correction provides a good approximation of the long-term dynamics even for a large core-hole lifetime.

Apart from accurately describing the wavefunction entanglement encoded in the pump-probe spectrum, the self-consistent Eq.~\eqref{eq:qfi_tay} also provides an alternative way to interpret the light-induced spin fluctuation dynamics. We note that the leading order $m=1$ term in the series of Eq.~\eqref{eq:qfi_tay} is the second-order time derivative of $f_\mathrm{Q}$, which reflects an inertia of the underlying wavefunction toward the change. This inertia enables self-driving of the spin fluctuations and explains why the QFI in Fig.~\ref{fig:lifetime} continues to oscillate for $t \gtrsim 5 t_\mathrm{h}^{-1}$ even when the pump field vanishes. In contrast, trRIXS spectra for non-interacting fermions (and accordingly any quantities extracted from them) completely recover to the initial equilibrium spectra after the pump is gone\,\cite{chen2019theory, chen2020observing}. Therefore, the self-driving wavefunctions are a unique feature of correlated systems with interactions and including these high-order time-derivatives in Eq.~\eqref{eq:qfi_tay} is crucial to correctly capture the nonequilibrium QFI dynamics. The ``snapshot QFI'' $\bar{f}_\mathrm{Q}(q, t)$, on the other hand, serves as a good approximation for the $f_\mathrm{Q}$ at the pump arrival, but starts to deviate at later time delays where the pump tails off. 

Following the pump pulse, the trRIXS spectral weight at large momenta and high energies is transferred into small momenta [see Fig.~\ref{fig:RIXSMomentum}]. This transfer suggests the onset of long-wavelength spin fluctuations, although the overall magnetic moment decreases with the generation of doublon-hole fluctuations. While we focus on the smallest momentum $q=\pi/6$ of the cluster in this work, the efficiency of the self-consistent approach is not restricted to any specific momentum. As shown in Supplemental Note 3, the evaluated QFIs accurately match those calculated through the wavefunction evolution. Therefore, the time-dependent QFIs for different momenta witness the transfer of entanglement at different length scales.

\subsection{Light-Enhanced Entanglement}\label{sec:entanglement}

A reliable extraction of time- and momentum-resolved QFI $f_\mathrm{Q}(q, t)$ allows us to witness the entanglement depth of the driven EHM. As originally investigated in pure spin systems\,\cite{hyllus2012fisher, toth2012multipartite, strobel2014fisher}, the QFI informs us about the presence of an entangled many-body state if its value exceeds a minimum value derived from the quantum Cramér-Rao bound. To witness entanglement dynamics in the spin sector, we determine a quantum bound suited to diagnose multipartite entanglement after obtaining the time-dependent QFI. Since the doped fermionic model has a local magnetic moment $\langle m_z^2\rangle <1$, we normalize the QFI bound by reducing the total spin $S$ by the doping concentration. This implies that, for a $k$-producible state, the QFI obtained by a RIXS spectrum is bounded by
\begin{equation}\label{eq:qfi_ineq}
    {f_\mathrm{Q}(q, t)} \leq 4kn^2S^2\,
\end{equation}
where $n = \langle\sum_{i\sigma} n_{i\sigma}\rangle/N$ is the average electron density per site and $S=1/2$ in the single-band system. Here, we neglected the reduction of the normalization factor resulting from double occupation (which was discussed in Refs.~\,\onlinecite{lorenzana2005sum, laurell2022magnetic}) since the latter is minimal in our quarter-filled system [see Supplementary Note 3]. It follows that the upper bound of $f_Q$ for a 1-producible state ($k=1$) is 0.25, as the average density is $n=0.5$ at quarter filling, and any value of $f_Q>0.25$ signals the presence of at least bipartite entanglement (i.e., $k\geq 2$).

\begin{figure}[!t]
	\includegraphics[width=8.5cm]{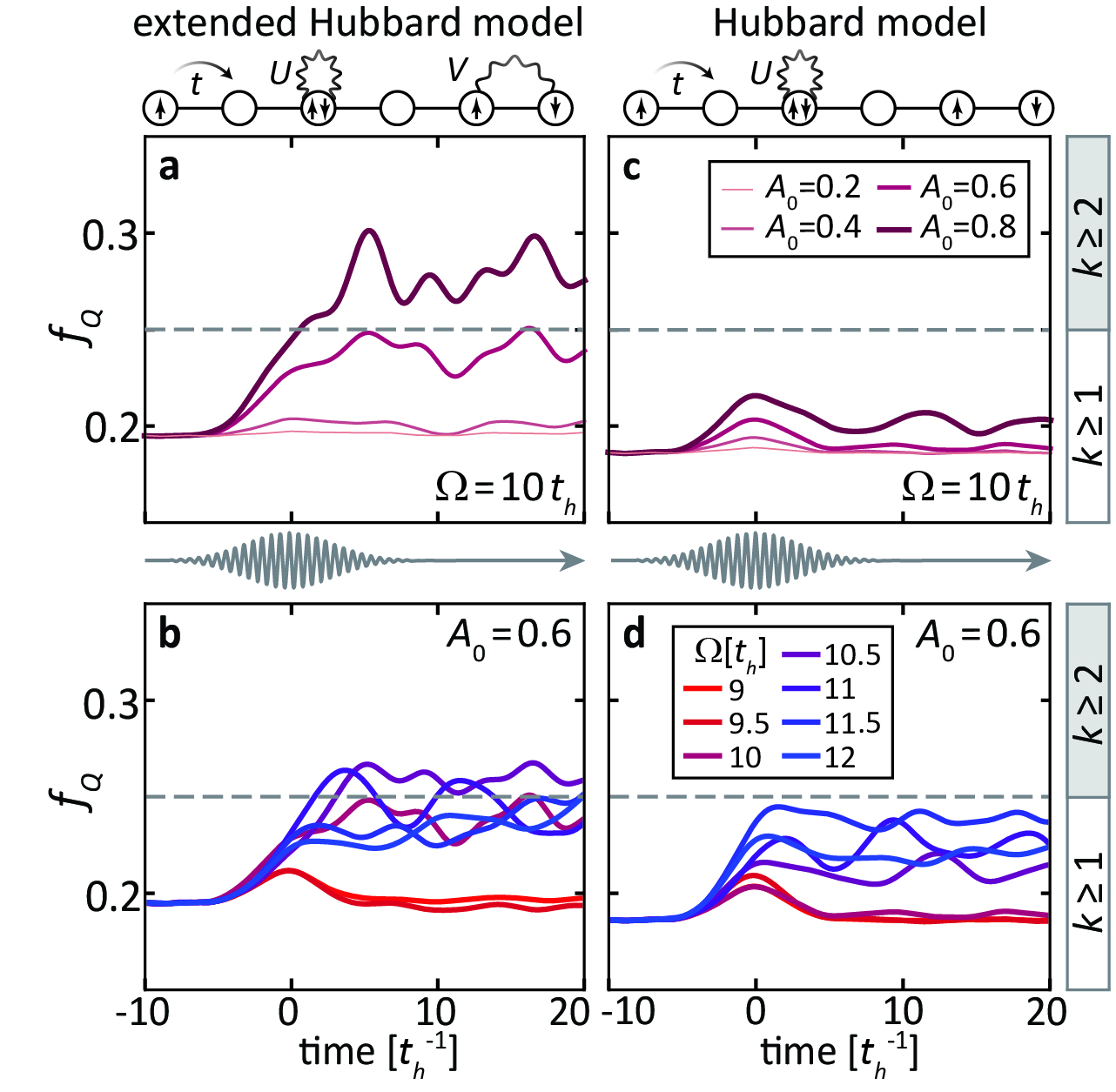}\vspace{-4mm}
	\caption{\textbf{Light-manipulation of the quantum Fisher information (QFI) for various pump conditions and models.} \textbf{a},\textbf{b} The transient QFI density $f_Q$ in a light-driven EHM [i.e., Eq.~\eqref{eq:EHM} with $V = -t_\mathrm{h}$] for selected \textbf{a} pump amplitudes at fixed energy $\Omega=10t_h$ and \textbf{b} pump energies at fixed amplitude $A_0=0.6$. \textbf{c},\textbf{d} Same as \textbf{a},\textbf{b} but for a Hubbard model without nonlocal interactions [i.e., $V = -t_\mathrm{h}$]. The bar to the right denotes the range and boundary for multiparticle entanglement with a $k$-producible state. The upper insets are cartoons of the Hamiltonian terms contained in each model, while the middle insets show the time evolution of the pump vector potential. }\label{fig:entanglement}
\end{figure}

As shown in Fig.~\ref{fig:entanglement}{a}, the QFI $f_Q$ increases following the excitation with a $\Omega=10t_h$ laser pulse. As the pump amplitude increases, $f_Q$ is monotonically pushed towards the boundary and exceeds 0.25 for $A>0.6$ (i.e., a 50\% enhancement), signaling that the pump-induced state is at least bipartite entangled. Note that time evolution is nonmonotonic, indicating the presence of oscillations of the many-body states which are not governed by thermalization. (This has been demonstrated by comparing nonequilibrium and finite-temperature spectra in Ref.~\onlinecite{wang2017producing}.)  	
We now investigate the pump frequency dependence of the light-enhanced entanglement. As shown in Fig.~\ref{fig:entanglement}{b}, the light-induced changes are weak for $\Omega < 10 t_h$. Since the particle-hole excitations in our EHM cost $U-2V=10t_h$, low-frequency pump pulses do not have enough energy to create doublon-hole pairs and scramble the spin configurations. Thus, the transient increase of $f_Q$, which disappears when the pump is over, can be attributed to Floquet engineering of the spin exchange interaction via virtual processes\,\cite{mentink2015ultrafast}. The maximal enhancement is achieved for $\Omega \sim 10 - 11t_h$, resonant with the doublon-holon excitation energy. In this case, long-wavelength spin fluctuations are generated through the creation and annihilation of doublon-hole pairs, which contribute to a more entangled many-body state. This process is slightly reduced when the pump photon energy is higher than the doublon-holon excitation energy.	
We argue that the light-enhanced entanglement likely reflects the proximity to a quantum phase transition. Although our trRIXS simulation is restricted to a small cluster and cannot rigorously determine phase boundaries, we compare the QFI dynamics under the same excitation conditions in the EHM and in a doped Hubbard model (without the attractive $V$). While the two Hamiltonians only differ by a $V\sim -t_h$ term, much smaller than the dominant on-site interaction $U=8t_h$, previous studies have shown the presence of a triplet superconducting phase for moderate attractive near-neighbor interactions\,\cite{lin1986condensation,lin1995phase,lin1997phase,xiang2019doping, shinjo2019machine}. This phase transition boundary has been explored through exact numerical methods in the thermodynamic limit, suggesting that, for $U=8t_h$, triplet superconductivity develops for $-1.7t_h\lesssim V\lesssim -1.1t_h$\,\cite{qu2021spin}. Therefore, the EHM with $V= -t_h$ is expected to display a stronger light-induced entanglement, compared with a pure Hubbard model (without $V$), due to the proximity to this phase boundary. As shown in Figs.~\ref{fig:entanglement}{c} and {d}, the QFI of the driven Hubbard model is much smaller than that of the EHM at same pump conditions, consistent with the absence of such quantum phase transition.  In other words, the nearest-neighbor interaction, whose feature was recently identified in 1D cuprate chains but the phonon-mediated mechanism widely exists in transition-metal oxides, is crucial towards achieving light control of entanglement in correlated materials.	
 	
While here we mainly aim to witness spin-mode multipartite entanglement, one might also explore entanglement depth and quantum bounds associated with fermions. In Supplementary Note 5, we report an extension of our calculations to the single-particle fermionic modes of the driven EHM following the approach in Ref.~\onlinecite{de2021entanglement}, with the $U(1)\times U(1)$ symmetry. The obtained bounds are weaker than Eq.~\eqref{eq:qfi_ineq} and they could be constrained by additional fermionic symmetries and more complex fermionic operators [see discussion in Supplementary Note 5]. A comprehensive study of basis-independent entanglement witnesses for indistinguishable fermions is beyond the scope of this work.

\section{Discussion}\label{sec:summary}

In this work, we have connected the trRIXS spectrum of a light-driven material to the entanglement depth of its time-dependent wavefunction. Our calculations explicitly account for the experimental time resolution and finite core-hole lifetime. We have developed a self-consistent procedure to extract the time-dependent QFI from a realistic pump-probe spectrum. With the full information about the x-ray probe pulse, one can reproduce the ultrafast dynamics of equal-time observables beyond the limitations set by the time resolution. The core-hole lifetime is an intrinsic property of materials and introduces a quantitative but not qualitative deviation between the extracted QFI and the exact values, which can be corrected by using equilibrium spectroscopy data.

Through the time-dependent QFI, we determine a small-wavevector enhancement of multipartite entanglement in the driven state of the EHM, which becomes at least bipartite entangled after the pump arrival. This finding is in contrast with the time evolution of a simple Hubbard model with the same pump parameters, in which the QFI never exceeds the minimum bound for a separable state. We interpret this difference in terms of light-enhanced quantum fluctuations due to the EHM proximity with a phase transition boundary between Luttinger liquid and triplet superconductivity. The predicted enhanced entanglement depth could be measured via future trRIXS experiments on doped quasi-1D cuprates (e.g. Ba$_{2-x}$Sr$_x$CuO$_3$) and other low-dimensional correlated oxides. Due to the crucial role of the attractive interactions on the light-enhanced entanglement, these experiments will in turn help detecting nonlocal interactions, usually challenging to characterize in quantum materials. Additionally, since the nonequilibrium control of entanglement was recently discussed also in the context of spintronic devices\,\cite{suresh2022electron}, our self-consistent trRIXS approach could eventually be applied to probe time-dependent entanglement in non-optically-driven systems.

Beyond experimentally diagnosing transient entanglement dynamics in driven quantum materials, we anticipate the need for further theoretical developments. Here, we mainly discuss multipartite entanglement and quantum bounds in the spin basis. However, it is also important to characterize the entanglement depth of fermionic degrees of freedom via observables specific to indistinguishable particles. The intrinsic entanglement of a many-body state is a basis-independent property\,\cite{eckert2002quantum}, i.e., it is invariant under a unitary transformation over all modes on the particle creation/annihilation operators. In a few-body system, the Slater rank number or fermionic concurrence\,\cite{schliemann2001quantum,schliemann2001double, eckert2002quantum} can serve as basis-independent quantification of entanglement. In a many-body material system, it is impractical to measure these quantities, but one can construct different witnesses to set tight bounds among various entangled states and to fit observables accessible by solid-state measurements. A promising witness for indistinguishable fermions is operators sensitive to paired states\,\cite{kraus2009pairing} and constructed by at most two creation and two annihilation operators.

\section{Methods}
\subsection{Time-Resolved Resonant Inelastic X-Ray Scattering}\label{method:trRIXS}
The trRIXS is a photon-in-photon-out scattering process with a resonant intermediate state, whose cross-section reads as\,\cite{chen2019theory, wang2021x}
\begin{eqnarray}\label{eq:cross}
\mathcal{I}({q}, \wout, \win, t) = &&\frac1{2\pi N}\!\iiiint\!  dt_1dt_2 dt_1'dt_2'e^{i\win (t_2 - t_1) -i\wout(t_2' - t_1')} \nonumber \\
&& \times \, g(t_1; t)g(t_2; t) l(t_1' - t_1) l(t_2' - t_2) \\
&& \times \langle \hat{\mathcal{D}}^\dagger_{{q}_i{\epsilon}_i}(t_2) 
\hat{\mathcal{D}}_{{q}_s{\epsilon}_s}(t_2') 
\hat{\mathcal{D}}^\dagger_{{q}_s{\epsilon}_s}(t_1') 
\hat{\mathcal{D}}_{{q}_i{\epsilon}_i}(t_1) 
 \rangle \nonumber 
\end{eqnarray}
where $q=\qin-\qout$ ($\omega=\win-\wout$) is the momentum (energy) transfer between incident and scattered photons, and $l(t)\! =\!  e^{- t/\tau_{\rm core}}\theta(\tau)$ the core-hole decay lifetime. For a direct transition, the dipole operator reads as \begin{eqnarray}\label{eq:dipole}
\mathcal{D}_{\qbf\eps}=\sum_{\ibf\alpha\sigma}e^{-i\qbf\cdot \rbf_\ibf}(M_{\alpha\eps} c_{\ibf\sigma}^\dagger p_{\ibf\alpha\sigma}  + h.c.)\,,
\end{eqnarray}
where $c_{\ibf\sigma}^\dagger$ ($c_{\ibf\sigma}$) and $p_{\ibf\alpha\sigma}^\dagger$ ($p_{\ibf\alpha\sigma}$) denote the creation (annihilation) operators for valence and core-level electrons at site $i$ with spin $\sigma = \uparrow, \downarrow$. Since the transition-metal $L$-edge usually involves multiple core-level $p$ orbitals, we label them by $\alpha$, for which $M_{\alpha\eps}$ is the matrix element of the dipole transition between each core level and the valence band via an $\eps$-polarized photon. For a transition-metal $L$-edge trRIXS, the full derivation of the matrix elements is reported in Ref.~\onlinecite{wang2021x}. While these dipole transitions preserve the total spin, the pair of photon absorption and emission events, described by $\hat{\mathcal{D}}^\dagger_{{q}_s{\epsilon}_s}(t_1') \hat{\mathcal{D}}_{{q}_i{\epsilon}_i}(t_1) $ in Eq.~\eqref{eq:cross}, may flip a spin due to the spin-orbit coupling of the core levels. This spin-flip process is maximized for the $\pi-\sigma$ polarization and, therefore, provides a good estimate of the dynamical structure factor\,\cite{ament2009theoretical,Braicovich2010magnetic,Haverkort2010theory,jia2016using,Robarts2021dynamical}. Throughout this paper we exclusively employ such a polarization configuration.

\subsection{Extended Hubbard Model}
The extended Hubbard model Hamiltonian reads as
\begin{eqnarray}\label{eq:EHM}
{\mathcal{H}} &=& -t_\mathrm{h} \sum_{i\sigma} \left[c_{i\sigma}^\dagger c_{i+1, \sigma} + \mathrm{H.c} \right]  + \, U \sum_{i} n_{i\uparrow}n_{i\downarrow}\nonumber  \\
&& + V\sum_{ i,\sigma,\sigma^\prime} n_{i\sigma}n_{i+1,\sigma^\prime} + \mathcal{H}_{\rm core},
\end{eqnarray}
where $c_{i\sigma}$ ($c_{i\sigma}^\dagger$) annihilates (creates) a valence electron and $n_{i\sigma} = c_{i\sigma}^\dagger c_{i\sigma}$ is the number operator. The valence electrons form a single band with nearest-neighbor hopping amplitude $t_\mathrm{h}$, on-site Coulomb repulsion $U$, and nearest-neighbor interaction $V$. Although the formalism is general, we choose here model parameters capturing the physics of cuprate chain compounds such as Ba$_{2-x}$Sr$_x$CuO$_{3+\delta}$, namely $U=8t_{\rm h}$ and $V=-t_{\rm h}$\,\cite{chen2019theory, wang2021phonon}. The ground state of EHM with these parameters at quarter filling was suggested to reside in proximity to a spin-triplet superconducting state\,\cite{lin1986condensation,lin1995phase,lin1997phase,xiang2019doping, shinjo2019machine,qu2021spin}.

To account for the x-ray absorption and emission processes, the full Hamiltonian in Eq.~\eqref{eq:EHM} also contains terms involving the core holes
\begin{eqnarray}\label{eq:HubbardCore}
	\mathcal{H}_{\rm core} &=&  \sum_{i\alpha\sigma} E_{\rm edge}(1-n^{(p)}_{i\alpha\sigma}) - U_c \sum_{i\alpha\sigma\sigma^\prime} n_{i\sigma}(1-n^{(p)}_{i\alpha\sigma^\prime})\nonumber\\ &&+\lambda\sum_{i\alpha\alpha^\prime\atop \sigma\sigma^\prime}p_{i\alpha\sigma}^\dagger \chi_{\alpha\alpha^\prime}^{\sigma\sigma^\prime}p_{i\alpha^\prime\sigma^\prime}.
\end{eqnarray}
Here, $p_{i \alpha \sigma}$ ($p_{i \alpha \sigma}^\dagger$) annihilates (creates) a core-level electron with multiple degenerate orbitals labeled by $\alpha$, corresponding to the $2p_{x,y,z}$ orbitals in transition-metal $L$-edge RIXS, and $n^{(p)}_{i\alpha\sigma}= p_{i\alpha \sigma}^\dagger p_{i\alpha \sigma}$ is the core-level electronic number operator. The potential $U_c$ describes the attractive interaction between the core-hole holes and valence-level electrons and is fixed at 4$t_h$ for all the 2$p$ orbitals\,\cite{tsutsui2000resonant,jia2014persistent,jia2016using}. The edge energy $E_{\rm edge}$ is selected as 938\,eV to represent the Cu $L$-edge x-ray absorption and the spin-orbit coupling $\lambda$ of the core states is set to 13\,eV\,\cite{tsutsui2000resonant,kourtis2012exact}.

\section*{Data Availability}
The numerical data that support the findings of this study are available from the corresponding authors upon reasonable request. The data generated in this study are provided in the \href{https://figshare.com/articles/dataset/Witnessing_Entanglement_using_trRIXS/22761959}{figshare repository}

\section*{Code Availability}
The relevant scripts of this study to reporoduce all figures are available at the \href{https://figshare.com/articles/dataset/Witnessing_Entanglement_using_trRIXS/22761959}{figshare repository}. Other codes are available from the corresponding authors upon reasonable request.

\section*{Acknowledgements}
We acknowledge insightful discussions with R.C. de Almeida, S. Ding, P. Hauke, and P. Laurell. J.H. and Y.W. acknowledge support from U.S. Department of Energy, Office of Science, Basic Energy Sciences, under Early Career Award No.~DE-SC0022874. D.R.B. and M.M. are primarily supported by from U.S. Department of Energy, Office of Science, Basic Energy Sciences, under Early Career Award No.~DE-SC0022883. D.R.B. also acknowledges funding by the Swiss National Science Foundation through Project No. P400P2$\_$194343. T.L. and M.L. acknowledge the support from U.S. Department of Energy (DOE), Office of Science, Basic Energy Sciences (BES), award No.~DE-SC0021940. This research used resources of the National Energy Research Scientific Computing Center (NERSC), a U.S. Department of Energy Office of Science User Facility located at Lawrence Berkeley National Laboratory, operated under Contract No. DE-AC02-05CH11231 using NERSC award BES-ERCAP0020159.

\section*{Author contributions} 
Y.W. and M.M. conceived the project. J.H., U.B, D.R.B., and Y.W. performed the calculations and data analysis. All the authors contributed to the interpretation of the results and the manuscript writing.

\section*{Competing interests} The authors declare no competing interests.

\appendix

\section{Derivation of the Self-Consistent Equation}\label{app:derivation}
The nonequilibrium dynamic spin structure factor  $S({q}, \omega, t)$ defined in Eq.~(2) of the main text reads as
\begin{eqnarray}\label{eq:Sqw2}
S({q}, \omega, t) 
&= & 
 \frac{1}{4\pi^2 {\sigma_\mathrm{pr}^2} N}\iint\displaylimits_{-\infty}^{+\infty} d\tau d\bar{\tau}\,
 e^{-(\bar{\tau} - t)^2/\sigma_{\rm pr}^2}e^{-{\tau}^2/4\sigma_{\rm pr}^2}
 e^{i\omega \tau}\nonumber\\
&& \left\langle\hat{\rho}_{-q}^\mathrm{s}\left(\bar{\tau} + \frac{\tau}{2}\right)
 \hat{\rho}_{q}^\mathrm{s}\left(\bar{\tau} - \frac{\tau}{2}\right) \right\rangle, 
\end{eqnarray}
where we use the Wigner transformation of the time variables $\bar{\tau} = (t_1 + t_2)/2$ and $\tau = t_1 - t_2$. Here, the average notation is not restricted to a pure state, but can also be generalized to a thermal ensemble
\begin{eqnarray}
&&\left\langle\hat{O}_{1}\left(t_1\right)\hat{O}_{2}\left(t_2\right) \right\rangle \nonumber\\
&=& \textrm{Tr}\left[ \frac{e^{-\beta \Ham}}{\mathcal{Z}} \hat{U}(-\infty, t_1)\hat{O}_{1}\hat{U}(t_1,t_2)\hat{O}_{2} \hat{U}(t_2, -\infty)\right]\,.
\end{eqnarray}
The $\hat{U}(t_1,t_2)$ is the unitary time evolution operator, $\mathcal{Z}$ is the partition function of the equilibrium state (at $t=-\infty$), and $\beta$ is the inverse temperature.
To simplify the derivation, we define the momentum-space spin excitation operator 
\begin{eqnarray}\label{eq:rhos}
\hat{\rho}^\mathrm{s}_{{q}} =  \sum_i \hat{S}^z_i e^{iq\cdot r_i}\,.
\end{eqnarray}
Note that the $\hat{\rho}_{q}$ in Eq.~\eqref{eq:Sqw2} is written in the interaction picture as defined in the main text Sec.~II and evolves according to the driven EHM Hamiltonian after the Peierls substitution. Therefore, Eq.~\eqref{eq:Sqw2} does not obey time-translational invariance and the $\bar{\tau}$ cannot be separated. Integrating in $\omega$ the left- and right-hand sides of Eq.~\eqref{eq:Sqw2} and Taylor expanding $\hat{\rho}^s_q$ in $\tau$ lead to the identity
\begin{equation}\label{eq:sqw_qfi}
\int\displaylimits_{-\infty}^{+\infty} d\omega\, S({q}, \omega, t)  =  \frac{1}{2\pi N\sigma_{\mathrm{pr}}^2}\int\displaylimits_{-\infty}^{+\infty} e^{ -\frac{(t - \tau)^2}{\sigma_\mathrm{pr}^2}}\langle \hat{\rho}^\mathrm{s}_{-{q}}({\tau}) \hat{\rho}^\mathrm{s}_{{q}}({\tau}) \rangle\, d{\tau}.
\end{equation}
This equality states that the energy integral of the nonequilibrium dynamic spin structure factor at a given momentum $q$ is a convolution of the two-time spin correlation functions with the time-dependent profile of a probe with finite pulse duration. Note that the QFI $f_\mathrm{Q}({q}, t)$ is defined as $4\langle \hat{\rho}^\mathrm{s}_{-{q}}(t) \hat{\rho}^\mathrm{s}_{{q}}(t) \rangle /N$, which is an equal-time measurement for an instantaneous wavefunction at time $t$\,\cite{pezze2009entanglement,hyllus2012fisher,toth2012multipartite}. Therefore, without time-translational invariance in nonequilibrium systems, one cannot obtain the correct time evolution of the QFI by just integrating the dynamic spin structure factor.

However, the information about the time sequence of the QFI $\{f_Q(q,t) | -\infty < t< \infty\}$ is encoded in the entire time evolution of $S(q,\omega,t)$, although without snapshot-to-snapshot correspondence. Therefore, one can apply self-consistent iterations to deconvolve $f_Q$ from the integral equation. By introducing a change of variable $x = \tau - t$ we can Taylor expand $f_Q(\tau) = f_Q(t + x)$ around $t$ in Eq.~\eqref{eq:sqw_qfi} to get
\begin{equation}
\int\displaylimits_{-\infty}^{+\infty} d\omega S({q}, \omega, t) = \frac{1}{8\pi \sigma_\mathrm{pr}^2} \sum_{m=0}^{\infty} \frac{1}{m!}\frac{\partial^m f_\mathrm{Q}}{\partial t^m}\int\displaylimits_{-\infty}^{+\infty} e^{-x^2/\sigma_\mathrm{pr}^2} x^m dx.
\end{equation}
Noting that the integral on the right-hand side is nonvanishing only for even values of $m$, we can rewrite this equation in the form of Eq.~\ref{eq:qfi_tay} of the main text
\begin{equation}\label{eq:sc_qfi}
f_\mathrm{Q}({q}, t) = 8\sigma_\mathrm{pr} \sqrt{\pi} \int\displaylimits_{-\infty}^{+\infty} d\omega\, S( {q}, \omega, t) + \sum_{m=1}^\infty \frac{\mathcal{C}_m}{2m!} \frac{\partial^{2m}f_\mathrm{Q}}{\partial t^{2m}},
\end{equation}
where $\mathcal{C}_m=-(1/\sigma_\mathrm{pr}\sqrt{\pi}) \int_{-\infty}^{\infty}e^{-x^2/\sigma_\mathrm{pr}^2}x^{2m} dx = -(\sigma_\mathrm{pr}^{m - 1/2}/\sqrt{\pi}) \Gamma(m + 1/2)$. Since simulating correlated electrons using the Krylov-subspace requires very small time steps, $S(q,\omega,t)$ is evaluated over a fine time grid and enables the reliable calculation of high-order derivatives.

Eq.~\eqref{eq:sc_qfi} can be numerically solved by a self-consistent iteration scheme where we truncate the derivative series to some finite $m = M$ and start with the snapshot QFI as 
\begin{equation}
f_\mathrm{Q}^{(0)}({q}, t) = \bar{f}_\mathrm{Q}({q}, t) = 8\sigma_\mathrm{pr} \sqrt{\pi} \int d\omega\, S( {q}, \omega, t)\,.
\end{equation}
Then the $k$-step iteration depends on the lower orders as 
\begin{equation}
f_\mathrm{Q}^{(k)} ({q}, t) = 8\sigma_\mathrm{pr} \sqrt{\pi} \int\displaylimits_{-\infty}^{+\infty} d\omega\, S( {q}, \omega, t) + \sum_{m=1}^{M} \frac{\mathcal{C}_m}{2m!} \frac{\partial^{2m}f_\mathrm{Q}^{(k-1)}}{\partial t^{2m}},
\end{equation}
The convergence is reached by satisfying the criterion $|f^{(k)}_\mathrm{Q} - f^{(k-1)}_\mathrm{Q}|< \delta$, where $\delta$ is a small number. Once convergence is reached, time-dependent QFI is reasonably well approximated by $f_\mathrm{Q}({q}, t) \approx f^{(k)}_\mathrm{Q}({q}, t)$. In practice, we employ the Fourier method to solve this iterative equation, which does not require an artificial termination.

\section{Correction of trRIXS with Long Core-Hole Lifetime}\label{app:trRIXSCorrection}

\begin{figure}[!t]
\centering
	\includegraphics[width=8.5cm]{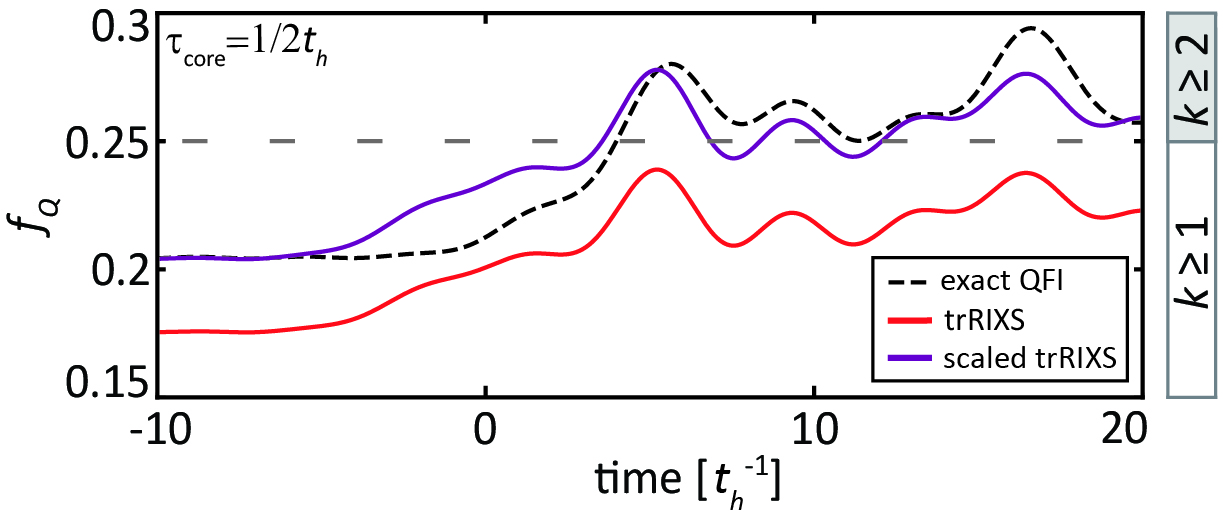}\vspace{-3mm}
	\caption{QFI dynamics evaluated using the simulated trRIXS with large core-hole lifetime ($\tau_{\rm core} = 1/2t_h$, red curve) and the corrected trRIXS spectra scaled by an overall factor of 1.163 (purple curve) determined by the ratio of equilibrium $S(q,\omega)$ and RIXS, compared with the exact QFI evolution obtained by instantaneous wavefunctions (black dashed curve) .}
	\label{fig:scaledtrRIXS}
\end{figure}

Due to the finite core-hole lifetime, RIXS does not reflect the exact $S(q,\omega)$. As shown in the Fig.~5 of the main text, trRIXS captures the oscillation of QFI for the wavefunction dynamics, but there is a finite offset between the values extracted from trRIXS and those evaluated exactly using the wavefunctions. It is important to note that this offset does not increase in time compared with that at equilibrium. This phenomenon indicates that finite lifetime effect, which causes the deviation between RIXS and $S(q,\omega)$, is insensitive to whether the initial state is equilibrium.

Therefore, we introduce a correction of the trRIXS intensity by a constant factor determined by comparing the equilibrium RIXS spectrum and the $S(q,\omega)$ (which can be measured with inelastic neutron scattering). As shown in Fig.~\ref{fig:scaledtrRIXS} for $\tau_{\rm core} = 1/2t_h$, the QFI extracted from equilibrium RIXS spectrum aligns with that from the $S(q,\omega)$ after scaling by an overall factor 1.163. After this rescaling, the QFI extracted from trRIXS is sufficiently accurate throughout the entire dynamics. The efficiency of this correction in turn demonstrates that trRIXS probes the spin excitations out of equilibrium with limited errors. This statement holds not only for the excitation energy, but also for the spectral intensity after accounting for this constant correction factor.

\section{Momentum Dependence and the Evolution of the Local Moment}\label{app:momentumDep}
\begin{figure}
\includegraphics[width=8.5cm]{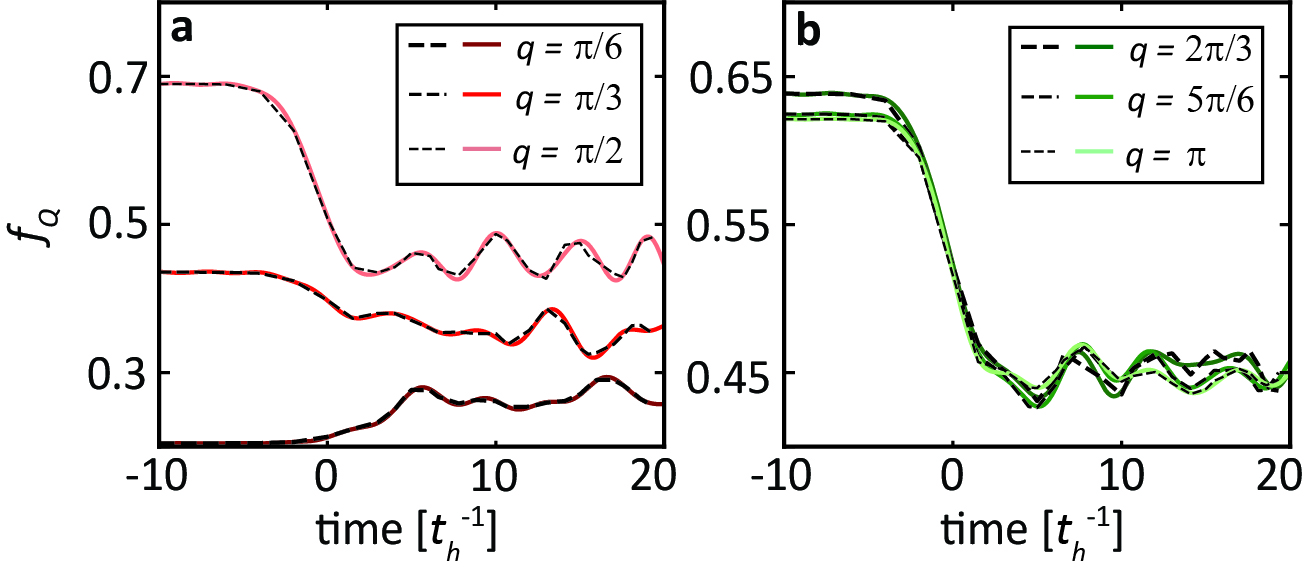}\vspace{-4mm}
	\caption{{QFI density for various momenta.} The colored curves represent $f_q$ evaluated from the trRIXS spectra using the self-consistent approach [i.e., main text Eq. (5)]. The corresponding dashed black lines represent the exact QFI density calculated using the instantaneous wavefunction in the main text Eq. (1).}
	\label{fig:momentumDependence}
\end{figure}

To further test the self-consistent approach, we apply Eq. (5) of the main text to all momenta. As reflected by the comparison between dashed and solid lines in Fig.~\ref{fig:momentumDependence}, the QFI evaluated from the trRIXS spectra and self-consistent iteration agrees well with that calculated using instantaneous wavefunctions, throughout the entire time evolution. This comparison verifies the general applicability of our approach discussed in Sec.~II.

\begin{figure}[!t]
	\includegraphics[width=8.5cm]{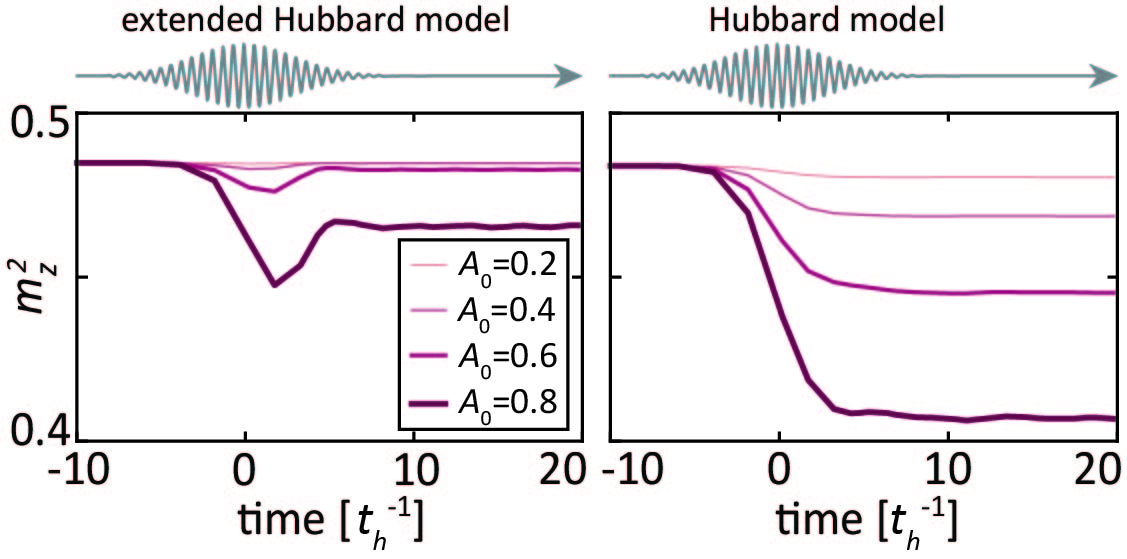}\vspace{-4mm}
	\caption{Time dependence of the local magnetic moment. Evolution of $\langle m_z^2\rangle$ in a light-driven EHM (left) and Hubbard model (right) for various pump conditions, as in main text Figs.~6\textbf{a},\textbf{b}. }
	\label{fig:mz2}
\end{figure}

The light-induced doublon-hole fluctuations are reflected by the time-dependent evolution of $\langle\psi(t)| m_z^2|\psi(t)\rangle$, defined as 
\begin{equation}
m_z^2 = \sum_i (n_{i\uparrow} - n_{i\downarrow})^2\,.
\end{equation}
As shown in Fig.~\ref{fig:mz2}, the equilibrium local moment at quarter filling is $\langle n\rangle \sim 0.5$. The computed value at negative time delays slightly differs from $0.5$ due to the presence of fluctuating double occupancies (0.0076 per site) in the quarter-filled system. At the pump arrival, the light-driven motion of charge carriers generates  additional doublons and holes and reduces the magnetic moments. Different from the Hubbard model ($V=0$), the magnetic moments are largely preserved in the EHM due to the presence of a nonlocal attractive interaction $V$, as shown in Fig.~\ref{fig:mz2}. Since this interaction favors adjacent singly-occupied states over doublons and holes, the local moment $\langle m_z^2\rangle$ quickly recovers after the pump for the EHM, in contrast to the persistent decrease observed in the Hubbard model.

This post-pump recovery of magnetic moments causes the transfer of spin fluctuations among different wavevectors. As shown in the black dashed curves of Fig.~\ref{fig:momentumDependence}, the compression of spin fluctuations at large momenta, inherent from the equilibrium Luttinger instability, is partially compensated by the enhancement at small ones ($q=\pi/6$ in this system) in the EHM. Such an enhancement is minor for the Hubbard model, as shown in the main text Fig.~6\textbf{c}. Therefore, the presence of the nonlocal interactions and the proximity to a phase boundary are crucial for the light-induced entanglement.

\section{Nonequilibrium Entanglement Entropy}\label{app:EE}
In this section, we compare the nonequilibrium QFI dynamics with the entanglement entropy, which has been widely employed as an entanglement measure in quantum information. We consider the 1D extended-Hubbard model with the same parameters as the main text. The R\'{e}nyi entropy for an instantaneous nonequilibrium state is defined as
\begin{equation}
\mathcal{S}_\mathrm{ent}(t; \alpha) = \frac{1}{1-\alpha}\ln\mathrm{Tr}_A\big[\mathrm{Tr}_{A^c}\ket{\psi(t)}\bra{\psi(t)}\big]^\alpha\,.
\end{equation}
Here we consider the case $\alpha=1$. The label $A$ refers to the label of the subsystem $A$, defined as half of the chain, and $A^c$ refers to the remaining subsystem.

\begin{figure}[!t]
\centering
	\includegraphics[width=8cm]{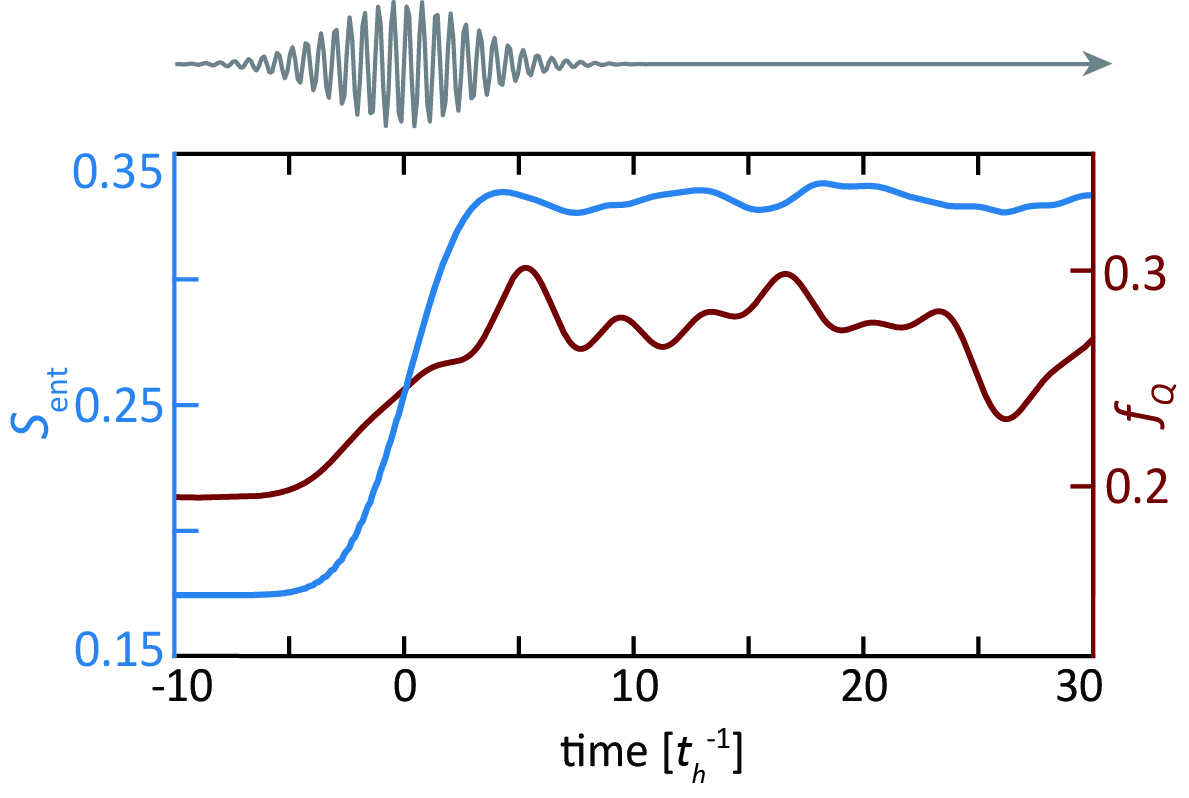}\vspace{-3mm}
	\caption{Comparison between transient entanglement entropy and QFI density for a light-driven extended-Hubbard model. Model parameters ($U=8t_h$ and $V=-t_h$) and pump conditions ($A_0 = 1$ and $\Omega = 10 t_h$) are the same as those in Fig.~4 of the main text.}	\label{fig:ent_entropy}
\end{figure}

In order to compare the time-dependent entanglement entropy with the QFI dynamics, we adopt the same pump condition as Fig.~4 of the main text, i.e.~$A_0=1$ and $\Omega=10t_h$. As shown in Fig.~\ref{fig:ent_entropy}, the nonequilibrium entanglement entropy increases at the same time of the light-driven QFI, further supporting the notion that the nonequilibrium state exhibits light-enhanced entanglement. It is, however, worth noting that, unlike the QFI, the entanglement entropy is not experimentally accessible in spectroscopic measurements of quantum materials.

\section{Upper Bound on the QFI based on the Single-Particle Fermionic Modes}\label{app:bound}
In this appendix, we follow Ref.~\onlinecite{de2021entanglement} and derive the QFI bounds based on single-particle fermionic modes (SPFM).
For the 1D extended Hubbard model of $N$ sites (with periodic boundary conditions) considered in the main text, the operator ${\rho}^\mathrm{s}_q$ in Eq.~\eqref{eq:rhos} obtained as a linear superposition of local operators $S_i^z$ is non-hermitian at all momenta (${\rho}^{\mathrm{s}\dagger}_{q}={\rho}^\mathrm{s}_{-q}$ except $q=0,\pi$). The corresponding Hermitian operator can be chosen as ${O}_q = ({\rho}_{q}^\mathrm{s} + {\rho}_{-q}^\mathrm{s})/2N$, such that
\begin{equation}\label{eq:odiag}
{O}_q =  \frac1{N}\sum_{j,\sigma} s_{\sigma} \cos(jq)\,
{c}_{j\sigma}^\dagger{c}_{j\sigma}.
\end{equation}
where $s_{\uparrow}=+, s_{\downarrow}=-$. The selection of local density operators in ${O}_q$ avoids ambiguities related to fermion anticommutation and sign. The fluctuation of this ${O}_q$ operator is related to the (instantaneous) QFI of the spin operator ${\rho}^\mathrm{s}_q$ through 
\begin{equation}\label{eq:op_ineq}
4\langle {O}_q^2 \rangle = \frac1N\langle\, ({\rho}^\mathrm{s}_{q})^2  + ({\rho}^\mathrm{s}_{-q})^2 \rangle + \frac{f_\mathrm{Q}(q, t)}{2}.
\end{equation}
For the simulated translational-symmetric system in Sec. IV, the time-dependent wavefunction $|\psi(t) \rangle$ has conserved momentum quantum number. This implies that $\langle ({\rho}^\mathrm{s}_{q})^2 \rangle \neq 0$ only when $q = 0,\pi$. At the same wavevectors $q = 0,\pi$, ${\rho}^\mathrm{s}_q$ is Hermitian, therefore Eq.~\eqref{eq:op_ineq} leads to
\begin{equation}\label{eq:rel}
f_\mathrm{Q}(q, t) =  \frac{8 \langle {O}_q^2 \rangle}{ (1 + \delta_{q0} + \delta_{q\pi})} .
\end{equation}

To estimate the upper bound on $f_\mathrm{Q}(q, t)$, we consider the upper bound on $\langle {O}^2_q \rangle$ as derived in Ref.~\onlinecite{de2021entanglement} for a fermionic many-body $k$-producible pure quantum state. For the creation (annihilation) operators ${c}_{n\sigma}^\dagger$ (${c}_{n\sigma}$), all SPFMs are spanned by $n = 1, 2, \cdots N$ and spin $\sigma = \uparrow, \downarrow$. If $P_m = [(n_1^m, \sigma_1^m), (n_2^m, \sigma_2^m), \cdots, (n_{N_m}^m, \sigma_{N_m}^m)]$ denotes a unique partition that contains $N_m$ SPFMs, then \mbox{$\mathcal{P} = P_1 \cup P_2 \cup \cdots \cup P_{N_\mathrm{parts}}$} (where $N_\mathrm{parts} \leq 2N$ is the number of partitions) forms the complete set of SPFMs.  With this notation, Ref.~\onlinecite{de2021entanglement} shows that if $|\Psi (t)\rangle_{k-\textrm{prod}}$ has a fixed number of electrons $N_e$, then the lower bound on  $\langle {O}_q^2\rangle = {}_{k-\textrm{prod}}\langle \Psi(t)| {O}_q^2 |\Psi(t)\rangle_{k-\textrm{prod}}$ can be determined by choosing a partition \mbox{$\bar{\mathcal{P}} = \bar{P}_1\cup \bar{P}_2\cup \cdots \bar{P}_{N_\mathrm{parts}} $} which maximizes the right-hand side of the following inequality
\begin{eqnarray}\label{eq:ineq}
 \langle {O}_q^2\rangle &\leq& \frac{1}{4}\sum_{m= 1}^{N_\mathrm{parts}} \left[ \sum_{\beta \in P_m^\mathrm{high}} A_\beta(q) - \sum_{\beta \in P_m^\mathrm{low}} A_\beta(q) \right]^2,
\end{eqnarray}
where $A_\beta(q) \in \{ \sigma \cos(q n) | \sigma = \pm1, n=1,2,\cdots,N\}$ such that $A_1(q) \geq A_2(q) \geq \cdots \geq A_{2N}(q)$ and $\bar{P}_{m} = P^\mathrm{high}_m \cup P^\mathrm{mid}_m\cup P^\mathrm{low}_m$ represents  a decomposition of the partition $\bar{P}_m$. The right-hand side in Eq.~\eqref{eq:ineq} can be numerically evaluated via the algorithm presented in Ref.~\onlinecite{de2021entanglement}, assuming conserved particle number $N_e$. Thus, the upper bound of QFI based on SPFM can be obtained by substituting Eq.~\eqref{eq:ineq} into Eq.~\eqref{eq:rel}, which is the fermionic bound mentioned at the end of Sec. V.   

We note that Eq.~\eqref{eq:ineq} does not automatically impose SU(2), translational, and point-group symmetries, which are challenging to implement in sizeable fermionic systems. Therefore, this bound is usually not strict for doped fermionic systems and for 2-producible SPFM states we find it to be $\sim 1.5$, i.e. much higher than the value obtained for the spin modes in the main text Fig. 6. Crucially, since the QFI bounds represent a sufficient but not necessary condition for the presence of entanglement, the overall scale of the QFI cannot exclude the presence of other highly-entangled states and the entanglement depth should be estimated via the highest available $k$ witnessed by different QFIs. A more accurate and basis-independent definition of entanglement in fermionic systems requires considering indistinguishable particles subject to anticommutation relations, and the superselection rule\,\cite{wiseman2003entanglement,banuls2007entanglement}. Entanglement measures satisfying these conditions, such as the the Slater rank\,\cite{schliemann2001quantum,eckert2002quantum}, usually require high-order correlation functions beyond the ones accessible with spectra in the form of the main text Eq.~(2). Therefore, distilling entanglement in the fermionic basis from spectroscopic measurements remains an open and interesting question.

\bibliography{references}

\begin{thebibliography}{109}%
\makeatletter
\providecommand \@ifxundefined [1]{%
 \@ifx{#1\undefined}
}%
\providecommand \@ifnum [1]{%
 \ifnum #1\expandafter \@firstoftwo
 \else \expandafter \@secondoftwo
 \fi
}%
\providecommand \@ifx [1]{%
 \ifx #1\expandafter \@firstoftwo
 \else \expandafter \@secondoftwo
 \fi
}%
\providecommand \natexlab [1]{#1}%
\providecommand \enquote  [1]{``#1''}%
\providecommand \bibnamefont  [1]{#1}%
\providecommand \bibfnamefont [1]{#1}%
\providecommand \citenamefont [1]{#1}%
\providecommand \href@noop [0]{\@secondoftwo}%
\providecommand \href [0]{\begingroup \@sanitize@url \@href}%
\providecommand \@href[1]{\@@startlink{#1}\@@href}%
\providecommand \@@href[1]{\endgroup#1\@@endlink}%
\providecommand \@sanitize@url [0]{\catcode `\\12\catcode `\$12\catcode
  `\&12\catcode `\#12\catcode `\^12\catcode `\_12\catcode `\%12\relax}%
\providecommand \@@startlink[1]{}%
\providecommand \@@endlink[0]{}%
\providecommand \url  [0]{\begingroup\@sanitize@url \@url }%
\providecommand \@url [1]{\endgroup\@href {#1}{\urlprefix }}%
\providecommand \urlprefix  [0]{URL }%
\providecommand \Eprint [0]{\href }%
\providecommand \doibase [0]{https://doi.org/}%
\providecommand \selectlanguage [0]{\@gobble}%
\providecommand \bibinfo  [0]{\@secondoftwo}%
\providecommand \bibfield  [0]{\@secondoftwo}%
\providecommand \translation [1]{[#1]}%
\providecommand \BibitemOpen [0]{}%
\providecommand \bibitemStop [0]{}%
\providecommand \bibitemNoStop [0]{.\EOS\space}%
\providecommand \EOS [0]{\spacefactor3000\relax}%
\providecommand \BibitemShut  [1]{\csname bibitem#1\endcsname}%
\let\auto@bib@innerbib\@empty
\bibitem [{\citenamefont {Keimer}\ and\ \citenamefont
  {Moore}(2017)}]{keimer2017physics}%
  \BibitemOpen
  \bibfield  {author} {\bibinfo {author} {\bibfnamefont {B.}~\bibnamefont
  {Keimer}}\ and\ \bibinfo {author} {\bibfnamefont {J.}~\bibnamefont {Moore}},\
  }\bibfield  {title} {\bibinfo {title} {\textit{The Physics of Quantum
  Materials}},\ }\href@noop {} {\bibfield  {journal} {\bibinfo  {journal} {Nat.
  Phys.}\ }\textbf {\bibinfo {volume} {13}},\ \bibinfo {pages} {1045} (\bibinfo
  {year} {2017})}\BibitemShut {NoStop}%
\bibitem [{\citenamefont {Matthias}\ \emph {et~al.}(1963)\citenamefont
  {Matthias}, \citenamefont {Geballe},\ and\ \citenamefont
  {Compton}}]{matthias1963superconductivity}%
  \BibitemOpen
  \bibfield  {author} {\bibinfo {author} {\bibfnamefont {B.~T.}\ \bibnamefont
  {Matthias}}, \bibinfo {author} {\bibfnamefont {T.~H.}\ \bibnamefont
  {Geballe}},\ and\ \bibinfo {author} {\bibfnamefont {V.~B.}\ \bibnamefont
  {Compton}},\ }\bibfield  {title} {\bibinfo {title}
  {\textit{Superconductivity}},\ }\href@noop {} {\bibfield  {journal} {\bibinfo
   {journal} {Rev. Mod. Phys.}\ }\textbf {\bibinfo {volume} {35}},\ \bibinfo
  {pages} {1} (\bibinfo {year} {1963})}\BibitemShut {NoStop}%
\bibitem [{\citenamefont {Wen}(2017)}]{wen2017colloquium}%
  \BibitemOpen
  \bibfield  {author} {\bibinfo {author} {\bibfnamefont {X.-G.}\ \bibnamefont
  {Wen}},\ }\bibfield  {title} {\bibinfo {title} {\textit{Colloquium: Zoo of
  Quantum-Topological Phases of Matter}},\ }\href@noop {} {\bibfield  {journal}
  {\bibinfo  {journal} {Rev. Mod. Phys.}\ }\textbf {\bibinfo {volume} {89}},\
  \bibinfo {pages} {041004} (\bibinfo {year} {2017})}\BibitemShut {NoStop}%
\bibitem [{\citenamefont {Balents}(2010)}]{balents2010spin}%
  \BibitemOpen
  \bibfield  {author} {\bibinfo {author} {\bibfnamefont {L.}~\bibnamefont
  {Balents}},\ }\bibfield  {title} {\bibinfo {title} {\textit{Spin Liquids in
  Frustrated Magnets}},\ }\href@noop {} {\bibfield  {journal} {\bibinfo
  {journal} {Nature}\ }\textbf {\bibinfo {volume} {464}},\ \bibinfo {pages}
  {199} (\bibinfo {year} {2010})}\BibitemShut {NoStop}%
\bibitem [{\citenamefont {Broholm}\ \emph {et~al.}(2020)\citenamefont
  {Broholm}, \citenamefont {Cava}, \citenamefont {Kivelson}, \citenamefont
  {Nocera}, \citenamefont {Norman},\ and\ \citenamefont
  {Senthil}}]{broholm2020quantum}%
  \BibitemOpen
  \bibfield  {author} {\bibinfo {author} {\bibfnamefont {C.}~\bibnamefont
  {Broholm}}, \bibinfo {author} {\bibfnamefont {R.}~\bibnamefont {Cava}},
  \bibinfo {author} {\bibfnamefont {S.}~\bibnamefont {Kivelson}}, \bibinfo
  {author} {\bibfnamefont {D.}~\bibnamefont {Nocera}}, \bibinfo {author}
  {\bibfnamefont {M.}~\bibnamefont {Norman}},\ and\ \bibinfo {author}
  {\bibfnamefont {T.}~\bibnamefont {Senthil}},\ }\bibfield  {title} {\bibinfo
  {title} {\textit{Quantum Spin Liquids}},\ }\href@noop {} {\bibfield
  {journal} {\bibinfo  {journal} {Science}\ }\textbf {\bibinfo {volume}
  {367}},\ \bibinfo {pages} {eaay0668} (\bibinfo {year} {2020})}\BibitemShut
  {NoStop}%
\bibitem [{\citenamefont {Amico}\ \emph {et~al.}(2008)\citenamefont {Amico},
  \citenamefont {Fazio}, \citenamefont {Osterloh},\ and\ \citenamefont
  {Vedral}}]{amico2008entanglement}%
  \BibitemOpen
  \bibfield  {author} {\bibinfo {author} {\bibfnamefont {L.}~\bibnamefont
  {Amico}}, \bibinfo {author} {\bibfnamefont {R.}~\bibnamefont {Fazio}},
  \bibinfo {author} {\bibfnamefont {A.}~\bibnamefont {Osterloh}},\ and\
  \bibinfo {author} {\bibfnamefont {V.}~\bibnamefont {Vedral}},\ }\bibfield
  {title} {\bibinfo {title} {\textit{Entanglement in Many-body Systems}},\
  }\href@noop {} {\bibfield  {journal} {\bibinfo  {journal} {Rev. Mod. Phys.}\
  }\textbf {\bibinfo {volume} {80}},\ \bibinfo {pages} {517} (\bibinfo {year}
  {2008})}\BibitemShut {NoStop}%
\bibitem [{\citenamefont {Horodecki}\ \emph {et~al.}(2009)\citenamefont
  {Horodecki}, \citenamefont {Horodecki}, \citenamefont {Horodecki},\ and\
  \citenamefont {Horodecki}}]{horodecki2009quantum}%
  \BibitemOpen
  \bibfield  {author} {\bibinfo {author} {\bibfnamefont {R.}~\bibnamefont
  {Horodecki}}, \bibinfo {author} {\bibfnamefont {P.}~\bibnamefont
  {Horodecki}}, \bibinfo {author} {\bibfnamefont {M.}~\bibnamefont
  {Horodecki}},\ and\ \bibinfo {author} {\bibfnamefont {K.}~\bibnamefont
  {Horodecki}},\ }\bibfield  {title} {\bibinfo {title} {\textit{Quantum
  Entanglement}},\ }\href@noop {} {\bibfield  {journal} {\bibinfo  {journal}
  {Rev. Mod. Phys.}\ }\textbf {\bibinfo {volume} {81}},\ \bibinfo {pages} {865}
  (\bibinfo {year} {2009})}\BibitemShut {NoStop}%
\bibitem [{\citenamefont {Imada}\ \emph {et~al.}(1998)\citenamefont {Imada},
  \citenamefont {Fujimori},\ and\ \citenamefont {Tokura}}]{imada1998metal}%
  \BibitemOpen
  \bibfield  {author} {\bibinfo {author} {\bibfnamefont {M.}~\bibnamefont
  {Imada}}, \bibinfo {author} {\bibfnamefont {A.}~\bibnamefont {Fujimori}},\
  and\ \bibinfo {author} {\bibfnamefont {Y.}~\bibnamefont {Tokura}},\
  }\bibfield  {title} {\bibinfo {title} {\textit{Metal-Insulator
  Transitions}},\ }\href@noop {} {\bibfield  {journal} {\bibinfo  {journal}
  {Rev. Mod. Phys.}\ }\textbf {\bibinfo {volume} {70}},\ \bibinfo {pages}
  {1039} (\bibinfo {year} {1998})}\BibitemShut {NoStop}%
\bibitem [{\citenamefont {Kondo}\ \emph {et~al.}(2015)\citenamefont {Kondo},
  \citenamefont {Malaeb}, \citenamefont {Ishida}, \citenamefont {Sasagawa},
  \citenamefont {Sakamoto}, \citenamefont {Takeuchi}, \citenamefont {Tohyama},\
  and\ \citenamefont {Shin}}]{kondo2015point}%
  \BibitemOpen
  \bibfield  {author} {\bibinfo {author} {\bibfnamefont {T.}~\bibnamefont
  {Kondo}}, \bibinfo {author} {\bibfnamefont {W.}~\bibnamefont {Malaeb}},
  \bibinfo {author} {\bibfnamefont {Y.}~\bibnamefont {Ishida}}, \bibinfo
  {author} {\bibfnamefont {T.}~\bibnamefont {Sasagawa}}, \bibinfo {author}
  {\bibfnamefont {H.}~\bibnamefont {Sakamoto}}, \bibinfo {author}
  {\bibfnamefont {T.}~\bibnamefont {Takeuchi}}, \bibinfo {author}
  {\bibfnamefont {T.}~\bibnamefont {Tohyama}},\ and\ \bibinfo {author}
  {\bibfnamefont {S.}~\bibnamefont {Shin}},\ }\bibfield  {title} {\bibinfo
  {title} {\textit{Point Nodes Persisting Far Beyond T$_c$ in Bi2212}},\
  }\href@noop {} {\bibfield  {journal} {\bibinfo  {journal} {Nat. Commun.}\
  }\textbf {\bibinfo {volume} {6}},\ \bibinfo {pages} {7699} (\bibinfo {year}
  {2015})}\BibitemShut {NoStop}%
\bibitem [{\citenamefont {Faeth}\ \emph {et~al.}(2021)\citenamefont {Faeth},
  \citenamefont {Yang}, \citenamefont {Kawasaki}, \citenamefont {Nelson},
  \citenamefont {Mishra}, \citenamefont {Chen}, \citenamefont {Schlom},\ and\
  \citenamefont {Shen}}]{faeth2020incoherent}%
  \BibitemOpen
  \bibfield  {author} {\bibinfo {author} {\bibfnamefont {B.~D.}\ \bibnamefont
  {Faeth}}, \bibinfo {author} {\bibfnamefont {S.}~\bibnamefont {Yang}},
  \bibinfo {author} {\bibfnamefont {J.~K.}\ \bibnamefont {Kawasaki}}, \bibinfo
  {author} {\bibfnamefont {J.~N.}\ \bibnamefont {Nelson}}, \bibinfo {author}
  {\bibfnamefont {P.}~\bibnamefont {Mishra}}, \bibinfo {author} {\bibfnamefont
  {L.}~\bibnamefont {Chen}}, \bibinfo {author} {\bibfnamefont {D.~G.}\
  \bibnamefont {Schlom}},\ and\ \bibinfo {author} {\bibfnamefont {K.~M.}\
  \bibnamefont {Shen}},\ }\bibfield  {title} {\bibinfo {title}
  {\textit{Incoherent Cooper Pairing and Pseudogap Behavior in Single-Layer
  FeSe/SrTiO$_3$}},\ }\href@noop {} {\bibfield  {journal} {\bibinfo  {journal}
  {Phys. Rev. X}\ }\textbf {\bibinfo {volume} {11}},\ \bibinfo {pages} {021054}
  (\bibinfo {year} {2021})}\BibitemShut {NoStop}%
\bibitem [{\citenamefont {Xu}\ \emph {et~al.}(2021)\citenamefont {Xu},
  \citenamefont {Rong}, \citenamefont {Wang}, \citenamefont {Wu}, \citenamefont
  {Hu}, \citenamefont {Cai}, \citenamefont {Gao}, \citenamefont {Yan},
  \citenamefont {Li}, \citenamefont {Yin}, \citenamefont {Chen}, \citenamefont
  {Huang}, \citenamefont {Zhu}, \citenamefont {Huang}, \citenamefont {Liu},
  \citenamefont {Xu}, \citenamefont {Lin},\ and\ \citenamefont
  {Zhou}}]{xu2020spectroscopic}%
  \BibitemOpen
  \bibfield  {author} {\bibinfo {author} {\bibfnamefont {Y.}~\bibnamefont
  {Xu}}, \bibinfo {author} {\bibfnamefont {H.}~\bibnamefont {Rong}}, \bibinfo
  {author} {\bibfnamefont {Q.}~\bibnamefont {Wang}}, \bibinfo {author}
  {\bibfnamefont {D.}~\bibnamefont {Wu}}, \bibinfo {author} {\bibfnamefont
  {Y.}~\bibnamefont {Hu}}, \bibinfo {author} {\bibfnamefont {Y.}~\bibnamefont
  {Cai}}, \bibinfo {author} {\bibfnamefont {Q.}~\bibnamefont {Gao}}, \bibinfo
  {author} {\bibfnamefont {H.}~\bibnamefont {Yan}}, \bibinfo {author}
  {\bibfnamefont {C.}~\bibnamefont {Li}}, \bibinfo {author} {\bibfnamefont
  {C.}~\bibnamefont {Yin}}, \bibinfo {author} {\bibfnamefont {H.}~\bibnamefont
  {Chen}}, \bibinfo {author} {\bibfnamefont {J.}~\bibnamefont {Huang}},
  \bibinfo {author} {\bibfnamefont {Z.}~\bibnamefont {Zhu}}, \bibinfo {author}
  {\bibfnamefont {Y.}~\bibnamefont {Huang}}, \bibinfo {author} {\bibfnamefont
  {G.}~\bibnamefont {Liu}}, \bibinfo {author} {\bibfnamefont {Z.}~\bibnamefont
  {Xu}}, \bibinfo {author} {\bibfnamefont {Z.}~\bibnamefont {Lin}},\ and\
  \bibinfo {author} {\bibfnamefont {X.~J.}\ \bibnamefont {Zhou}},\ }\bibfield
  {title} {\bibinfo {title} {\textit{Spectroscopic Evidence of
  Superconductivity Pairing at 83 K in Single-Layer FeSe/SrTiO$_3$ Films}},\
  }\href@noop {} {\bibfield  {journal} {\bibinfo  {journal} {Nat. Commun.}\
  }\textbf {\bibinfo {volume} {12}},\ \bibinfo {pages} {2840} (\bibinfo {year}
  {2021})}\BibitemShut {NoStop}%
\bibitem [{\citenamefont {He}\ \emph {et~al.}(2021)\citenamefont {He},
  \citenamefont {Chen}, \citenamefont {Li}, \citenamefont {Zhao}, \citenamefont
  {Song}, \citenamefont {Yoshida}, \citenamefont {Eisaki}, \citenamefont {Wu},
  \citenamefont {Chen}, \citenamefont {Lu}, \citenamefont {Meingast},
  \citenamefont {Devereaux}, \citenamefont {Birgeneau}, \citenamefont
  {Hashimoto}, \citenamefont {Lee},\ and\ \citenamefont
  {Shen}}]{he2021superconducting}%
  \BibitemOpen
  \bibfield  {author} {\bibinfo {author} {\bibfnamefont {Y.}~\bibnamefont
  {He}}, \bibinfo {author} {\bibfnamefont {S.-D.}\ \bibnamefont {Chen}},
  \bibinfo {author} {\bibfnamefont {Z.-X.}\ \bibnamefont {Li}}, \bibinfo
  {author} {\bibfnamefont {D.}~\bibnamefont {Zhao}}, \bibinfo {author}
  {\bibfnamefont {D.}~\bibnamefont {Song}}, \bibinfo {author} {\bibfnamefont
  {Y.}~\bibnamefont {Yoshida}}, \bibinfo {author} {\bibfnamefont
  {H.}~\bibnamefont {Eisaki}}, \bibinfo {author} {\bibfnamefont
  {T.}~\bibnamefont {Wu}}, \bibinfo {author} {\bibfnamefont {X.-H.}\
  \bibnamefont {Chen}}, \bibinfo {author} {\bibfnamefont {D.-H.}\ \bibnamefont
  {Lu}}, \bibinfo {author} {\bibfnamefont {C.}~\bibnamefont {Meingast}},
  \bibinfo {author} {\bibfnamefont {T.~P.}\ \bibnamefont {Devereaux}}, \bibinfo
  {author} {\bibfnamefont {R.~J.}\ \bibnamefont {Birgeneau}}, \bibinfo {author}
  {\bibfnamefont {M.}~\bibnamefont {Hashimoto}}, \bibinfo {author}
  {\bibfnamefont {D.-H.}\ \bibnamefont {Lee}},\ and\ \bibinfo {author}
  {\bibfnamefont {Z.-X.}\ \bibnamefont {Shen}},\ }\bibfield  {title} {\bibinfo
  {title} {\textit{Superconducting Fluctuations in Overdoped
  $Bi_2Sr_2CaCu_2O_{8+\delta}$}},\ }\href@noop {} {\bibfield  {journal}
  {\bibinfo  {journal} {Phys. Rev. X}\ }\textbf {\bibinfo {volume} {11}},\
  \bibinfo {pages} {031068} (\bibinfo {year} {2021})}\BibitemShut {NoStop}%
\bibitem [{\citenamefont {Chen}\ \emph {et~al.}(2022)\citenamefont {Chen},
  \citenamefont {Chen}, \citenamefont {Tang}, \citenamefont {Li}, \citenamefont
  {Wang}, \citenamefont {Ding}, \citenamefont {Jozwiak}, \citenamefont
  {Bostwick}, \citenamefont {Rotenberg}, \citenamefont {Hashimoto},
  \citenamefont {Lu}, \citenamefont {P.C.~Ruff}, \citenamefont {Louie},
  \citenamefont {Birgenau}, \citenamefont {Chen}, \citenamefont {Wang},\ and\
  \citenamefont {He}}]{chen2022lattice}%
  \BibitemOpen
  \bibfield  {author} {\bibinfo {author} {\bibfnamefont {C.}~\bibnamefont
  {Chen}}, \bibinfo {author} {\bibfnamefont {X.}~\bibnamefont {Chen}}, \bibinfo
  {author} {\bibfnamefont {W.}~\bibnamefont {Tang}}, \bibinfo {author}
  {\bibfnamefont {Z.}~\bibnamefont {Li}}, \bibinfo {author} {\bibfnamefont
  {S.}~\bibnamefont {Wang}}, \bibinfo {author} {\bibfnamefont {S.}~\bibnamefont
  {Ding}}, \bibinfo {author} {\bibfnamefont {C.}~\bibnamefont {Jozwiak}},
  \bibinfo {author} {\bibfnamefont {A.}~\bibnamefont {Bostwick}}, \bibinfo
  {author} {\bibfnamefont {E.}~\bibnamefont {Rotenberg}}, \bibinfo {author}
  {\bibfnamefont {M.}~\bibnamefont {Hashimoto}}, \bibinfo {author}
  {\bibfnamefont {D.}~\bibnamefont {Lu}}, \bibinfo {author} {\bibfnamefont
  {J.}~\bibnamefont {P.C.~Ruff}}, \bibinfo {author} {\bibfnamefont {S.~G.}\
  \bibnamefont {Louie}}, \bibinfo {author} {\bibfnamefont {R.}~\bibnamefont
  {Birgenau}}, \bibinfo {author} {\bibfnamefont {Y.}~\bibnamefont {Chen}},
  \bibinfo {author} {\bibfnamefont {Y.}~\bibnamefont {Wang}},\ and\ \bibinfo
  {author} {\bibfnamefont {Y.}~\bibnamefont {He}},\ }\bibfield  {title}
  {\bibinfo {title} {\textit{Lattice Fluctuation Induced Pseudogap in
  Quasi-one-dimensional Ta$_2$NiSe$_5$}},\ }\href@noop {} {\bibfield  {journal}
  {\bibinfo  {journal} {arXiv:2203.06817}\ } (\bibinfo {year}
  {2022})}\BibitemShut {NoStop}%
\bibitem [{\citenamefont {Vedral}(2008)}]{vedral2008quantifying}%
  \BibitemOpen
  \bibfield  {author} {\bibinfo {author} {\bibfnamefont {V.}~\bibnamefont
  {Vedral}},\ }\bibfield  {title} {\bibinfo {title} {\textit{Quantifying
  Entanglement in Macroscopic Systems}},\ }\href@noop {} {\bibfield  {journal}
  {\bibinfo  {journal} {Nature}\ }\textbf {\bibinfo {volume} {453}},\ \bibinfo
  {pages} {1004} (\bibinfo {year} {2008})}\BibitemShut {NoStop}%
\bibitem [{\citenamefont {G{\"u}hne}\ and\ \citenamefont
  {T{\'o}th}(2009)}]{guhne2009entanglement}%
  \BibitemOpen
  \bibfield  {author} {\bibinfo {author} {\bibfnamefont {O.}~\bibnamefont
  {G{\"u}hne}}\ and\ \bibinfo {author} {\bibfnamefont {G.}~\bibnamefont
  {T{\'o}th}},\ }\bibfield  {title} {\bibinfo {title} {\textit{Entanglement
  Detection}},\ }\href@noop {} {\bibfield  {journal} {\bibinfo  {journal}
  {Phys. Rep.}\ }\textbf {\bibinfo {volume} {474}},\ \bibinfo {pages} {1}
  (\bibinfo {year} {2009})}\BibitemShut {NoStop}%
\bibitem [{\citenamefont {de~Leon}\ \emph {et~al.}(2021)\citenamefont
  {de~Leon}, \citenamefont {Itoh}, \citenamefont {Kim}, \citenamefont {Mehta},
  \citenamefont {Northup}, \citenamefont {Paik}, \citenamefont {Palmer},
  \citenamefont {Samarth}, \citenamefont {Sangtawesin},\ and\ \citenamefont
  {Steuerman}}]{de2021materials}%
  \BibitemOpen
  \bibfield  {author} {\bibinfo {author} {\bibfnamefont {N.~P.}\ \bibnamefont
  {de~Leon}}, \bibinfo {author} {\bibfnamefont {K.~M.}\ \bibnamefont {Itoh}},
  \bibinfo {author} {\bibfnamefont {D.}~\bibnamefont {Kim}}, \bibinfo {author}
  {\bibfnamefont {K.~K.}\ \bibnamefont {Mehta}}, \bibinfo {author}
  {\bibfnamefont {T.~E.}\ \bibnamefont {Northup}}, \bibinfo {author}
  {\bibfnamefont {H.}~\bibnamefont {Paik}}, \bibinfo {author} {\bibfnamefont
  {B.}~\bibnamefont {Palmer}}, \bibinfo {author} {\bibfnamefont
  {N.}~\bibnamefont {Samarth}}, \bibinfo {author} {\bibfnamefont
  {S.}~\bibnamefont {Sangtawesin}},\ and\ \bibinfo {author} {\bibfnamefont
  {D.}~\bibnamefont {Steuerman}},\ }\bibfield  {title} {\bibinfo {title}
  {\textit{Materials challenges and opportunities for quantum computing
  hardware}},\ }\href@noop {} {\bibfield  {journal} {\bibinfo  {journal}
  {Science}\ }\textbf {\bibinfo {volume} {372}},\ \bibinfo {pages} {eabb2823}
  (\bibinfo {year} {2021})}\BibitemShut {NoStop}%
\bibitem [{\citenamefont {H{\"a}ffner}\ \emph {et~al.}(2005)\citenamefont
  {H{\"a}ffner}, \citenamefont {H{\"a}nsel}, \citenamefont {Roos},
  \citenamefont {Benhelm}, \citenamefont {Chek-al Kar}, \citenamefont
  {Chwalla}, \citenamefont {K{\"o}rber}, \citenamefont {Rapol}, \citenamefont
  {Riebe}, \citenamefont {Schmidt}, \citenamefont {Becher}, \citenamefont
  {G{\"u}hne}, \citenamefont {D{\"u}r},\ and\ \citenamefont
  {Blatt}}]{haffner2005scalable}%
  \BibitemOpen
  \bibfield  {author} {\bibinfo {author} {\bibfnamefont {H.}~\bibnamefont
  {H{\"a}ffner}}, \bibinfo {author} {\bibfnamefont {W.}~\bibnamefont
  {H{\"a}nsel}}, \bibinfo {author} {\bibfnamefont {C.}~\bibnamefont {Roos}},
  \bibinfo {author} {\bibfnamefont {J.}~\bibnamefont {Benhelm}}, \bibinfo
  {author} {\bibfnamefont {D.}~\bibnamefont {Chek-al Kar}}, \bibinfo {author}
  {\bibfnamefont {M.}~\bibnamefont {Chwalla}}, \bibinfo {author} {\bibfnamefont
  {T.}~\bibnamefont {K{\"o}rber}}, \bibinfo {author} {\bibfnamefont
  {U.}~\bibnamefont {Rapol}}, \bibinfo {author} {\bibfnamefont
  {M.}~\bibnamefont {Riebe}}, \bibinfo {author} {\bibfnamefont
  {P.}~\bibnamefont {Schmidt}}, \bibinfo {author} {\bibfnamefont
  {C.}~\bibnamefont {Becher}}, \bibinfo {author} {\bibfnamefont
  {O.}~\bibnamefont {G{\"u}hne}}, \bibinfo {author} {\bibfnamefont
  {W.}~\bibnamefont {D{\"u}r}},\ and\ \bibinfo {author} {\bibfnamefont
  {R.}~\bibnamefont {Blatt}},\ }\bibfield  {title} {\bibinfo {title}
  {\textit{Scalable Multiparticle Entanglement of Trapped Ions}},\ }\href@noop
  {} {\bibfield  {journal} {\bibinfo  {journal} {Nature}\ }\textbf {\bibinfo
  {volume} {438}},\ \bibinfo {pages} {643} (\bibinfo {year}
  {2005})}\BibitemShut {NoStop}%
\bibitem [{\citenamefont {Esteve}\ \emph {et~al.}(2008)\citenamefont {Esteve},
  \citenamefont {Gross}, \citenamefont {Weller}, \citenamefont {Giovanazzi},\
  and\ \citenamefont {Oberthaler}}]{esteve2008squeezing}%
  \BibitemOpen
  \bibfield  {author} {\bibinfo {author} {\bibfnamefont {J.}~\bibnamefont
  {Esteve}}, \bibinfo {author} {\bibfnamefont {C.}~\bibnamefont {Gross}},
  \bibinfo {author} {\bibfnamefont {A.}~\bibnamefont {Weller}}, \bibinfo
  {author} {\bibfnamefont {S.}~\bibnamefont {Giovanazzi}},\ and\ \bibinfo
  {author} {\bibfnamefont {M.~K.}\ \bibnamefont {Oberthaler}},\ }\bibfield
  {title} {\bibinfo {title} {\textit{Squeezing and Entanglement in a
  Bose--Einstein Condensate}},\ }\href@noop {} {\bibfield  {journal} {\bibinfo
  {journal} {Nature}\ }\textbf {\bibinfo {volume} {455}},\ \bibinfo {pages}
  {1216} (\bibinfo {year} {2008})}\BibitemShut {NoStop}%
\bibitem [{\citenamefont {Gross}\ \emph {et~al.}(2010)\citenamefont {Gross},
  \citenamefont {Zibold}, \citenamefont {Nicklas}, \citenamefont {Esteve},\
  and\ \citenamefont {Oberthaler}}]{gross2010nonlinear}%
  \BibitemOpen
  \bibfield  {author} {\bibinfo {author} {\bibfnamefont {C.}~\bibnamefont
  {Gross}}, \bibinfo {author} {\bibfnamefont {T.}~\bibnamefont {Zibold}},
  \bibinfo {author} {\bibfnamefont {E.}~\bibnamefont {Nicklas}}, \bibinfo
  {author} {\bibfnamefont {J.}~\bibnamefont {Esteve}},\ and\ \bibinfo {author}
  {\bibfnamefont {M.~K.}\ \bibnamefont {Oberthaler}},\ }\bibfield  {title}
  {\bibinfo {title} {\textit{Nonlinear Atom Interferometer Surpasses Classical
  Precision Limit}},\ }\href@noop {} {\bibfield  {journal} {\bibinfo  {journal}
  {Nature}\ }\textbf {\bibinfo {volume} {464}},\ \bibinfo {pages} {1165}
  (\bibinfo {year} {2010})}\BibitemShut {NoStop}%
\bibitem [{\citenamefont {Van~Enk}\ and\ \citenamefont
  {Beenakker}(2012)}]{van2012measuring}%
  \BibitemOpen
  \bibfield  {author} {\bibinfo {author} {\bibfnamefont {S.}~\bibnamefont
  {Van~Enk}}\ and\ \bibinfo {author} {\bibfnamefont {C.}~\bibnamefont
  {Beenakker}},\ }\bibfield  {title} {\bibinfo {title} {\textit{Measuring
  Tr$\rho$n on Single Copies of $\rho$ Using Random Measurements}},\
  }\href@noop {} {\bibfield  {journal} {\bibinfo  {journal} {Phys. Rev. Lett.}\
  }\textbf {\bibinfo {volume} {108}},\ \bibinfo {pages} {110503} (\bibinfo
  {year} {2012})}\BibitemShut {NoStop}%
\bibitem [{\citenamefont {Islam}\ \emph {et~al.}(2015)\citenamefont {Islam},
  \citenamefont {Ma}, \citenamefont {Preiss}, \citenamefont {Tai},
  \citenamefont {Lukin}, \citenamefont {Rispoli},\ and\ \citenamefont
  {Greiner}}]{islam2015measuring}%
  \BibitemOpen
  \bibfield  {author} {\bibinfo {author} {\bibfnamefont {R.}~\bibnamefont
  {Islam}}, \bibinfo {author} {\bibfnamefont {R.}~\bibnamefont {Ma}}, \bibinfo
  {author} {\bibfnamefont {P.~M.}\ \bibnamefont {Preiss}}, \bibinfo {author}
  {\bibfnamefont {M.~E.}\ \bibnamefont {Tai}}, \bibinfo {author} {\bibfnamefont
  {A.}~\bibnamefont {Lukin}}, \bibinfo {author} {\bibfnamefont
  {M.}~\bibnamefont {Rispoli}},\ and\ \bibinfo {author} {\bibfnamefont
  {M.}~\bibnamefont {Greiner}},\ }\bibfield  {title} {\bibinfo {title}
  {\textit{Measuring Entanglement Entropy in a Quantum Many-body System}},\
  }\href@noop {} {\bibfield  {journal} {\bibinfo  {journal} {Nature}\ }\textbf
  {\bibinfo {volume} {528}},\ \bibinfo {pages} {77} (\bibinfo {year}
  {2015})}\BibitemShut {NoStop}%
\bibitem [{\citenamefont {Kaufman}\ \emph {et~al.}(2016)\citenamefont
  {Kaufman}, \citenamefont {Tai}, \citenamefont {Lukin}, \citenamefont
  {Rispoli}, \citenamefont {Schittko}, \citenamefont {Preiss},\ and\
  \citenamefont {Greiner}}]{kaufman2016quantum}%
  \BibitemOpen
  \bibfield  {author} {\bibinfo {author} {\bibfnamefont {A.~M.}\ \bibnamefont
  {Kaufman}}, \bibinfo {author} {\bibfnamefont {M.~E.}\ \bibnamefont {Tai}},
  \bibinfo {author} {\bibfnamefont {A.}~\bibnamefont {Lukin}}, \bibinfo
  {author} {\bibfnamefont {M.}~\bibnamefont {Rispoli}}, \bibinfo {author}
  {\bibfnamefont {R.}~\bibnamefont {Schittko}}, \bibinfo {author}
  {\bibfnamefont {P.~M.}\ \bibnamefont {Preiss}},\ and\ \bibinfo {author}
  {\bibfnamefont {M.}~\bibnamefont {Greiner}},\ }\bibfield  {title} {\bibinfo
  {title} {\textit{Quantum Thermalization Through Entanglement in an Isolated
  Many-body System}},\ }\href@noop {} {\bibfield  {journal} {\bibinfo
  {journal} {Science}\ }\textbf {\bibinfo {volume} {353}},\ \bibinfo {pages}
  {794} (\bibinfo {year} {2016})}\BibitemShut {NoStop}%
\bibitem [{\citenamefont {Tubman}(2016)}]{tubman2016measuring}%
  \BibitemOpen
  \bibfield  {author} {\bibinfo {author} {\bibfnamefont {N.~M.}\ \bibnamefont
  {Tubman}},\ }\bibfield  {title} {\bibinfo {title} {\textit{Measuring Quantum
  Entanglement, Machine Learning and Wave Function Tomography: Bridging Theory
  and Experiment with the Quantum Gas Microscope}},\ }\href@noop {} {\bibfield
  {journal} {\bibinfo  {journal} {arXiv:1609.08142}\ } (\bibinfo {year}
  {2016})}\BibitemShut {NoStop}%
\bibitem [{\citenamefont {Linke}\ \emph {et~al.}(2018)\citenamefont {Linke},
  \citenamefont {Johri}, \citenamefont {Figgatt}, \citenamefont {Landsman},
  \citenamefont {Matsuura},\ and\ \citenamefont {Monroe}}]{linke2018measuring}%
  \BibitemOpen
  \bibfield  {author} {\bibinfo {author} {\bibfnamefont {N.~M.}\ \bibnamefont
  {Linke}}, \bibinfo {author} {\bibfnamefont {S.}~\bibnamefont {Johri}},
  \bibinfo {author} {\bibfnamefont {C.}~\bibnamefont {Figgatt}}, \bibinfo
  {author} {\bibfnamefont {K.~A.}\ \bibnamefont {Landsman}}, \bibinfo {author}
  {\bibfnamefont {A.~Y.}\ \bibnamefont {Matsuura}},\ and\ \bibinfo {author}
  {\bibfnamefont {C.}~\bibnamefont {Monroe}},\ }\bibfield  {title} {\bibinfo
  {title} {\textit{Measuring the R{\'e}nyi Entropy of a Two-Site Fermi-Hubbard
  Model on a Trapped Ion Quantum Computer}},\ }\href@noop {} {\bibfield
  {journal} {\bibinfo  {journal} {Phys. Rev. A}\ }\textbf {\bibinfo {volume}
  {98}},\ \bibinfo {pages} {052334} (\bibinfo {year} {2018})}\BibitemShut
  {NoStop}%
\bibitem [{\citenamefont {Brydges}\ \emph {et~al.}(2019)\citenamefont
  {Brydges}, \citenamefont {Elben}, \citenamefont {Jurcevic}, \citenamefont
  {Vermersch}, \citenamefont {Maier}, \citenamefont {Lanyon}, \citenamefont
  {Zoller}, \citenamefont {Blatt},\ and\ \citenamefont
  {Roos}}]{brydges2019probing}%
  \BibitemOpen
  \bibfield  {author} {\bibinfo {author} {\bibfnamefont {T.}~\bibnamefont
  {Brydges}}, \bibinfo {author} {\bibfnamefont {A.}~\bibnamefont {Elben}},
  \bibinfo {author} {\bibfnamefont {P.}~\bibnamefont {Jurcevic}}, \bibinfo
  {author} {\bibfnamefont {B.}~\bibnamefont {Vermersch}}, \bibinfo {author}
  {\bibfnamefont {C.}~\bibnamefont {Maier}}, \bibinfo {author} {\bibfnamefont
  {B.~P.}\ \bibnamefont {Lanyon}}, \bibinfo {author} {\bibfnamefont
  {P.}~\bibnamefont {Zoller}}, \bibinfo {author} {\bibfnamefont
  {R.}~\bibnamefont {Blatt}},\ and\ \bibinfo {author} {\bibfnamefont {C.~F.}\
  \bibnamefont {Roos}},\ }\bibfield  {title} {\bibinfo {title} {\textit{Probing
  R{\'e}nyi Entanglement Entropy via Randomized Measurements}},\ }\href@noop {}
  {\bibfield  {journal} {\bibinfo  {journal} {Science}\ }\textbf {\bibinfo
  {volume} {364}},\ \bibinfo {pages} {260} (\bibinfo {year}
  {2019})}\BibitemShut {NoStop}%
\bibitem [{\citenamefont {Schweigler}\ \emph {et~al.}(2017)\citenamefont
  {Schweigler}, \citenamefont {Kasper}, \citenamefont {Erne}, \citenamefont
  {Mazets}, \citenamefont {Rauer}, \citenamefont {Cataldini}, \citenamefont
  {Langen}, \citenamefont {Gasenzer}, \citenamefont {Berges},\ and\
  \citenamefont {Schmiedmayer}}]{schweigler2017experimental}%
  \BibitemOpen
  \bibfield  {author} {\bibinfo {author} {\bibfnamefont {T.}~\bibnamefont
  {Schweigler}}, \bibinfo {author} {\bibfnamefont {V.}~\bibnamefont {Kasper}},
  \bibinfo {author} {\bibfnamefont {S.}~\bibnamefont {Erne}}, \bibinfo {author}
  {\bibfnamefont {I.}~\bibnamefont {Mazets}}, \bibinfo {author} {\bibfnamefont
  {B.}~\bibnamefont {Rauer}}, \bibinfo {author} {\bibfnamefont
  {F.}~\bibnamefont {Cataldini}}, \bibinfo {author} {\bibfnamefont
  {T.}~\bibnamefont {Langen}}, \bibinfo {author} {\bibfnamefont
  {T.}~\bibnamefont {Gasenzer}}, \bibinfo {author} {\bibfnamefont
  {J.}~\bibnamefont {Berges}},\ and\ \bibinfo {author} {\bibfnamefont
  {J.}~\bibnamefont {Schmiedmayer}},\ }\bibfield  {title} {\bibinfo {title}
  {\textit{Experimental Characterization of a Quantum Many-body System via
  Higher-order Correlations}},\ }\href@noop {} {\bibfield  {journal} {\bibinfo
  {journal} {Nature}\ }\textbf {\bibinfo {volume} {545}},\ \bibinfo {pages}
  {323} (\bibinfo {year} {2017})}\BibitemShut {NoStop}%
\bibitem [{\citenamefont {Hilker}\ \emph {et~al.}(2017)\citenamefont {Hilker},
  \citenamefont {Salomon}, \citenamefont {Grusdt}, \citenamefont {Omran},
  \citenamefont {Boll}, \citenamefont {Demler}, \citenamefont {Bloch},\ and\
  \citenamefont {Gross}}]{hilker2017revealing}%
  \BibitemOpen
  \bibfield  {author} {\bibinfo {author} {\bibfnamefont {T.~A.}\ \bibnamefont
  {Hilker}}, \bibinfo {author} {\bibfnamefont {G.}~\bibnamefont {Salomon}},
  \bibinfo {author} {\bibfnamefont {F.}~\bibnamefont {Grusdt}}, \bibinfo
  {author} {\bibfnamefont {A.}~\bibnamefont {Omran}}, \bibinfo {author}
  {\bibfnamefont {M.}~\bibnamefont {Boll}}, \bibinfo {author} {\bibfnamefont
  {E.}~\bibnamefont {Demler}}, \bibinfo {author} {\bibfnamefont
  {I.}~\bibnamefont {Bloch}},\ and\ \bibinfo {author} {\bibfnamefont
  {C.}~\bibnamefont {Gross}},\ }\bibfield  {title} {\bibinfo {title}
  {\textit{Revealing Hidden Antiferromagnetic Correlations in Doped Hubbard
  Chains via String Correlators}},\ }\href@noop {} {\bibfield  {journal}
  {\bibinfo  {journal} {Science}\ }\textbf {\bibinfo {volume} {357}},\ \bibinfo
  {pages} {484} (\bibinfo {year} {2017})}\BibitemShut {NoStop}%
\bibitem [{\citenamefont {Salomon}\ \emph {et~al.}(2019)\citenamefont
  {Salomon}, \citenamefont {Koepsell}, \citenamefont {Vijayan}, \citenamefont
  {Hilker}, \citenamefont {Nespolo}, \citenamefont {Pollet}, \citenamefont
  {Bloch},\ and\ \citenamefont {Gross}}]{salomon2019direct}%
  \BibitemOpen
  \bibfield  {author} {\bibinfo {author} {\bibfnamefont {G.}~\bibnamefont
  {Salomon}}, \bibinfo {author} {\bibfnamefont {J.}~\bibnamefont {Koepsell}},
  \bibinfo {author} {\bibfnamefont {J.}~\bibnamefont {Vijayan}}, \bibinfo
  {author} {\bibfnamefont {T.~A.}\ \bibnamefont {Hilker}}, \bibinfo {author}
  {\bibfnamefont {J.}~\bibnamefont {Nespolo}}, \bibinfo {author} {\bibfnamefont
  {L.}~\bibnamefont {Pollet}}, \bibinfo {author} {\bibfnamefont
  {I.}~\bibnamefont {Bloch}},\ and\ \bibinfo {author} {\bibfnamefont
  {C.}~\bibnamefont {Gross}},\ }\bibfield  {title} {\bibinfo {title}
  {\textit{Direct Observation of Incommensurate Magnetism in Hubbard Chains}},\
  }\href@noop {} {\bibfield  {journal} {\bibinfo  {journal} {Nature}\ }\textbf
  {\bibinfo {volume} {565}},\ \bibinfo {pages} {56} (\bibinfo {year}
  {2019})}\BibitemShut {NoStop}%
\bibitem [{\citenamefont {Koepsell}\ \emph {et~al.}(2019)\citenamefont
  {Koepsell}, \citenamefont {Vijayan}, \citenamefont {Sompet}, \citenamefont
  {Grusdt}, \citenamefont {Hilker}, \citenamefont {Demler}, \citenamefont
  {Salomon}, \citenamefont {Bloch},\ and\ \citenamefont
  {Gross}}]{koepsell2019imaging}%
  \BibitemOpen
  \bibfield  {author} {\bibinfo {author} {\bibfnamefont {J.}~\bibnamefont
  {Koepsell}}, \bibinfo {author} {\bibfnamefont {J.}~\bibnamefont {Vijayan}},
  \bibinfo {author} {\bibfnamefont {P.}~\bibnamefont {Sompet}}, \bibinfo
  {author} {\bibfnamefont {F.}~\bibnamefont {Grusdt}}, \bibinfo {author}
  {\bibfnamefont {T.~A.}\ \bibnamefont {Hilker}}, \bibinfo {author}
  {\bibfnamefont {E.}~\bibnamefont {Demler}}, \bibinfo {author} {\bibfnamefont
  {G.}~\bibnamefont {Salomon}}, \bibinfo {author} {\bibfnamefont
  {I.}~\bibnamefont {Bloch}},\ and\ \bibinfo {author} {\bibfnamefont
  {C.}~\bibnamefont {Gross}},\ }\bibfield  {title} {\bibinfo {title}
  {\textit{Imaging Magnetic Polarons in the Doped Fermi-Hubbard Model}},\
  }\href@noop {} {\bibfield  {journal} {\bibinfo  {journal} {Nature}\ }\textbf
  {\bibinfo {volume} {572}},\ \bibinfo {pages} {358} (\bibinfo {year}
  {2019})}\BibitemShut {NoStop}%
\bibitem [{\citenamefont {Vijayan}\ \emph {et~al.}(2020)\citenamefont
  {Vijayan}, \citenamefont {Sompet}, \citenamefont {Salomon}, \citenamefont
  {Koepsell}, \citenamefont {Hirthe}, \citenamefont {Bohrdt}, \citenamefont
  {Grusdt}, \citenamefont {Bloch},\ and\ \citenamefont
  {Gross}}]{vijayan2020time}%
  \BibitemOpen
  \bibfield  {author} {\bibinfo {author} {\bibfnamefont {J.}~\bibnamefont
  {Vijayan}}, \bibinfo {author} {\bibfnamefont {P.}~\bibnamefont {Sompet}},
  \bibinfo {author} {\bibfnamefont {G.}~\bibnamefont {Salomon}}, \bibinfo
  {author} {\bibfnamefont {J.}~\bibnamefont {Koepsell}}, \bibinfo {author}
  {\bibfnamefont {S.}~\bibnamefont {Hirthe}}, \bibinfo {author} {\bibfnamefont
  {A.}~\bibnamefont {Bohrdt}}, \bibinfo {author} {\bibfnamefont
  {F.}~\bibnamefont {Grusdt}}, \bibinfo {author} {\bibfnamefont
  {I.}~\bibnamefont {Bloch}},\ and\ \bibinfo {author} {\bibfnamefont
  {C.}~\bibnamefont {Gross}},\ }\bibfield  {title} {\bibinfo {title}
  {\textit{Time-Resolved Observation of Spin-Charge Deconfinement in Fermionic
  Hubbard Chains}},\ }\href@noop {} {\bibfield  {journal} {\bibinfo  {journal}
  {Science}\ }\textbf {\bibinfo {volume} {367}},\ \bibinfo {pages} {186}
  (\bibinfo {year} {2020})}\BibitemShut {NoStop}%
\bibitem [{\citenamefont {Pr{\"u}fer}\ \emph {et~al.}(2020)\citenamefont
  {Pr{\"u}fer}, \citenamefont {Zache}, \citenamefont {Kunkel}, \citenamefont
  {Lannig}, \citenamefont {Bonnin}, \citenamefont {Strobel}, \citenamefont
  {Berges},\ and\ \citenamefont {Oberthaler}}]{prufer2020experimental}%
  \BibitemOpen
  \bibfield  {author} {\bibinfo {author} {\bibfnamefont {M.}~\bibnamefont
  {Pr{\"u}fer}}, \bibinfo {author} {\bibfnamefont {T.~V.}\ \bibnamefont
  {Zache}}, \bibinfo {author} {\bibfnamefont {P.}~\bibnamefont {Kunkel}},
  \bibinfo {author} {\bibfnamefont {S.}~\bibnamefont {Lannig}}, \bibinfo
  {author} {\bibfnamefont {A.}~\bibnamefont {Bonnin}}, \bibinfo {author}
  {\bibfnamefont {H.}~\bibnamefont {Strobel}}, \bibinfo {author} {\bibfnamefont
  {J.}~\bibnamefont {Berges}},\ and\ \bibinfo {author} {\bibfnamefont {M.~K.}\
  \bibnamefont {Oberthaler}},\ }\bibfield  {title} {\bibinfo {title}
  {\textit{Experimental Extraction of the Quantum Effective Action for a
  Non-Equilibrium Many-Body System}},\ }\href@noop {} {\bibfield  {journal}
  {\bibinfo  {journal} {Nat. Phys.}\ }\textbf {\bibinfo {volume} {16}},\
  \bibinfo {pages} {1012} (\bibinfo {year} {2020})}\BibitemShut {NoStop}%
\bibitem [{\citenamefont {Zache}\ \emph {et~al.}(2020)\citenamefont {Zache},
  \citenamefont {Schweigler}, \citenamefont {Erne}, \citenamefont
  {Schmiedmayer},\ and\ \citenamefont {Berges}}]{zache2020extracting}%
  \BibitemOpen
  \bibfield  {author} {\bibinfo {author} {\bibfnamefont {T.~V.}\ \bibnamefont
  {Zache}}, \bibinfo {author} {\bibfnamefont {T.}~\bibnamefont {Schweigler}},
  \bibinfo {author} {\bibfnamefont {S.}~\bibnamefont {Erne}}, \bibinfo {author}
  {\bibfnamefont {J.}~\bibnamefont {Schmiedmayer}},\ and\ \bibinfo {author}
  {\bibfnamefont {J.}~\bibnamefont {Berges}},\ }\bibfield  {title} {\bibinfo
  {title} {\textit{Extracting the Field Theory Description of a Quantum
  Many-Body System from Experimental Data}},\ }\href@noop {} {\bibfield
  {journal} {\bibinfo  {journal} {Phys. Rev. X}\ }\textbf {\bibinfo {volume}
  {10}},\ \bibinfo {pages} {011020} (\bibinfo {year} {2020})}\BibitemShut
  {NoStop}%
\bibitem [{\citenamefont {Koepsell}\ \emph {et~al.}(2020)\citenamefont
  {Koepsell}, \citenamefont {Bourgund}, \citenamefont {Sompet}, \citenamefont
  {Hirthe}, \citenamefont {Bohrdt}, \citenamefont {Wang}, \citenamefont
  {Grusdt}, \citenamefont {Demler}, \citenamefont {Salomon}, \citenamefont
  {Gross},\ and\ \citenamefont {Bloch}}]{koepsell2020microscopic}%
  \BibitemOpen
  \bibfield  {author} {\bibinfo {author} {\bibfnamefont {J.}~\bibnamefont
  {Koepsell}}, \bibinfo {author} {\bibfnamefont {D.}~\bibnamefont {Bourgund}},
  \bibinfo {author} {\bibfnamefont {P.}~\bibnamefont {Sompet}}, \bibinfo
  {author} {\bibfnamefont {S.}~\bibnamefont {Hirthe}}, \bibinfo {author}
  {\bibfnamefont {A.}~\bibnamefont {Bohrdt}}, \bibinfo {author} {\bibfnamefont
  {Y.}~\bibnamefont {Wang}}, \bibinfo {author} {\bibfnamefont {F.}~\bibnamefont
  {Grusdt}}, \bibinfo {author} {\bibfnamefont {E.}~\bibnamefont {Demler}},
  \bibinfo {author} {\bibfnamefont {G.}~\bibnamefont {Salomon}}, \bibinfo
  {author} {\bibfnamefont {C.}~\bibnamefont {Gross}},\ and\ \bibinfo {author}
  {\bibfnamefont {I.}~\bibnamefont {Bloch}},\ }\bibfield  {title} {\bibinfo
  {title} {\textit{Microscopic Evolution of Doped Mott Insulators from
  Polaronic Metal to Fermi Liquid}},\ }\href@noop {} {\bibfield  {journal}
  {\bibinfo  {journal} {Science}\ }\textbf {\bibinfo {volume} {374}},\ \bibinfo
  {pages} {82} (\bibinfo {year} {2020})}\BibitemShut {NoStop}%
\bibitem [{\citenamefont {Wootters}(1981)}]{wootters1981statistical}%
  \BibitemOpen
  \bibfield  {author} {\bibinfo {author} {\bibfnamefont {W.~K.}\ \bibnamefont
  {Wootters}},\ }\bibfield  {title} {\bibinfo {title} {\textit{Statistical
  Distance and Hilbert Space}},\ }\href@noop {} {\bibfield  {journal} {\bibinfo
   {journal} {Phys. Rev. D}\ }\textbf {\bibinfo {volume} {23}},\ \bibinfo
  {pages} {357} (\bibinfo {year} {1981})}\BibitemShut {NoStop}%
\bibitem [{\citenamefont {Carvalho}\ \emph {et~al.}(2004)\citenamefont
  {Carvalho}, \citenamefont {Mintert},\ and\ \citenamefont
  {Buchleitner}}]{carvalho2004decoherence}%
  \BibitemOpen
  \bibfield  {author} {\bibinfo {author} {\bibfnamefont {A.~R.}\ \bibnamefont
  {Carvalho}}, \bibinfo {author} {\bibfnamefont {F.}~\bibnamefont {Mintert}},\
  and\ \bibinfo {author} {\bibfnamefont {A.}~\bibnamefont {Buchleitner}},\
  }\bibfield  {title} {\bibinfo {title} {\textit{Decoherence and Multipartite
  Entanglement}},\ }\href@noop {} {\bibfield  {journal} {\bibinfo  {journal}
  {Phys. Rev. Lett.}\ }\textbf {\bibinfo {volume} {93}},\ \bibinfo {pages}
  {230501} (\bibinfo {year} {2004})}\BibitemShut {NoStop}%
\bibitem [{\citenamefont {Mintert}\ \emph {et~al.}(2005)\citenamefont
  {Mintert}, \citenamefont {Ku{\'s}},\ and\ \citenamefont
  {Buchleitner}}]{mintert2005concurrence}%
  \BibitemOpen
  \bibfield  {author} {\bibinfo {author} {\bibfnamefont {F.}~\bibnamefont
  {Mintert}}, \bibinfo {author} {\bibfnamefont {M.}~\bibnamefont {Ku{\'s}}},\
  and\ \bibinfo {author} {\bibfnamefont {A.}~\bibnamefont {Buchleitner}},\
  }\bibfield  {title} {\bibinfo {title} {\textit{Concurrence of Mixed
  Multipartite Quantum States}},\ }\href@noop {} {\bibfield  {journal}
  {\bibinfo  {journal} {Phys. Rev. Lett.}\ }\textbf {\bibinfo {volume} {95}},\
  \bibinfo {pages} {260502} (\bibinfo {year} {2005})}\BibitemShut {NoStop}%
\bibitem [{\citenamefont {Aolita}\ and\ \citenamefont
  {Mintert}(2006)}]{aolita2006measuring}%
  \BibitemOpen
  \bibfield  {author} {\bibinfo {author} {\bibfnamefont {L.}~\bibnamefont
  {Aolita}}\ and\ \bibinfo {author} {\bibfnamefont {F.}~\bibnamefont
  {Mintert}},\ }\bibfield  {title} {\bibinfo {title} {\textit{Measuring
  Multipartite Concurrence with a Single Factorizable Observable}},\
  }\href@noop {} {\bibfield  {journal} {\bibinfo  {journal} {Phys. Rev. Lett.}\
  }\textbf {\bibinfo {volume} {97}},\ \bibinfo {pages} {050501} (\bibinfo
  {year} {2006})}\BibitemShut {NoStop}%
\bibitem [{\citenamefont {S{\o}rensen}\ and\ \citenamefont
  {M{\o}lmer}(2001)}]{sorensen2001entanglement}%
  \BibitemOpen
  \bibfield  {author} {\bibinfo {author} {\bibfnamefont {A.~S.}\ \bibnamefont
  {S{\o}rensen}}\ and\ \bibinfo {author} {\bibfnamefont {K.}~\bibnamefont
  {M{\o}lmer}},\ }\bibfield  {title} {\bibinfo {title} {\textit{Entanglement
  and Extreme Spin Squeezing}},\ }\href@noop {} {\bibfield  {journal} {\bibinfo
   {journal} {Phys. Rev. Lett.}\ }\textbf {\bibinfo {volume} {86}},\ \bibinfo
  {pages} {4431} (\bibinfo {year} {2001})}\BibitemShut {NoStop}%
\bibitem [{\citenamefont {G{\"u}hne}\ \emph {et~al.}(2005)\citenamefont
  {G{\"u}hne}, \citenamefont {T{\'o}th},\ and\ \citenamefont
  {Briegel}}]{guhne2005multipartite}%
  \BibitemOpen
  \bibfield  {author} {\bibinfo {author} {\bibfnamefont {O.}~\bibnamefont
  {G{\"u}hne}}, \bibinfo {author} {\bibfnamefont {G.}~\bibnamefont
  {T{\'o}th}},\ and\ \bibinfo {author} {\bibfnamefont {H.~J.}\ \bibnamefont
  {Briegel}},\ }\bibfield  {title} {\bibinfo {title} {\textit{Multipartite
  Entanglement in Spin Chains}},\ }\href@noop {} {\bibfield  {journal}
  {\bibinfo  {journal} {New J. Phys.}\ }\textbf {\bibinfo {volume} {7}},\
  \bibinfo {pages} {229} (\bibinfo {year} {2005})}\BibitemShut {NoStop}%
\bibitem [{\citenamefont {Ac{\'\i}n}\ \emph {et~al.}(2001)\citenamefont
  {Ac{\'\i}n}, \citenamefont {Bru{\ss}}, \citenamefont {Lewenstein},\ and\
  \citenamefont {Sanpera}}]{acin2001classification}%
  \BibitemOpen
  \bibfield  {author} {\bibinfo {author} {\bibfnamefont {A.}~\bibnamefont
  {Ac{\'\i}n}}, \bibinfo {author} {\bibfnamefont {D.}~\bibnamefont {Bru{\ss}}},
  \bibinfo {author} {\bibfnamefont {M.}~\bibnamefont {Lewenstein}},\ and\
  \bibinfo {author} {\bibfnamefont {A.}~\bibnamefont {Sanpera}},\ }\bibfield
  {title} {\bibinfo {title} {\textit{Classification of Mixed Three-qubit
  States}},\ }\href@noop {} {\bibfield  {journal} {\bibinfo  {journal} {Phys.
  Rev. Lett.}\ }\textbf {\bibinfo {volume} {87}},\ \bibinfo {pages} {040401}
  (\bibinfo {year} {2001})}\BibitemShut {NoStop}%
\bibitem [{\citenamefont {Friis}\ \emph {et~al.}(2019)\citenamefont {Friis},
  \citenamefont {Vitagliano}, \citenamefont {Malik},\ and\ \citenamefont
  {Huber}}]{friis2019entanglement}%
  \BibitemOpen
  \bibfield  {author} {\bibinfo {author} {\bibfnamefont {N.}~\bibnamefont
  {Friis}}, \bibinfo {author} {\bibfnamefont {G.}~\bibnamefont {Vitagliano}},
  \bibinfo {author} {\bibfnamefont {M.}~\bibnamefont {Malik}},\ and\ \bibinfo
  {author} {\bibfnamefont {M.}~\bibnamefont {Huber}},\ }\bibfield  {title}
  {\bibinfo {title} {\textit{Entanglement Certification from Theory to
  Experiment}},\ }\href@noop {} {\bibfield  {journal} {\bibinfo  {journal}
  {Nat. Rev. Phys.}\ }\textbf {\bibinfo {volume} {1}},\ \bibinfo {pages} {72}
  (\bibinfo {year} {2019})}\BibitemShut {NoStop}%
\bibitem [{\citenamefont {Terhal}(2000)}]{terhal2000bell}%
  \BibitemOpen
  \bibfield  {author} {\bibinfo {author} {\bibfnamefont {B.~M.}\ \bibnamefont
  {Terhal}},\ }\bibfield  {title} {\bibinfo {title} {\textit{Bell Inequalities
  and the Separability Criterion}},\ }\href@noop {} {\bibfield  {journal}
  {\bibinfo  {journal} {Phys. Lett. A}\ }\textbf {\bibinfo {volume} {271}},\
  \bibinfo {pages} {319} (\bibinfo {year} {2000})}\BibitemShut {NoStop}%
\bibitem [{\citenamefont {Coffman}\ \emph {et~al.}(2000)\citenamefont
  {Coffman}, \citenamefont {Kundu},\ and\ \citenamefont
  {Wootters}}]{coffman2000distributed}%
  \BibitemOpen
  \bibfield  {author} {\bibinfo {author} {\bibfnamefont {V.}~\bibnamefont
  {Coffman}}, \bibinfo {author} {\bibfnamefont {J.}~\bibnamefont {Kundu}},\
  and\ \bibinfo {author} {\bibfnamefont {W.~K.}\ \bibnamefont {Wootters}},\
  }\bibfield  {title} {\bibinfo {title} {\textit{Distributed Entanglement}},\
  }\href@noop {} {\bibfield  {journal} {\bibinfo  {journal} {Phys. Rev. A}\
  }\textbf {\bibinfo {volume} {61}},\ \bibinfo {pages} {052306} (\bibinfo
  {year} {2000})}\BibitemShut {NoStop}%
\bibitem [{\citenamefont {Amico}\ \emph {et~al.}(2004)\citenamefont {Amico},
  \citenamefont {Osterloh}, \citenamefont {Plastina}, \citenamefont {Fazio},\
  and\ \citenamefont {Palma}}]{amico2004dynamics}%
  \BibitemOpen
  \bibfield  {author} {\bibinfo {author} {\bibfnamefont {L.}~\bibnamefont
  {Amico}}, \bibinfo {author} {\bibfnamefont {A.}~\bibnamefont {Osterloh}},
  \bibinfo {author} {\bibfnamefont {F.}~\bibnamefont {Plastina}}, \bibinfo
  {author} {\bibfnamefont {R.}~\bibnamefont {Fazio}},\ and\ \bibinfo {author}
  {\bibfnamefont {G.~M.}\ \bibnamefont {Palma}},\ }\bibfield  {title} {\bibinfo
  {title} {\textit{Dynamics of Entanglement in One-Dimensional Spin Systems}},\
  }\href@noop {} {\bibfield  {journal} {\bibinfo  {journal} {Phys. Rev. A}\
  }\textbf {\bibinfo {volume} {69}},\ \bibinfo {pages} {022304} (\bibinfo
  {year} {2004})}\BibitemShut {NoStop}%
\bibitem [{\citenamefont {Roscilde}\ \emph {et~al.}(2004)\citenamefont
  {Roscilde}, \citenamefont {Verrucchi}, \citenamefont {Fubini}, \citenamefont
  {Haas},\ and\ \citenamefont {Tognetti}}]{roscilde2004studying}%
  \BibitemOpen
  \bibfield  {author} {\bibinfo {author} {\bibfnamefont {T.}~\bibnamefont
  {Roscilde}}, \bibinfo {author} {\bibfnamefont {P.}~\bibnamefont {Verrucchi}},
  \bibinfo {author} {\bibfnamefont {A.}~\bibnamefont {Fubini}}, \bibinfo
  {author} {\bibfnamefont {S.}~\bibnamefont {Haas}},\ and\ \bibinfo {author}
  {\bibfnamefont {V.}~\bibnamefont {Tognetti}},\ }\bibfield  {title} {\bibinfo
  {title} {\textit{Studying Quantum Spin Systems Through Entanglement
  Estimators}},\ }\href@noop {} {\bibfield  {journal} {\bibinfo  {journal}
  {Phys. Rev. Lett.}\ }\textbf {\bibinfo {volume} {93}},\ \bibinfo {pages}
  {167203} (\bibinfo {year} {2004})}\BibitemShut {NoStop}%
\bibitem [{\citenamefont {Brukner}\ \emph {et~al.}(2006)\citenamefont
  {Brukner}, \citenamefont {Vedral},\ and\ \citenamefont
  {Zeilinger}}]{brukner2006crucial}%
  \BibitemOpen
  \bibfield  {author} {\bibinfo {author} {\bibfnamefont {{\v{C}}.}~\bibnamefont
  {Brukner}}, \bibinfo {author} {\bibfnamefont {V.}~\bibnamefont {Vedral}},\
  and\ \bibinfo {author} {\bibfnamefont {A.}~\bibnamefont {Zeilinger}},\
  }\bibfield  {title} {\bibinfo {title} {\textit{Crucial Role of Quantum
  Entanglement in Bulk Properties of Solids}},\ }\href@noop {} {\bibfield
  {journal} {\bibinfo  {journal} {Phys. Rev. A}\ }\textbf {\bibinfo {volume}
  {73}},\ \bibinfo {pages} {012110} (\bibinfo {year} {2006})}\BibitemShut
  {NoStop}%
\bibitem [{\citenamefont {Pezz{\'e}}\ and\ \citenamefont
  {Smerzi}(2009)}]{pezze2009entanglement}%
  \BibitemOpen
  \bibfield  {author} {\bibinfo {author} {\bibfnamefont {L.}~\bibnamefont
  {Pezz{\'e}}}\ and\ \bibinfo {author} {\bibfnamefont {A.}~\bibnamefont
  {Smerzi}},\ }\bibfield  {title} {\bibinfo {title} {\textit{Entanglement,
  Nonlinear Dynamics, and the Heisenberg Limit}},\ }\href@noop {} {\bibfield
  {journal} {\bibinfo  {journal} {Phys. Rev. Lett.}\ }\textbf {\bibinfo
  {volume} {102}},\ \bibinfo {pages} {100401} (\bibinfo {year}
  {2009})}\BibitemShut {NoStop}%
\bibitem [{\citenamefont {Hyllus}\ \emph {et~al.}(2012)\citenamefont {Hyllus},
  \citenamefont {Laskowski}, \citenamefont {Krischek}, \citenamefont
  {Schwemmer}, \citenamefont {Wieczorek}, \citenamefont {Weinfurter},
  \citenamefont {Pezz{\'e}},\ and\ \citenamefont {Smerzi}}]{hyllus2012fisher}%
  \BibitemOpen
  \bibfield  {author} {\bibinfo {author} {\bibfnamefont {P.}~\bibnamefont
  {Hyllus}}, \bibinfo {author} {\bibfnamefont {W.}~\bibnamefont {Laskowski}},
  \bibinfo {author} {\bibfnamefont {R.}~\bibnamefont {Krischek}}, \bibinfo
  {author} {\bibfnamefont {C.}~\bibnamefont {Schwemmer}}, \bibinfo {author}
  {\bibfnamefont {W.}~\bibnamefont {Wieczorek}}, \bibinfo {author}
  {\bibfnamefont {H.}~\bibnamefont {Weinfurter}}, \bibinfo {author}
  {\bibfnamefont {L.}~\bibnamefont {Pezz{\'e}}},\ and\ \bibinfo {author}
  {\bibfnamefont {A.}~\bibnamefont {Smerzi}},\ }\bibfield  {title} {\bibinfo
  {title} {\textit{Fisher Information and Multiparticle Entanglement}},\
  }\href@noop {} {\bibfield  {journal} {\bibinfo  {journal} {Phys. Rev. A}\
  }\textbf {\bibinfo {volume} {85}},\ \bibinfo {pages} {022321} (\bibinfo
  {year} {2012})}\BibitemShut {NoStop}%
\bibitem [{\citenamefont {T{\'o}th}(2012)}]{toth2012multipartite}%
  \BibitemOpen
  \bibfield  {author} {\bibinfo {author} {\bibfnamefont {G.}~\bibnamefont
  {T{\'o}th}},\ }\bibfield  {title} {\bibinfo {title} {\textit{Multipartite
  Entanglement and High-precision Metrology}},\ }\href@noop {} {\bibfield
  {journal} {\bibinfo  {journal} {Phys. Rev. A}\ }\textbf {\bibinfo {volume}
  {85}},\ \bibinfo {pages} {022322} (\bibinfo {year} {2012})}\BibitemShut
  {NoStop}%
\bibitem [{\citenamefont {Hauke}\ \emph {et~al.}(2016)\citenamefont {Hauke},
  \citenamefont {Heyl}, \citenamefont {Tagliacozzo},\ and\ \citenamefont
  {Zoller}}]{hauke2016measuring}%
  \BibitemOpen
  \bibfield  {author} {\bibinfo {author} {\bibfnamefont {P.}~\bibnamefont
  {Hauke}}, \bibinfo {author} {\bibfnamefont {M.}~\bibnamefont {Heyl}},
  \bibinfo {author} {\bibfnamefont {L.}~\bibnamefont {Tagliacozzo}},\ and\
  \bibinfo {author} {\bibfnamefont {P.}~\bibnamefont {Zoller}},\ }\bibfield
  {title} {\bibinfo {title} {\textit{Measuring Multipartite Entanglement
  Through Dynamic Susceptibilities}},\ }\href@noop {} {\bibfield  {journal}
  {\bibinfo  {journal} {Nat. Phys.}\ }\textbf {\bibinfo {volume} {12}},\
  \bibinfo {pages} {778} (\bibinfo {year} {2016})}\BibitemShut {NoStop}%
\bibitem [{\citenamefont {Helstrom}(1969)}]{helstrom1969quantum}%
  \BibitemOpen
  \bibfield  {author} {\bibinfo {author} {\bibfnamefont {C.~W.}\ \bibnamefont
  {Helstrom}},\ }\bibfield  {title} {\bibinfo {title} {\textit{Quantum
  Detection and Estimation Theory}},\ }\href@noop {} {\bibfield  {journal}
  {\bibinfo  {journal} {J. Stat. Phys.}\ }\textbf {\bibinfo {volume} {1}},\
  \bibinfo {pages} {231} (\bibinfo {year} {1969})}\BibitemShut {NoStop}%
\bibitem [{\citenamefont {Braunstein}\ and\ \citenamefont
  {Caves}(1994)}]{braunstein1994statistical}%
  \BibitemOpen
  \bibfield  {author} {\bibinfo {author} {\bibfnamefont {S.~L.}\ \bibnamefont
  {Braunstein}}\ and\ \bibinfo {author} {\bibfnamefont {C.~M.}\ \bibnamefont
  {Caves}},\ }\bibfield  {title} {\bibinfo {title} {\textit{Statistical
  Distance and the Geometry of Quantum States}},\ }\href@noop {} {\bibfield
  {journal} {\bibinfo  {journal} {Phys. Rev. Lett.}\ }\textbf {\bibinfo
  {volume} {72}},\ \bibinfo {pages} {3439} (\bibinfo {year}
  {1994})}\BibitemShut {NoStop}%
\bibitem [{\citenamefont {Braunstein}\ \emph {et~al.}(1996)\citenamefont
  {Braunstein}, \citenamefont {Caves},\ and\ \citenamefont
  {Milburn}}]{braunstein1996generalized}%
  \BibitemOpen
  \bibfield  {author} {\bibinfo {author} {\bibfnamefont {S.~L.}\ \bibnamefont
  {Braunstein}}, \bibinfo {author} {\bibfnamefont {C.~M.}\ \bibnamefont
  {Caves}},\ and\ \bibinfo {author} {\bibfnamefont {G.~J.}\ \bibnamefont
  {Milburn}},\ }\bibfield  {title} {\bibinfo {title} {\textit{Generalized
  Uncertainty Relations: Theory, Examples, and Lorentz Invariance}},\
  }\href@noop {} {\bibfield  {journal} {\bibinfo  {journal} {Ann. Phys.}\
  }\textbf {\bibinfo {volume} {247}},\ \bibinfo {pages} {135} (\bibinfo {year}
  {1996})}\BibitemShut {NoStop}%
\bibitem [{\citenamefont {Mathew}\ \emph {et~al.}(2020)\citenamefont {Mathew},
  \citenamefont {Silva}, \citenamefont {Jain}, \citenamefont {Mohan},
  \citenamefont {Adroja}, \citenamefont {Sakai}, \citenamefont {Tomy},
  \citenamefont {Banerjee}, \citenamefont {Goreti}, \citenamefont {Singh} \emph
  {et~al.}}]{mathew2020experimental}%
  \BibitemOpen
  \bibfield  {author} {\bibinfo {author} {\bibfnamefont {G.}~\bibnamefont
  {Mathew}}, \bibinfo {author} {\bibfnamefont {S.~L.}\ \bibnamefont {Silva}},
  \bibinfo {author} {\bibfnamefont {A.}~\bibnamefont {Jain}}, \bibinfo {author}
  {\bibfnamefont {A.}~\bibnamefont {Mohan}}, \bibinfo {author} {\bibfnamefont
  {D.}~\bibnamefont {Adroja}}, \bibinfo {author} {\bibfnamefont
  {V.}~\bibnamefont {Sakai}}, \bibinfo {author} {\bibfnamefont
  {C.}~\bibnamefont {Tomy}}, \bibinfo {author} {\bibfnamefont {A.}~\bibnamefont
  {Banerjee}}, \bibinfo {author} {\bibfnamefont {R.}~\bibnamefont {Goreti}},
  \bibinfo {author} {\bibfnamefont {R.}~\bibnamefont {Singh}}, \emph {et~al.},\
  }\bibfield  {title} {\bibinfo {title} {\textit{Experimental Realization of
  Multipartite Entanglement via Quantum Fisher Information in a Uniform
  Antiferromagnetic Quantum Spin Chain}},\ }\href@noop {} {\bibfield  {journal}
  {\bibinfo  {journal} {Phys. Rev. Research}\ }\textbf {\bibinfo {volume}
  {2}},\ \bibinfo {pages} {043329} (\bibinfo {year} {2020})}\BibitemShut
  {NoStop}%
\bibitem [{\citenamefont {Laurell}\ \emph {et~al.}(2021)\citenamefont
  {Laurell}, \citenamefont {Scheie}, \citenamefont {Mukherjee}, \citenamefont
  {Koza}, \citenamefont {Enderle}, \citenamefont {Tylczynski}, \citenamefont
  {Okamoto}, \citenamefont {Coldea}, \citenamefont {Tennant},\ and\
  \citenamefont {Alvarez}}]{laurell2021quantifying}%
  \BibitemOpen
  \bibfield  {author} {\bibinfo {author} {\bibfnamefont {P.}~\bibnamefont
  {Laurell}}, \bibinfo {author} {\bibfnamefont {A.}~\bibnamefont {Scheie}},
  \bibinfo {author} {\bibfnamefont {C.~J.}\ \bibnamefont {Mukherjee}}, \bibinfo
  {author} {\bibfnamefont {M.~M.}\ \bibnamefont {Koza}}, \bibinfo {author}
  {\bibfnamefont {M.}~\bibnamefont {Enderle}}, \bibinfo {author} {\bibfnamefont
  {Z.}~\bibnamefont {Tylczynski}}, \bibinfo {author} {\bibfnamefont
  {S.}~\bibnamefont {Okamoto}}, \bibinfo {author} {\bibfnamefont
  {R.}~\bibnamefont {Coldea}}, \bibinfo {author} {\bibfnamefont {D.~A.}\
  \bibnamefont {Tennant}},\ and\ \bibinfo {author} {\bibfnamefont
  {G.}~\bibnamefont {Alvarez}},\ }\bibfield  {title} {\bibinfo {title}
  {\textit{Quantifying and Controlling Entanglement in the Quantum Magnet
  Cs$_2$ CoCl$_4$}},\ }\href@noop {} {\bibfield  {journal} {\bibinfo  {journal}
  {Phys. Rev. Lett.}\ }\textbf {\bibinfo {volume} {127}},\ \bibinfo {pages}
  {037201} (\bibinfo {year} {2021})}\BibitemShut {NoStop}%
\bibitem [{\citenamefont {Scheie}\ \emph {et~al.}(2021)\citenamefont {Scheie},
  \citenamefont {Laurell}, \citenamefont {Samarakoon}, \citenamefont {Lake},
  \citenamefont {Nagler}, \citenamefont {Granroth}, \citenamefont {Okamoto},
  \citenamefont {Alvarez},\ and\ \citenamefont
  {Tennant}}]{scheie2021witnessing}%
  \BibitemOpen
  \bibfield  {author} {\bibinfo {author} {\bibfnamefont {A.}~\bibnamefont
  {Scheie}}, \bibinfo {author} {\bibfnamefont {P.}~\bibnamefont {Laurell}},
  \bibinfo {author} {\bibfnamefont {A.}~\bibnamefont {Samarakoon}}, \bibinfo
  {author} {\bibfnamefont {B.}~\bibnamefont {Lake}}, \bibinfo {author}
  {\bibfnamefont {S.}~\bibnamefont {Nagler}}, \bibinfo {author} {\bibfnamefont
  {G.}~\bibnamefont {Granroth}}, \bibinfo {author} {\bibfnamefont
  {S.}~\bibnamefont {Okamoto}}, \bibinfo {author} {\bibfnamefont
  {G.}~\bibnamefont {Alvarez}},\ and\ \bibinfo {author} {\bibfnamefont
  {D.}~\bibnamefont {Tennant}},\ }\bibfield  {title} {\bibinfo {title}
  {\textit{Witnessing Entanglement in Quantum Magnets Using Neutron
  Scattering}},\ }\href@noop {} {\bibfield  {journal} {\bibinfo  {journal}
  {Phys. Rev. B}\ }\textbf {\bibinfo {volume} {103}},\ \bibinfo {pages}
  {224434} (\bibinfo {year} {2021})}\BibitemShut {NoStop}%
\bibitem [{\citenamefont {Zhang}\ and\ \citenamefont
  {Averitt}(2014)}]{zhang2014dynamics}%
  \BibitemOpen
  \bibfield  {author} {\bibinfo {author} {\bibfnamefont {J.}~\bibnamefont
  {Zhang}}\ and\ \bibinfo {author} {\bibfnamefont {R.}~\bibnamefont
  {Averitt}},\ }\bibfield  {title} {\bibinfo {title} {\textit{Dynamics and
  Control in Complex Transition Metal Oxides}},\ }\href@noop {} {\bibfield
  {journal} {\bibinfo  {journal} {Annu. Rev. Mater. Res.}\ }\textbf {\bibinfo
  {volume} {44}},\ \bibinfo {pages} {19} (\bibinfo {year} {2014})}\BibitemShut
  {NoStop}%
\bibitem [{\citenamefont {Basov}\ \emph {et~al.}(2017)\citenamefont {Basov},
  \citenamefont {Averitt},\ and\ \citenamefont {Hsieh}}]{basov2017towards}%
  \BibitemOpen
  \bibfield  {author} {\bibinfo {author} {\bibfnamefont {D.}~\bibnamefont
  {Basov}}, \bibinfo {author} {\bibfnamefont {R.}~\bibnamefont {Averitt}},\
  and\ \bibinfo {author} {\bibfnamefont {D.}~\bibnamefont {Hsieh}},\ }\bibfield
   {title} {\bibinfo {title} {\textit{Towards Properties on Demand in Quantum
  Materials}},\ }\href@noop {} {\bibfield  {journal} {\bibinfo  {journal} {Nat.
  Mater.}\ }\textbf {\bibinfo {volume} {16}},\ \bibinfo {pages} {1077}
  (\bibinfo {year} {2017})}\BibitemShut {NoStop}%
\bibitem [{\citenamefont {de~la Torre}\ \emph {et~al.}(2021)\citenamefont
  {de~la Torre}, \citenamefont {Kennes}, \citenamefont {Claassen},
  \citenamefont {Gerber}, \citenamefont {McIver},\ and\ \citenamefont
  {Sentef}}]{delatorre2021nonthermal}%
  \BibitemOpen
  \bibfield  {author} {\bibinfo {author} {\bibfnamefont {A.}~\bibnamefont
  {de~la Torre}}, \bibinfo {author} {\bibfnamefont {D.~M.}\ \bibnamefont
  {Kennes}}, \bibinfo {author} {\bibfnamefont {M.}~\bibnamefont {Claassen}},
  \bibinfo {author} {\bibfnamefont {S.}~\bibnamefont {Gerber}}, \bibinfo
  {author} {\bibfnamefont {J.~W.}\ \bibnamefont {McIver}},\ and\ \bibinfo
  {author} {\bibfnamefont {M.~A.}\ \bibnamefont {Sentef}},\ }\bibfield  {title}
  {\bibinfo {title} {\textit{Colloquium: Nonthermal Pathways to Ultrafast
  Control in Quantum Materials}},\ }\href@noop {} {\bibfield  {journal}
  {\bibinfo  {journal} {Rev. Mod. Phys.}\ }\textbf {\bibinfo {volume} {93}},\
  \bibinfo {pages} {041002} (\bibinfo {year} {2021})}\BibitemShut {NoStop}%
\bibitem [{\citenamefont {Alba}\ and\ \citenamefont
  {Calabrese}(2017)}]{alba2017entanglement}%
  \BibitemOpen
  \bibfield  {author} {\bibinfo {author} {\bibfnamefont {V.}~\bibnamefont
  {Alba}}\ and\ \bibinfo {author} {\bibfnamefont {P.}~\bibnamefont
  {Calabrese}},\ }\bibfield  {title} {\bibinfo {title} {\textit{Entanglement
  and Thermodynamics after a Quantum Quench in Integrable Systems}},\
  }\href@noop {} {\bibfield  {journal} {\bibinfo  {journal} {Proc. Natl. Acad.
  Sci. U.S.A.}\ }\textbf {\bibinfo {volume} {114}},\ \bibinfo {pages} {7947}
  (\bibinfo {year} {2017})}\BibitemShut {NoStop}%
\bibitem [{\citenamefont {Nahum}\ \emph {et~al.}(2017)\citenamefont {Nahum},
  \citenamefont {Ruhman}, \citenamefont {Vijay},\ and\ \citenamefont
  {Haah}}]{nahum2017quantum}%
  \BibitemOpen
  \bibfield  {author} {\bibinfo {author} {\bibfnamefont {A.}~\bibnamefont
  {Nahum}}, \bibinfo {author} {\bibfnamefont {J.}~\bibnamefont {Ruhman}},
  \bibinfo {author} {\bibfnamefont {S.}~\bibnamefont {Vijay}},\ and\ \bibinfo
  {author} {\bibfnamefont {J.}~\bibnamefont {Haah}},\ }\bibfield  {title}
  {\bibinfo {title} {\textit{Quantum Entanglement Growth under Random Unitary
  Dynamics}},\ }\href@noop {} {\bibfield  {journal} {\bibinfo  {journal} {Phys.
  Rev. X}\ }\textbf {\bibinfo {volume} {7}},\ \bibinfo {pages} {031016}
  (\bibinfo {year} {2017})}\BibitemShut {NoStop}%
\bibitem [{\citenamefont {Dean}\ \emph {et~al.}(2016)\citenamefont {Dean},
  \citenamefont {Cao}, \citenamefont {Liu}, \citenamefont {Wall}, \citenamefont
  {Zhu}, \citenamefont {Mankowsky}, \citenamefont {Thampy}, \citenamefont
  {Chen}, \citenamefont {Vale}, \citenamefont {Casa}, \citenamefont {Kim},
  \citenamefont {Said}, \citenamefont {Juhas}, \citenamefont {Alonso-Mori},
  \citenamefont {Glownia}, \citenamefont {Robert}, \citenamefont {Robinson},
  \citenamefont {Sikorski}, \citenamefont {Song}, \citenamefont {Kozina},
  \citenamefont {Lemke}, \citenamefont {Patthey}, \citenamefont {Owada},
  \citenamefont {Katayama}, \citenamefont {Yabashi}, \citenamefont {Tanaka},
  \citenamefont {Togashi}, \citenamefont {Liu}, \citenamefont {Serrao},
  \citenamefont {Kim}, \citenamefont {Huber}, \citenamefont {Chang},
  \citenamefont {McMorrow}, \citenamefont {F{\"o}rst},\ and\ \citenamefont
  {Hill}}]{dean2016ultrafast}%
  \BibitemOpen
  \bibfield  {author} {\bibinfo {author} {\bibfnamefont {M.~P.}\ \bibnamefont
  {Dean}}, \bibinfo {author} {\bibfnamefont {Y.}~\bibnamefont {Cao}}, \bibinfo
  {author} {\bibfnamefont {X.}~\bibnamefont {Liu}}, \bibinfo {author}
  {\bibfnamefont {S.}~\bibnamefont {Wall}}, \bibinfo {author} {\bibfnamefont
  {D.}~\bibnamefont {Zhu}}, \bibinfo {author} {\bibfnamefont {R.}~\bibnamefont
  {Mankowsky}}, \bibinfo {author} {\bibfnamefont {V.}~\bibnamefont {Thampy}},
  \bibinfo {author} {\bibfnamefont {X.}~\bibnamefont {Chen}}, \bibinfo {author}
  {\bibfnamefont {J.~G.}\ \bibnamefont {Vale}}, \bibinfo {author}
  {\bibfnamefont {D.}~\bibnamefont {Casa}}, \bibinfo {author} {\bibfnamefont
  {J.}~\bibnamefont {Kim}}, \bibinfo {author} {\bibfnamefont {A.}~\bibnamefont
  {Said}}, \bibinfo {author} {\bibfnamefont {P.}~\bibnamefont {Juhas}},
  \bibinfo {author} {\bibfnamefont {R.}~\bibnamefont {Alonso-Mori}}, \bibinfo
  {author} {\bibfnamefont {J.}~\bibnamefont {Glownia}}, \bibinfo {author}
  {\bibfnamefont {A.}~\bibnamefont {Robert}}, \bibinfo {author} {\bibfnamefont
  {J.}~\bibnamefont {Robinson}}, \bibinfo {author} {\bibfnamefont
  {M.}~\bibnamefont {Sikorski}}, \bibinfo {author} {\bibfnamefont
  {S.}~\bibnamefont {Song}}, \bibinfo {author} {\bibfnamefont {M.}~\bibnamefont
  {Kozina}}, \bibinfo {author} {\bibfnamefont {H.}~\bibnamefont {Lemke}},
  \bibinfo {author} {\bibfnamefont {L.}~\bibnamefont {Patthey}}, \bibinfo
  {author} {\bibfnamefont {S.}~\bibnamefont {Owada}}, \bibinfo {author}
  {\bibfnamefont {T.}~\bibnamefont {Katayama}}, \bibinfo {author}
  {\bibfnamefont {M.}~\bibnamefont {Yabashi}}, \bibinfo {author} {\bibfnamefont
  {Y.}~\bibnamefont {Tanaka}}, \bibinfo {author} {\bibfnamefont
  {T.}~\bibnamefont {Togashi}}, \bibinfo {author} {\bibfnamefont
  {J.}~\bibnamefont {Liu}}, \bibinfo {author} {\bibfnamefont {C.~R.}\
  \bibnamefont {Serrao}}, \bibinfo {author} {\bibfnamefont {B.~J.}\
  \bibnamefont {Kim}}, \bibinfo {author} {\bibfnamefont {L.}~\bibnamefont
  {Huber}}, \bibinfo {author} {\bibfnamefont {C.-L.}\ \bibnamefont {Chang}},
  \bibinfo {author} {\bibfnamefont {D.~F.}\ \bibnamefont {McMorrow}}, \bibinfo
  {author} {\bibfnamefont {M.}~\bibnamefont {F{\"o}rst}},\ and\ \bibinfo
  {author} {\bibfnamefont {J.~P.}\ \bibnamefont {Hill}},\ }\bibfield  {title}
  {\bibinfo {title} {\textit{Ultrafast Energy-and Momentum-resolved Dynamics of
  Magnetic Correlations in the Photo-doped Mott Insulator $Sr_2 IrO_4$}},\
  }\href@noop {} {\bibfield  {journal} {\bibinfo  {journal} {Nat. Mater.}\
  }\textbf {\bibinfo {volume} {15}},\ \bibinfo {pages} {601} (\bibinfo {year}
  {2016})}\BibitemShut {NoStop}%
\bibitem [{\citenamefont {Cao}\ \emph {et~al.}(2019)\citenamefont {Cao},
  \citenamefont {Mazzone}, \citenamefont {Meyers}, \citenamefont {Hill},
  \citenamefont {Liu}, \citenamefont {Wall},\ and\ \citenamefont
  {Dean}}]{cao2019ultrafast}%
  \BibitemOpen
  \bibfield  {author} {\bibinfo {author} {\bibfnamefont {Y.}~\bibnamefont
  {Cao}}, \bibinfo {author} {\bibfnamefont {D.}~\bibnamefont {Mazzone}},
  \bibinfo {author} {\bibfnamefont {D.}~\bibnamefont {Meyers}}, \bibinfo
  {author} {\bibfnamefont {J.}~\bibnamefont {Hill}}, \bibinfo {author}
  {\bibfnamefont {X.}~\bibnamefont {Liu}}, \bibinfo {author} {\bibfnamefont
  {S.}~\bibnamefont {Wall}},\ and\ \bibinfo {author} {\bibfnamefont
  {M.}~\bibnamefont {Dean}},\ }\bibfield  {title} {\bibinfo {title}
  {\textit{Ultrafast Dynamics of Spin and Orbital Correlations in Quantum
  Materials: an Energy-and Momentum-resolved Perspective}},\ }\href@noop {}
  {\bibfield  {journal} {\bibinfo  {journal} {Philos. Trans. R. Soc. A}\
  }\textbf {\bibinfo {volume} {377}},\ \bibinfo {pages} {20170480} (\bibinfo
  {year} {2019})}\BibitemShut {NoStop}%
\bibitem [{\citenamefont {Mitrano}\ and\ \citenamefont
  {Wang}(2020)}]{mitrano2020probing}%
  \BibitemOpen
  \bibfield  {author} {\bibinfo {author} {\bibfnamefont {M.}~\bibnamefont
  {Mitrano}}\ and\ \bibinfo {author} {\bibfnamefont {Y.}~\bibnamefont {Wang}},\
  }\bibfield  {title} {\bibinfo {title} {\textit{Probing Light-driven Quantum
  Materials with Ultrafast Resonant Inelastic X-ray Scattering}},\ }\href@noop
  {} {\bibfield  {journal} {\bibinfo  {journal} {Commun. Phys.}\ }\textbf
  {\bibinfo {volume} {3}},\ \bibinfo {pages} {184} (\bibinfo {year}
  {2020})}\BibitemShut {NoStop}%
\bibitem [{\citenamefont {Mitrano}\ \emph
  {et~al.}(2019{\natexlab{a}})\citenamefont {Mitrano}, \citenamefont {Lee},
  \citenamefont {Husain}, \citenamefont {Delacretaz}, \citenamefont {Zhu},
  \citenamefont {de~la Pe{\~n}a~Munoz}, \citenamefont {Sun}, \citenamefont
  {Joe}, \citenamefont {Reid}, \citenamefont {Wandel}, \citenamefont
  {Coslovich}, \citenamefont {Schlotter}, \citenamefont {VanDriel},
  \citenamefont {Schneeloch}, \citenamefont {Gu}, \citenamefont {Hartnoll},
  \citenamefont {Goldenfeld},\ and\ \citenamefont
  {Abbamonte}}]{mitrano2019ultrafast}%
  \BibitemOpen
  \bibfield  {author} {\bibinfo {author} {\bibfnamefont {M.}~\bibnamefont
  {Mitrano}}, \bibinfo {author} {\bibfnamefont {S.}~\bibnamefont {Lee}},
  \bibinfo {author} {\bibfnamefont {A.~A.}\ \bibnamefont {Husain}}, \bibinfo
  {author} {\bibfnamefont {L.}~\bibnamefont {Delacretaz}}, \bibinfo {author}
  {\bibfnamefont {M.}~\bibnamefont {Zhu}}, \bibinfo {author} {\bibfnamefont
  {G.}~\bibnamefont {de~la Pe{\~n}a~Munoz}}, \bibinfo {author} {\bibfnamefont
  {S.~X.-L.}\ \bibnamefont {Sun}}, \bibinfo {author} {\bibfnamefont {Y.~I.}\
  \bibnamefont {Joe}}, \bibinfo {author} {\bibfnamefont {A.~H.}\ \bibnamefont
  {Reid}}, \bibinfo {author} {\bibfnamefont {S.~F.}\ \bibnamefont {Wandel}},
  \bibinfo {author} {\bibfnamefont {G.}~\bibnamefont {Coslovich}}, \bibinfo
  {author} {\bibfnamefont {W.}~\bibnamefont {Schlotter}}, \bibinfo {author}
  {\bibfnamefont {T.}~\bibnamefont {VanDriel}}, \bibinfo {author}
  {\bibfnamefont {J.}~\bibnamefont {Schneeloch}}, \bibinfo {author}
  {\bibfnamefont {G.}~\bibnamefont {Gu}}, \bibinfo {author} {\bibfnamefont
  {S.}~\bibnamefont {Hartnoll}}, \bibinfo {author} {\bibfnamefont
  {N.}~\bibnamefont {Goldenfeld}},\ and\ \bibinfo {author} {\bibfnamefont
  {P.}~\bibnamefont {Abbamonte}},\ }\bibfield  {title} {\bibinfo {title}
  {\textit{Ultrafast Time-Resolved X-ray Scattering Reveals Diffusive Charge
  Order Dynamics in La2--xBaxCuO4}},\ }\href@noop {} {\bibfield  {journal}
  {\bibinfo  {journal} {Sci. Adv.}\ }\textbf {\bibinfo {volume} {5}},\ \bibinfo
  {pages} {eaax3346} (\bibinfo {year} {2019}{\natexlab{a}})}\BibitemShut
  {NoStop}%
\bibitem [{\citenamefont {Mazzone}\ \emph {et~al.}(2021)\citenamefont
  {Mazzone}, \citenamefont {Meyers}, \citenamefont {Cao}, \citenamefont {Vale},
  \citenamefont {Dashwood}, \citenamefont {Shi}, \citenamefont {James},
  \citenamefont {Robinson}, \citenamefont {Lin}, \citenamefont {Thampy},
  \citenamefont {Tanaka}, \citenamefont {Johnson}, \citenamefont {Miao},
  \citenamefont {Wang}, \citenamefont {Assefa}, \citenamefont {Kim},
  \citenamefont {Casa}, \citenamefont {Mankowsky}, \citenamefont {Zhu},
  \citenamefont {Alonso-Mori}, \citenamefont {Song}, \citenamefont {Yavas},
  \citenamefont {Katayama}, \citenamefont {Yabashi}, \citenamefont {Kubota},
  \citenamefont {Owada}, \citenamefont {Liu}, \citenamefont {Yang},
  \citenamefont {Konik}, \citenamefont {Robinson}, \citenamefont {Hill},
  \citenamefont {McMorrow}, \citenamefont {F{\"o}rst}, \citenamefont {Wall},
  \citenamefont {Liu},\ and\ \citenamefont {Dean}}]{mazzone2021laser}%
  \BibitemOpen
  \bibfield  {author} {\bibinfo {author} {\bibfnamefont {D.~G.}\ \bibnamefont
  {Mazzone}}, \bibinfo {author} {\bibfnamefont {D.}~\bibnamefont {Meyers}},
  \bibinfo {author} {\bibfnamefont {Y.}~\bibnamefont {Cao}}, \bibinfo {author}
  {\bibfnamefont {J.~G.}\ \bibnamefont {Vale}}, \bibinfo {author}
  {\bibfnamefont {C.~D.}\ \bibnamefont {Dashwood}}, \bibinfo {author}
  {\bibfnamefont {Y.}~\bibnamefont {Shi}}, \bibinfo {author} {\bibfnamefont
  {A.~J.~A.}\ \bibnamefont {James}}, \bibinfo {author} {\bibfnamefont {N.~J.}\
  \bibnamefont {Robinson}}, \bibinfo {author} {\bibfnamefont {J.}~\bibnamefont
  {Lin}}, \bibinfo {author} {\bibfnamefont {V.}~\bibnamefont {Thampy}},
  \bibinfo {author} {\bibfnamefont {Y.}~\bibnamefont {Tanaka}}, \bibinfo
  {author} {\bibfnamefont {A.~S.}\ \bibnamefont {Johnson}}, \bibinfo {author}
  {\bibfnamefont {H.}~\bibnamefont {Miao}}, \bibinfo {author} {\bibfnamefont
  {R.}~\bibnamefont {Wang}}, \bibinfo {author} {\bibfnamefont {T.~A.}\
  \bibnamefont {Assefa}}, \bibinfo {author} {\bibfnamefont {J.}~\bibnamefont
  {Kim}}, \bibinfo {author} {\bibfnamefont {D.}~\bibnamefont {Casa}}, \bibinfo
  {author} {\bibfnamefont {R.}~\bibnamefont {Mankowsky}}, \bibinfo {author}
  {\bibfnamefont {D.}~\bibnamefont {Zhu}}, \bibinfo {author} {\bibfnamefont
  {R.}~\bibnamefont {Alonso-Mori}}, \bibinfo {author} {\bibfnamefont
  {S.}~\bibnamefont {Song}}, \bibinfo {author} {\bibfnamefont {H.}~\bibnamefont
  {Yavas}}, \bibinfo {author} {\bibfnamefont {T.}~\bibnamefont {Katayama}},
  \bibinfo {author} {\bibfnamefont {M.}~\bibnamefont {Yabashi}}, \bibinfo
  {author} {\bibfnamefont {Y.}~\bibnamefont {Kubota}}, \bibinfo {author}
  {\bibfnamefont {S.}~\bibnamefont {Owada}}, \bibinfo {author} {\bibfnamefont
  {J.}~\bibnamefont {Liu}}, \bibinfo {author} {\bibfnamefont {J.}~\bibnamefont
  {Yang}}, \bibinfo {author} {\bibfnamefont {R.~M.}\ \bibnamefont {Konik}},
  \bibinfo {author} {\bibfnamefont {I.~K.}\ \bibnamefont {Robinson}}, \bibinfo
  {author} {\bibfnamefont {J.~P.}\ \bibnamefont {Hill}}, \bibinfo {author}
  {\bibfnamefont {D.~F.}\ \bibnamefont {McMorrow}}, \bibinfo {author}
  {\bibfnamefont {M.}~\bibnamefont {F{\"o}rst}}, \bibinfo {author}
  {\bibfnamefont {S.}~\bibnamefont {Wall}}, \bibinfo {author} {\bibfnamefont
  {X.}~\bibnamefont {Liu}},\ and\ \bibinfo {author} {\bibfnamefont {M.~P.~M.}\
  \bibnamefont {Dean}},\ }\bibfield  {title} {\bibinfo {title}
  {\textit{Laser-Induced Transient Magnons in Sr$_3$Ir$_2$O$_7$ Throughout the
  Brillouin Zone}},\ }\href@noop {} {\bibfield  {journal} {\bibinfo  {journal}
  {Proc. Natl. Acad. Sci. U.S.A.}\ }\textbf {\bibinfo {volume} {118}} (\bibinfo
  {year} {2021})}\BibitemShut {NoStop}%
\bibitem [{\citenamefont {Mitrano}\ \emph
  {et~al.}(2019{\natexlab{b}})\citenamefont {Mitrano}, \citenamefont {Lee},
  \citenamefont {Husain}, \citenamefont {Zhu}, \citenamefont {de~la
  Pe{\~n}a~Munoz}, \citenamefont {Sun}, \citenamefont {Joe}, \citenamefont
  {Reid}, \citenamefont {Wandel}, \citenamefont {Coslovich}, \citenamefont
  {Schlotter}, \citenamefont {vanDriel}, \citenamefont {Schneeloch},
  \citenamefont {Gu}, \citenamefont {Goldenfeld},\ and\ \citenamefont
  {Abbamonte}}]{mitrano2019evidence}%
  \BibitemOpen
  \bibfield  {author} {\bibinfo {author} {\bibfnamefont {M.}~\bibnamefont
  {Mitrano}}, \bibinfo {author} {\bibfnamefont {S.}~\bibnamefont {Lee}},
  \bibinfo {author} {\bibfnamefont {A.~A.}\ \bibnamefont {Husain}}, \bibinfo
  {author} {\bibfnamefont {M.}~\bibnamefont {Zhu}}, \bibinfo {author}
  {\bibfnamefont {G.}~\bibnamefont {de~la Pe{\~n}a~Munoz}}, \bibinfo {author}
  {\bibfnamefont {S.~X.-L.}\ \bibnamefont {Sun}}, \bibinfo {author}
  {\bibfnamefont {Y.~I.}\ \bibnamefont {Joe}}, \bibinfo {author} {\bibfnamefont
  {A.~H.}\ \bibnamefont {Reid}}, \bibinfo {author} {\bibfnamefont {S.~F.}\
  \bibnamefont {Wandel}}, \bibinfo {author} {\bibfnamefont {G.}~\bibnamefont
  {Coslovich}}, \bibinfo {author} {\bibfnamefont {W.}~\bibnamefont
  {Schlotter}}, \bibinfo {author} {\bibfnamefont {T.}~\bibnamefont {vanDriel}},
  \bibinfo {author} {\bibfnamefont {J.}~\bibnamefont {Schneeloch}}, \bibinfo
  {author} {\bibfnamefont {G.}~\bibnamefont {Gu}}, \bibinfo {author}
  {\bibfnamefont {N.}~\bibnamefont {Goldenfeld}},\ and\ \bibinfo {author}
  {\bibfnamefont {P.}~\bibnamefont {Abbamonte}},\ }\bibfield  {title} {\bibinfo
  {title} {\textit{Evidence for Photoinduced Sliding of the Charge-order
  Condensate in La 1.875 Ba 0.125 CuO$_ 4$}},\ }\href@noop {} {\bibfield
  {journal} {\bibinfo  {journal} {Phys. Rev. B}\ }\textbf {\bibinfo {volume}
  {100}},\ \bibinfo {pages} {205125} (\bibinfo {year}
  {2019}{\natexlab{b}})}\BibitemShut {NoStop}%
\bibitem [{\citenamefont {Parchenko}\ \emph {et~al.}(2020)\citenamefont
  {Parchenko}, \citenamefont {Paris}, \citenamefont {McNally}, \citenamefont
  {Abreu}, \citenamefont {Dantz}, \citenamefont {Bothschafter}, \citenamefont
  {Reid}, \citenamefont {Schlotter}, \citenamefont {Lin}, \citenamefont
  {Wandel}, \citenamefont {Coslovich}, \citenamefont {Zohar}, \citenamefont
  {Dakovski}, \citenamefont {Turner}, \citenamefont {Moeller}, \citenamefont
  {Tseng}, \citenamefont {Radovic}, \citenamefont {Saathe}, \citenamefont
  {Agaaker}, \citenamefont {Nordgren}, \citenamefont {Johnson}, \citenamefont
  {Schmitt},\ and\ \citenamefont {Staub}}]{parchenko2020orbital}%
  \BibitemOpen
  \bibfield  {author} {\bibinfo {author} {\bibfnamefont {S.}~\bibnamefont
  {Parchenko}}, \bibinfo {author} {\bibfnamefont {E.}~\bibnamefont {Paris}},
  \bibinfo {author} {\bibfnamefont {D.}~\bibnamefont {McNally}}, \bibinfo
  {author} {\bibfnamefont {E.}~\bibnamefont {Abreu}}, \bibinfo {author}
  {\bibfnamefont {M.}~\bibnamefont {Dantz}}, \bibinfo {author} {\bibfnamefont
  {E.~M.}\ \bibnamefont {Bothschafter}}, \bibinfo {author} {\bibfnamefont
  {A.~H.}\ \bibnamefont {Reid}}, \bibinfo {author} {\bibfnamefont {W.~F.}\
  \bibnamefont {Schlotter}}, \bibinfo {author} {\bibfnamefont {M.-F.}\
  \bibnamefont {Lin}}, \bibinfo {author} {\bibfnamefont {S.~F.}\ \bibnamefont
  {Wandel}}, \bibinfo {author} {\bibfnamefont {G.}~\bibnamefont {Coslovich}},
  \bibinfo {author} {\bibfnamefont {S.}~\bibnamefont {Zohar}}, \bibinfo
  {author} {\bibfnamefont {G.~L.}\ \bibnamefont {Dakovski}}, \bibinfo {author}
  {\bibfnamefont {J.~J.}\ \bibnamefont {Turner}}, \bibinfo {author}
  {\bibfnamefont {S.}~\bibnamefont {Moeller}}, \bibinfo {author} {\bibfnamefont
  {Y.}~\bibnamefont {Tseng}}, \bibinfo {author} {\bibfnamefont
  {M.}~\bibnamefont {Radovic}}, \bibinfo {author} {\bibfnamefont
  {C.}~\bibnamefont {Saathe}}, \bibinfo {author} {\bibfnamefont
  {M.}~\bibnamefont {Agaaker}}, \bibinfo {author} {\bibfnamefont {J.~E.}\
  \bibnamefont {Nordgren}}, \bibinfo {author} {\bibfnamefont {S.~L.}\
  \bibnamefont {Johnson}}, \bibinfo {author} {\bibfnamefont {T.}~\bibnamefont
  {Schmitt}},\ and\ \bibinfo {author} {\bibfnamefont {U.}~\bibnamefont
  {Staub}},\ }\bibfield  {title} {\bibinfo {title} {\textit{Orbital Dynamics
  during an Ultrafast Insulator to Metal Transition}},\ }\href@noop {}
  {\bibfield  {journal} {\bibinfo  {journal} {Phys. Rev. Research}\ }\textbf
  {\bibinfo {volume} {2}},\ \bibinfo {pages} {023110} (\bibinfo {year}
  {2020})}\BibitemShut {NoStop}%
\bibitem [{\citenamefont {Pang}\ and\ \citenamefont
  {Jordan}(2017)}]{pang2017optimal}%
  \BibitemOpen
  \bibfield  {author} {\bibinfo {author} {\bibfnamefont {S.}~\bibnamefont
  {Pang}}\ and\ \bibinfo {author} {\bibfnamefont {A.~N.}\ \bibnamefont
  {Jordan}},\ }\bibfield  {title} {\bibinfo {title} {\textit{Optimal Adaptive
  Control for Quantum Metrology with Time-dependent Hamiltonians}},\
  }\href@noop {} {\bibfield  {journal} {\bibinfo  {journal} {Nat. Commun.}\
  }\textbf {\bibinfo {volume} {8}},\ \bibinfo {pages} {14695} (\bibinfo {year}
  {2017})}\BibitemShut {NoStop}%
\bibitem [{\citenamefont {de~Almeida}\ and\ \citenamefont
  {Hauke}(2021)}]{de2021entanglement}%
  \BibitemOpen
  \bibfield  {author} {\bibinfo {author} {\bibfnamefont {R.~C.}\ \bibnamefont
  {de~Almeida}}\ and\ \bibinfo {author} {\bibfnamefont {P.}~\bibnamefont
  {Hauke}},\ }\bibfield  {title} {\bibinfo {title} {\textit{From Entanglement
  Certification with Quench Dynamics to Multipartite Entanglement of
  Interacting Fermions}},\ }\href@noop {} {\bibfield  {journal} {\bibinfo
  {journal} {Phys. Rev. Research}\ }\textbf {\bibinfo {volume} {3}},\ \bibinfo
  {pages} {L032051} (\bibinfo {year} {2021})}\BibitemShut {NoStop}%
\bibitem [{\citenamefont {Chen}\ \emph {et~al.}(2021)\citenamefont {Chen},
  \citenamefont {Wang}, \citenamefont {Rebec}, \citenamefont {Jia},
  \citenamefont {Hashimoto}, \citenamefont {Lu}, \citenamefont {Moritz},
  \citenamefont {Moore}, \citenamefont {Devereaux},\ and\ \citenamefont
  {Shen}}]{chen2021anomalously}%
  \BibitemOpen
  \bibfield  {author} {\bibinfo {author} {\bibfnamefont {Z.}~\bibnamefont
  {Chen}}, \bibinfo {author} {\bibfnamefont {Y.}~\bibnamefont {Wang}}, \bibinfo
  {author} {\bibfnamefont {S.~N.}\ \bibnamefont {Rebec}}, \bibinfo {author}
  {\bibfnamefont {T.}~\bibnamefont {Jia}}, \bibinfo {author} {\bibfnamefont
  {M.}~\bibnamefont {Hashimoto}}, \bibinfo {author} {\bibfnamefont
  {D.}~\bibnamefont {Lu}}, \bibinfo {author} {\bibfnamefont {B.}~\bibnamefont
  {Moritz}}, \bibinfo {author} {\bibfnamefont {R.~G.}\ \bibnamefont {Moore}},
  \bibinfo {author} {\bibfnamefont {T.~P.}\ \bibnamefont {Devereaux}},\ and\
  \bibinfo {author} {\bibfnamefont {Z.-X.}\ \bibnamefont {Shen}},\ }\bibfield
  {title} {\bibinfo {title} {\textit{Anomalously Strong Near-neighbor
  Attraction in Doped 1D Cuprate Chains}},\ }\href@noop {} {\bibfield
  {journal} {\bibinfo  {journal} {Science}\ }\textbf {\bibinfo {volume}
  {373}},\ \bibinfo {pages} {1235} (\bibinfo {year} {2021})}\BibitemShut
  {NoStop}%
\bibitem [{\citenamefont {Wang}\ \emph {et~al.}(2017)\citenamefont {Wang},
  \citenamefont {Claassen}, \citenamefont {Moritz},\ and\ \citenamefont
  {Devereaux}}]{wang2017producing}%
  \BibitemOpen
  \bibfield  {author} {\bibinfo {author} {\bibfnamefont {Y.}~\bibnamefont
  {Wang}}, \bibinfo {author} {\bibfnamefont {M.}~\bibnamefont {Claassen}},
  \bibinfo {author} {\bibfnamefont {B.}~\bibnamefont {Moritz}},\ and\ \bibinfo
  {author} {\bibfnamefont {T.}~\bibnamefont {Devereaux}},\ }\bibfield  {title}
  {\bibinfo {title} {\textit{Producing Coherent Excitations in Pumped Mott
  Antiferromagnetic Insulators}},\ }\href@noop {} {\bibfield  {journal}
  {\bibinfo  {journal} {Phys. Rev. B}\ }\textbf {\bibinfo {volume} {96}},\
  \bibinfo {pages} {235142} (\bibinfo {year} {2017})}\BibitemShut {NoStop}%
\bibitem [{\citenamefont {Freericks}\ \emph {et~al.}(2009)\citenamefont
  {Freericks}, \citenamefont {Krishnamurthy},\ and\ \citenamefont
  {Pruschke}}]{freericks2009theoretical}%
  \BibitemOpen
  \bibfield  {author} {\bibinfo {author} {\bibfnamefont {J.}~\bibnamefont
  {Freericks}}, \bibinfo {author} {\bibfnamefont {H.}~\bibnamefont
  {Krishnamurthy}},\ and\ \bibinfo {author} {\bibfnamefont {T.}~\bibnamefont
  {Pruschke}},\ }\bibfield  {title} {\bibinfo {title} {\textit{Theoretical
  Description of Time-resolved Photoemission Spectroscopy: Application to
  Pump-probe Experiments}},\ }\href@noop {} {\bibfield  {journal} {\bibinfo
  {journal} {Phys. Rev. Lett.}\ }\textbf {\bibinfo {volume} {102}},\ \bibinfo
  {pages} {136401} (\bibinfo {year} {2009})}\BibitemShut {NoStop}%
\bibitem [{\citenamefont {Lovesey}(1984)}]{Lovesey1984theory}%
  \BibitemOpen
  \bibfield  {author} {\bibinfo {author} {\bibfnamefont {S.~W.}\ \bibnamefont
  {Lovesey}},\ }\href@noop {} {\emph {\bibinfo {title} {Theory of Neutron
  Scattering from Condensed Matter}}}\ (\bibinfo  {publisher} {Oxford
  University Press},\ \bibinfo {year} {1984})\BibitemShut {NoStop}%
\bibitem [{\citenamefont {Ament}\ \emph {et~al.}(2009)\citenamefont {Ament},
  \citenamefont {Ghiringhelli}, \citenamefont {Sala}, \citenamefont
  {Braicovich},\ and\ \citenamefont {van~den Brink}}]{ament2009theoretical}%
  \BibitemOpen
  \bibfield  {author} {\bibinfo {author} {\bibfnamefont {L.~J.}\ \bibnamefont
  {Ament}}, \bibinfo {author} {\bibfnamefont {G.}~\bibnamefont {Ghiringhelli}},
  \bibinfo {author} {\bibfnamefont {M.~M.}\ \bibnamefont {Sala}}, \bibinfo
  {author} {\bibfnamefont {L.}~\bibnamefont {Braicovich}},\ and\ \bibinfo
  {author} {\bibfnamefont {J.}~\bibnamefont {van~den Brink}},\ }\bibfield
  {title} {\bibinfo {title} {\textit{Theoretical Demonstration of How the
  Dispersion of Magnetic Excitations in Cuprate Compounds Can Be Determined
  Using Resonant Inelastic X-Ray Scattering}},\ }\href@noop {} {\bibfield
  {journal} {\bibinfo  {journal} {Phys. Rev. Lett.}\ }\textbf {\bibinfo
  {volume} {103}},\ \bibinfo {pages} {117003} (\bibinfo {year}
  {2009})}\BibitemShut {NoStop}%
\bibitem [{\citenamefont {Wang}\ \emph
  {et~al.}(2021{\natexlab{a}})\citenamefont {Wang}, \citenamefont {Chen},
  \citenamefont {Devereaux}, \citenamefont {Moritz},\ and\ \citenamefont
  {Mitrano}}]{wang2021x}%
  \BibitemOpen
  \bibfield  {author} {\bibinfo {author} {\bibfnamefont {Y.}~\bibnamefont
  {Wang}}, \bibinfo {author} {\bibfnamefont {Y.}~\bibnamefont {Chen}}, \bibinfo
  {author} {\bibfnamefont {T.~P.}\ \bibnamefont {Devereaux}}, \bibinfo {author}
  {\bibfnamefont {B.}~\bibnamefont {Moritz}},\ and\ \bibinfo {author}
  {\bibfnamefont {M.}~\bibnamefont {Mitrano}},\ }\bibfield  {title} {\bibinfo
  {title} {\textit{X-Ray Scattering from Light-Driven Spin Fluctuations in a
  Doped Mott Insulator}},\ }\href@noop {} {\bibfield  {journal} {\bibinfo
  {journal} {Commun. Phys.}\ }\textbf {\bibinfo {volume} {4}},\ \bibinfo
  {pages} {212} (\bibinfo {year} {2021}{\natexlab{a}})}\BibitemShut {NoStop}%
\bibitem [{\citenamefont {Jia}\ \emph {et~al.}(2016)\citenamefont {Jia},
  \citenamefont {Wohlfeld}, \citenamefont {Wang}, \citenamefont {Moritz},\ and\
  \citenamefont {Devereaux}}]{jia2016using}%
  \BibitemOpen
  \bibfield  {author} {\bibinfo {author} {\bibfnamefont {C.}~\bibnamefont
  {Jia}}, \bibinfo {author} {\bibfnamefont {K.}~\bibnamefont {Wohlfeld}},
  \bibinfo {author} {\bibfnamefont {Y.}~\bibnamefont {Wang}}, \bibinfo {author}
  {\bibfnamefont {B.}~\bibnamefont {Moritz}},\ and\ \bibinfo {author}
  {\bibfnamefont {T.~P.}\ \bibnamefont {Devereaux}},\ }\bibfield  {title}
  {\bibinfo {title} {\textit{Using RIXS to Uncover Elementary Charge and Spin
  Excitations}},\ }\href@noop {} {\bibfield  {journal} {\bibinfo  {journal}
  {Phy. Rev. X}\ }\textbf {\bibinfo {volume} {6}},\ \bibinfo {pages} {021020}
  (\bibinfo {year} {2016})}\BibitemShut {NoStop}%
\bibitem [{\citenamefont {Schlappa}\ \emph {et~al.}(2012)\citenamefont
  {Schlappa}, \citenamefont {Wohlfeld}, \citenamefont {Zhou}, \citenamefont
  {Mourigal}, \citenamefont {Haverkort}, \citenamefont {Strocov}, \citenamefont
  {Hozoi}, \citenamefont {Monney}, \citenamefont {Nishimoto}, \citenamefont
  {Singh}, \citenamefont {Revcolevschi}, \citenamefont {Caux}, \citenamefont
  {Patthey}, \citenamefont {R{\o}nnow}, \citenamefont {van~den Brink},\ and\
  \citenamefont {Schmitt}}]{Schlappa2012spin}%
  \BibitemOpen
  \bibfield  {author} {\bibinfo {author} {\bibfnamefont {J.}~\bibnamefont
  {Schlappa}}, \bibinfo {author} {\bibfnamefont {K.}~\bibnamefont {Wohlfeld}},
  \bibinfo {author} {\bibfnamefont {K.~J.}\ \bibnamefont {Zhou}}, \bibinfo
  {author} {\bibfnamefont {M.}~\bibnamefont {Mourigal}}, \bibinfo {author}
  {\bibfnamefont {M.~W.}\ \bibnamefont {Haverkort}}, \bibinfo {author}
  {\bibfnamefont {V.~N.}\ \bibnamefont {Strocov}}, \bibinfo {author}
  {\bibfnamefont {L.}~\bibnamefont {Hozoi}}, \bibinfo {author} {\bibfnamefont
  {C.}~\bibnamefont {Monney}}, \bibinfo {author} {\bibfnamefont
  {S.}~\bibnamefont {Nishimoto}}, \bibinfo {author} {\bibfnamefont
  {S.}~\bibnamefont {Singh}}, \bibinfo {author} {\bibfnamefont
  {A.}~\bibnamefont {Revcolevschi}}, \bibinfo {author} {\bibfnamefont {J.-S.}\
  \bibnamefont {Caux}}, \bibinfo {author} {\bibfnamefont {L.}~\bibnamefont
  {Patthey}}, \bibinfo {author} {\bibfnamefont {H.~M.}\ \bibnamefont
  {R{\o}nnow}}, \bibinfo {author} {\bibfnamefont {J.}~\bibnamefont {van~den
  Brink}},\ and\ \bibinfo {author} {\bibfnamefont {T.}~\bibnamefont
  {Schmitt}},\ }\bibfield  {title} {\bibinfo {title} {Spin--orbital separation
  in the quasi-one-dimensional mott insulator sr$_2$cuo$_3$},\ }\href@noop {}
  {\bibfield  {journal} {\bibinfo  {journal} {Nature}\ }\textbf {\bibinfo
  {volume} {485}},\ \bibinfo {pages} {82} (\bibinfo {year} {2012})}\BibitemShut
  {NoStop}%
\bibitem [{\citenamefont {Wang}\ \emph
  {et~al.}(2021{\natexlab{b}})\citenamefont {Wang}, \citenamefont {Chen},
  \citenamefont {Shi}, \citenamefont {Moritz},\ and\ \citenamefont
  {Devereaux}}]{wang2021phonon}%
  \BibitemOpen
  \bibfield  {author} {\bibinfo {author} {\bibfnamefont {Y.}~\bibnamefont
  {Wang}}, \bibinfo {author} {\bibfnamefont {Z.}~\bibnamefont {Chen}}, \bibinfo
  {author} {\bibfnamefont {T.}~\bibnamefont {Shi}}, \bibinfo {author}
  {\bibfnamefont {B.}~\bibnamefont {Moritz}},\ and\ \bibinfo {author}
  {\bibfnamefont {T.~P.}\ \bibnamefont {Devereaux}},\ }\bibfield  {title}
  {\bibinfo {title} {\textit{Phonon-Mediated Long-Range Attractive Interaction
  in One-Dimensional Cuprates}},\ }\href@noop {} {\bibfield  {journal}
  {\bibinfo  {journal} {Phys. Rev. Lett.}\ }\textbf {\bibinfo {volume} {127}},\
  \bibinfo {pages} {197003} (\bibinfo {year} {2021}{\natexlab{b}})}\BibitemShut
  {NoStop}%
\bibitem [{\citenamefont {Lehoucq}\ \emph {et~al.}(1998)\citenamefont
  {Lehoucq}, \citenamefont {Sorensen},\ and\ \citenamefont
  {Yang}}]{lehoucq1998arpack}%
  \BibitemOpen
  \bibfield  {author} {\bibinfo {author} {\bibfnamefont {R.~B.}\ \bibnamefont
  {Lehoucq}}, \bibinfo {author} {\bibfnamefont {D.~C.}\ \bibnamefont
  {Sorensen}},\ and\ \bibinfo {author} {\bibfnamefont {C.}~\bibnamefont
  {Yang}},\ }\href@noop {} {\emph {\bibinfo {title} {\textit{ARPACK Users'
  Guide: Solution of Large-scale Eigenvalue Problems with Implicitly Restarted
  Arnoldi Methods}}}}\ (\bibinfo  {publisher} {SIAM},\ \bibinfo {year}
  {1998})\BibitemShut {NoStop}%
\bibitem [{\citenamefont {Jia}\ \emph {et~al.}(2018)\citenamefont {Jia},
  \citenamefont {Wang}, \citenamefont {Mendl}, \citenamefont {Moritz},\ and\
  \citenamefont {Devereaux}}]{jia2017paradeisos}%
  \BibitemOpen
  \bibfield  {author} {\bibinfo {author} {\bibfnamefont {C.}~\bibnamefont
  {Jia}}, \bibinfo {author} {\bibfnamefont {Y.}~\bibnamefont {Wang}}, \bibinfo
  {author} {\bibfnamefont {C.}~\bibnamefont {Mendl}}, \bibinfo {author}
  {\bibfnamefont {B.}~\bibnamefont {Moritz}},\ and\ \bibinfo {author}
  {\bibfnamefont {T.}~\bibnamefont {Devereaux}},\ }\bibfield  {title} {\bibinfo
  {title} {\textit{Paradeisos: a Perfect Hashing Algorithm for Many-body
  Eigenvalue Problems}},\ }\href@noop {} {\bibfield  {journal} {\bibinfo
  {journal} {Comput. Phys. Commun.}\ }\textbf {\bibinfo {volume} {224}},\
  \bibinfo {pages} {81} (\bibinfo {year} {2018})}\BibitemShut {NoStop}%
\bibitem [{\citenamefont {Manmana}\ \emph {et~al.}(2007)\citenamefont
  {Manmana}, \citenamefont {Wessel}, \citenamefont {Noack},\ and\ \citenamefont
  {Muramatsu}}]{manmana2007strongly}%
  \BibitemOpen
  \bibfield  {author} {\bibinfo {author} {\bibfnamefont {S.~R.}\ \bibnamefont
  {Manmana}}, \bibinfo {author} {\bibfnamefont {S.}~\bibnamefont {Wessel}},
  \bibinfo {author} {\bibfnamefont {R.~M.}\ \bibnamefont {Noack}},\ and\
  \bibinfo {author} {\bibfnamefont {A.}~\bibnamefont {Muramatsu}},\ }\bibfield
  {title} {\bibinfo {title} {\textit{Strongly Correlated Fermions After a
  Quantum Quench}},\ }\href@noop {} {\bibfield  {journal} {\bibinfo  {journal}
  {Phys. Rev. Lett.}\ }\textbf {\bibinfo {volume} {98}},\ \bibinfo {pages}
  {210405} (\bibinfo {year} {2007})}\BibitemShut {NoStop}%
\bibitem [{\citenamefont {Balzer}\ \emph {et~al.}(2011)\citenamefont {Balzer},
  \citenamefont {Gdaniec},\ and\ \citenamefont {Potthoff}}]{balzer2011krylov}%
  \BibitemOpen
  \bibfield  {author} {\bibinfo {author} {\bibfnamefont {M.}~\bibnamefont
  {Balzer}}, \bibinfo {author} {\bibfnamefont {N.}~\bibnamefont {Gdaniec}},\
  and\ \bibinfo {author} {\bibfnamefont {M.}~\bibnamefont {Potthoff}},\
  }\bibfield  {title} {\bibinfo {title} {\textit{Krylov-space Approach to the
  Equilibrium and Nonequilibrium Single-particle Green’s Function}},\
  }\href@noop {} {\bibfield  {journal} {\bibinfo  {journal} {J. Phys. Condens.
  Matter}\ }\textbf {\bibinfo {volume} {24}},\ \bibinfo {pages} {035603}
  (\bibinfo {year} {2011})}\BibitemShut {NoStop}%
\bibitem [{\citenamefont {Lin}\ \emph {et~al.}(1995)\citenamefont {Lin},
  \citenamefont {Gagliano}, \citenamefont {Campbell}, \citenamefont {Fradkin},\
  and\ \citenamefont {Gubernatis}}]{lin1995phase}%
  \BibitemOpen
  \bibfield  {author} {\bibinfo {author} {\bibfnamefont {H.}~\bibnamefont
  {Lin}}, \bibinfo {author} {\bibfnamefont {E.}~\bibnamefont {Gagliano}},
  \bibinfo {author} {\bibfnamefont {D.}~\bibnamefont {Campbell}}, \bibinfo
  {author} {\bibfnamefont {E.}~\bibnamefont {Fradkin}},\ and\ \bibinfo {author}
  {\bibfnamefont {J.}~\bibnamefont {Gubernatis}},\ }\bibfield  {title}
  {\bibinfo {title} {The phase diagram of the one-dimensional extended hubbard
  model},\ }in\ \href@noop {} {\emph {\bibinfo {booktitle} {The Hubbard
  Model}}}\ (\bibinfo  {publisher} {Springer},\ \bibinfo {year} {1995})\ pp.\
  \bibinfo {pages} {315--326}\BibitemShut {NoStop}%
\bibitem [{\citenamefont {Qu}\ \emph {et~al.}(2021)\citenamefont {Qu},
  \citenamefont {Chen}, \citenamefont {Jiang}, \citenamefont {Wang},\ and\
  \citenamefont {Li}}]{qu2021spin}%
  \BibitemOpen
  \bibfield  {author} {\bibinfo {author} {\bibfnamefont {D.-W.}\ \bibnamefont
  {Qu}}, \bibinfo {author} {\bibfnamefont {B.-B.}\ \bibnamefont {Chen}},
  \bibinfo {author} {\bibfnamefont {H.-C.}\ \bibnamefont {Jiang}}, \bibinfo
  {author} {\bibfnamefont {Y.}~\bibnamefont {Wang}},\ and\ \bibinfo {author}
  {\bibfnamefont {W.}~\bibnamefont {Li}},\ }\bibfield  {title} {\bibinfo
  {title} {\textit{Spin-Triplet Pairing Induced by Near-Neighbor Attraction in
  the Extended Hubbard Model}},\ }\href@noop {} {\bibfield  {journal} {\bibinfo
   {journal} {arXiv:2110.00564}\ } (\bibinfo {year} {2021})}\BibitemShut
  {NoStop}%
\bibitem [{\citenamefont {P{\"a}rschke}\ \emph {et~al.}(2019)\citenamefont
  {P{\"a}rschke}, \citenamefont {Wang}, \citenamefont {Moritz}, \citenamefont
  {Devereaux}, \citenamefont {Chen},\ and\ \citenamefont
  {Wohlfeld}}]{parschke2019numerical}%
  \BibitemOpen
  \bibfield  {author} {\bibinfo {author} {\bibfnamefont {E.~M.}\ \bibnamefont
  {P{\"a}rschke}}, \bibinfo {author} {\bibfnamefont {Y.}~\bibnamefont {Wang}},
  \bibinfo {author} {\bibfnamefont {B.}~\bibnamefont {Moritz}}, \bibinfo
  {author} {\bibfnamefont {T.~P.}\ \bibnamefont {Devereaux}}, \bibinfo {author}
  {\bibfnamefont {C.-C.}\ \bibnamefont {Chen}},\ and\ \bibinfo {author}
  {\bibfnamefont {K.}~\bibnamefont {Wohlfeld}},\ }\bibfield  {title} {\bibinfo
  {title} {\textit{Numerical Investigation of Spin Excitations in a Doped Spin
  Chain}},\ }\href@noop {} {\bibfield  {journal} {\bibinfo  {journal} {Phys.
  Rev. B}\ }\textbf {\bibinfo {volume} {99}},\ \bibinfo {pages} {205102}
  (\bibinfo {year} {2019})}\BibitemShut {NoStop}%
\bibitem [{\citenamefont {Jia}\ \emph {et~al.}(2014)\citenamefont {Jia},
  \citenamefont {Nowadnick}, \citenamefont {Wohlfeld}, \citenamefont {Kung},
  \citenamefont {Chen}, \citenamefont {Johnston}, \citenamefont {Tohyama},
  \citenamefont {Moritz},\ and\ \citenamefont {Devereaux}}]{jia2014persistent}%
  \BibitemOpen
  \bibfield  {author} {\bibinfo {author} {\bibfnamefont {C.~J.}\ \bibnamefont
  {Jia}}, \bibinfo {author} {\bibfnamefont {E.~A.}\ \bibnamefont {Nowadnick}},
  \bibinfo {author} {\bibfnamefont {K.}~\bibnamefont {Wohlfeld}}, \bibinfo
  {author} {\bibfnamefont {Y.~F.}\ \bibnamefont {Kung}}, \bibinfo {author}
  {\bibfnamefont {C.~C.}\ \bibnamefont {Chen}}, \bibinfo {author}
  {\bibfnamefont {S.}~\bibnamefont {Johnston}}, \bibinfo {author}
  {\bibfnamefont {T.}~\bibnamefont {Tohyama}}, \bibinfo {author} {\bibfnamefont
  {B.}~\bibnamefont {Moritz}},\ and\ \bibinfo {author} {\bibfnamefont {T.~P.}\
  \bibnamefont {Devereaux}},\ }\bibfield  {title} {\bibinfo {title}
  {\textit{Persistent Spin Excitations in Doped Antiferromagnets Revealed by
  Resonant Inelastic Light Scattering}},\ }\href@noop {} {\bibfield  {journal}
  {\bibinfo  {journal} {Nat. Commun.}\ }\textbf {\bibinfo {volume} {5}},\
  \bibinfo {pages} {3314} (\bibinfo {year} {2014})}\BibitemShut {NoStop}%
\bibitem [{\citenamefont {Mentink}\ \emph {et~al.}(2015)\citenamefont
  {Mentink}, \citenamefont {Balzer},\ and\ \citenamefont
  {Eckstein}}]{mentink2015ultrafast}%
  \BibitemOpen
  \bibfield  {author} {\bibinfo {author} {\bibfnamefont {J.}~\bibnamefont
  {Mentink}}, \bibinfo {author} {\bibfnamefont {K.}~\bibnamefont {Balzer}},\
  and\ \bibinfo {author} {\bibfnamefont {M.}~\bibnamefont {Eckstein}},\
  }\bibfield  {title} {\bibinfo {title} {\textit{Ultrafast and Reversible
  Control of the Exchange Interaction in Mott Insulators}},\ }\href@noop {}
  {\bibfield  {journal} {\bibinfo  {journal} {Nat. Commun.}\ }\textbf {\bibinfo
  {volume} {6}},\ \bibinfo {pages} {1} (\bibinfo {year} {2015})}\BibitemShut
  {NoStop}%
\bibitem [{\citenamefont {Chen}\ \emph {et~al.}(2019)\citenamefont {Chen},
  \citenamefont {Wang}, \citenamefont {Jia}, \citenamefont {Moritz},
  \citenamefont {Shvaika}, \citenamefont {Freericks},\ and\ \citenamefont
  {Devereaux}}]{chen2019theory}%
  \BibitemOpen
  \bibfield  {author} {\bibinfo {author} {\bibfnamefont {Y.}~\bibnamefont
  {Chen}}, \bibinfo {author} {\bibfnamefont {Y.}~\bibnamefont {Wang}}, \bibinfo
  {author} {\bibfnamefont {C.}~\bibnamefont {Jia}}, \bibinfo {author}
  {\bibfnamefont {B.}~\bibnamefont {Moritz}}, \bibinfo {author} {\bibfnamefont
  {A.~M.}\ \bibnamefont {Shvaika}}, \bibinfo {author} {\bibfnamefont {J.~K.}\
  \bibnamefont {Freericks}},\ and\ \bibinfo {author} {\bibfnamefont {T.~P.}\
  \bibnamefont {Devereaux}},\ }\bibfield  {title} {\bibinfo {title}
  {\textit{Theory for Time-resolved Resonant Inelastic X-ray Scattering}},\
  }\href@noop {} {\bibfield  {journal} {\bibinfo  {journal} {Phys. Rev. B}\
  }\textbf {\bibinfo {volume} {99}},\ \bibinfo {pages} {104306} (\bibinfo
  {year} {2019})}\BibitemShut {NoStop}%
\bibitem [{\citenamefont {Chen}\ \emph {et~al.}(2020)\citenamefont {Chen},
  \citenamefont {Wang}, \citenamefont {Claassen}, \citenamefont {Moritz},\ and\
  \citenamefont {Devereaux}}]{chen2020observing}%
  \BibitemOpen
  \bibfield  {author} {\bibinfo {author} {\bibfnamefont {Y.}~\bibnamefont
  {Chen}}, \bibinfo {author} {\bibfnamefont {Y.}~\bibnamefont {Wang}}, \bibinfo
  {author} {\bibfnamefont {M.}~\bibnamefont {Claassen}}, \bibinfo {author}
  {\bibfnamefont {B.}~\bibnamefont {Moritz}},\ and\ \bibinfo {author}
  {\bibfnamefont {T.~P.}\ \bibnamefont {Devereaux}},\ }\bibfield  {title}
  {\bibinfo {title} {\textit{Observing Photo-induced Chiral Edge States of
  Graphene Nanoribbons in Pump-probe Spectroscopies}},\ }\href@noop {}
  {\bibfield  {journal} {\bibinfo  {journal} {npj Quantum Mater.}\ }\textbf
  {\bibinfo {volume} {5}},\ \bibinfo {pages} {84} (\bibinfo {year}
  {2020})}\BibitemShut {NoStop}%
\bibitem [{\citenamefont {Strobel}\ \emph {et~al.}(2014)\citenamefont
  {Strobel}, \citenamefont {Muessel}, \citenamefont {Linnemann}, \citenamefont
  {Zibold}, \citenamefont {Hume}, \citenamefont {Pezz{\`e}}, \citenamefont
  {Smerzi},\ and\ \citenamefont {Oberthaler}}]{strobel2014fisher}%
  \BibitemOpen
  \bibfield  {author} {\bibinfo {author} {\bibfnamefont {H.}~\bibnamefont
  {Strobel}}, \bibinfo {author} {\bibfnamefont {W.}~\bibnamefont {Muessel}},
  \bibinfo {author} {\bibfnamefont {D.}~\bibnamefont {Linnemann}}, \bibinfo
  {author} {\bibfnamefont {T.}~\bibnamefont {Zibold}}, \bibinfo {author}
  {\bibfnamefont {D.~B.}\ \bibnamefont {Hume}}, \bibinfo {author}
  {\bibfnamefont {L.}~\bibnamefont {Pezz{\`e}}}, \bibinfo {author}
  {\bibfnamefont {A.}~\bibnamefont {Smerzi}},\ and\ \bibinfo {author}
  {\bibfnamefont {M.~K.}\ \bibnamefont {Oberthaler}},\ }\bibfield  {title}
  {\bibinfo {title} {\textit{Fisher Information and Entanglement of
  Non-Gaussian Spin States}},\ }\href@noop {} {\bibfield  {journal} {\bibinfo
  {journal} {Science}\ }\textbf {\bibinfo {volume} {345}},\ \bibinfo {pages}
  {424} (\bibinfo {year} {2014})}\BibitemShut {NoStop}%
\bibitem [{\citenamefont {Lorenzana}\ \emph {et~al.}(2005)\citenamefont
  {Lorenzana}, \citenamefont {Seibold},\ and\ \citenamefont
  {Coldea}}]{lorenzana2005sum}%
  \BibitemOpen
  \bibfield  {author} {\bibinfo {author} {\bibfnamefont {J.}~\bibnamefont
  {Lorenzana}}, \bibinfo {author} {\bibfnamefont {G.}~\bibnamefont {Seibold}},\
  and\ \bibinfo {author} {\bibfnamefont {R.}~\bibnamefont {Coldea}},\
  }\bibfield  {title} {\bibinfo {title} {\textit{Sum rules and missing spectral
  weight in magnetic neutron scattering in the cuprates}},\ }\href@noop {}
  {\bibfield  {journal} {\bibinfo  {journal} {Physical Review B}\ }\textbf
  {\bibinfo {volume} {72}},\ \bibinfo {pages} {224511} (\bibinfo {year}
  {2005})}\BibitemShut {NoStop}%
\bibitem [{\citenamefont {Laurell}\ \emph {et~al.}(2022)\citenamefont
  {Laurell}, \citenamefont {Scheie}, \citenamefont {Tennant}, \citenamefont
  {Okamoto}, \citenamefont {Alvarez},\ and\ \citenamefont
  {Dagotto}}]{laurell2022magnetic}%
  \BibitemOpen
  \bibfield  {author} {\bibinfo {author} {\bibfnamefont {P.}~\bibnamefont
  {Laurell}}, \bibinfo {author} {\bibfnamefont {A.}~\bibnamefont {Scheie}},
  \bibinfo {author} {\bibfnamefont {D.~A.}\ \bibnamefont {Tennant}}, \bibinfo
  {author} {\bibfnamefont {S.}~\bibnamefont {Okamoto}}, \bibinfo {author}
  {\bibfnamefont {G.}~\bibnamefont {Alvarez}},\ and\ \bibinfo {author}
  {\bibfnamefont {E.}~\bibnamefont {Dagotto}},\ }\bibfield  {title} {\bibinfo
  {title} {\textit{Magnetic Excitations, Nonclassicality, and Quantum Wake Spin
  Dynamics in the Hubbard Chain}},\ }\href@noop {} {\bibfield  {journal}
  {\bibinfo  {journal} {Phys. Rev. B}\ }\textbf {\bibinfo {volume} {106}},\
  \bibinfo {pages} {085110} (\bibinfo {year} {2022})}\BibitemShut {NoStop}%
\bibitem [{\citenamefont {Lin}\ and\ \citenamefont
  {Hirsch}(1986)}]{lin1986condensation}%
  \BibitemOpen
  \bibfield  {author} {\bibinfo {author} {\bibfnamefont {H.~Q.}\ \bibnamefont
  {Lin}}\ and\ \bibinfo {author} {\bibfnamefont {J.~E.}\ \bibnamefont
  {Hirsch}},\ }\bibfield  {title} {\bibinfo {title} {\textit{Condensation
  Transition in the One-dimensional Extended {Hubbard} Model}},\ }\href@noop {}
  {\bibfield  {journal} {\bibinfo  {journal} {Phys. Rev. B}\ }\textbf {\bibinfo
  {volume} {33}},\ \bibinfo {pages} {8155} (\bibinfo {year}
  {1986})}\BibitemShut {NoStop}%
\bibitem [{\citenamefont {Lin}\ \emph {et~al.}(1997)\citenamefont {Lin},
  \citenamefont {Gagliano},\ and\ \citenamefont {Campbell}}]{lin1997phase}%
  \BibitemOpen
  \bibfield  {author} {\bibinfo {author} {\bibfnamefont {H.}~\bibnamefont
  {Lin}}, \bibinfo {author} {\bibfnamefont {E.}~\bibnamefont {Gagliano}},\ and\
  \bibinfo {author} {\bibfnamefont {D.}~\bibnamefont {Campbell}},\ }\bibfield
  {title} {\bibinfo {title} {Phase separation in the {1-D} extended hubbard
  model},\ }\href@noop {} {\bibfield  {journal} {\bibinfo  {journal} {Physica
  C: Superconductivity}\ }\textbf {\bibinfo {volume} {282}},\ \bibinfo {pages}
  {1875} (\bibinfo {year} {1997})}\BibitemShut {NoStop}%
\bibitem [{\citenamefont {Xiang}\ \emph {et~al.}(2019)\citenamefont {Xiang},
  \citenamefont {Liu}, \citenamefont {Yuan}, \citenamefont {Cao},\ and\
  \citenamefont {Tang}}]{xiang2019doping}%
  \BibitemOpen
  \bibfield  {author} {\bibinfo {author} {\bibfnamefont {Y.-Y.}\ \bibnamefont
  {Xiang}}, \bibinfo {author} {\bibfnamefont {X.-J.}\ \bibnamefont {Liu}},
  \bibinfo {author} {\bibfnamefont {Y.-H.}\ \bibnamefont {Yuan}}, \bibinfo
  {author} {\bibfnamefont {J.}~\bibnamefont {Cao}},\ and\ \bibinfo {author}
  {\bibfnamefont {C.-M.}\ \bibnamefont {Tang}},\ }\bibfield  {title} {\bibinfo
  {title} {\textit{Doping Dependence of the Phase Diagram in One-dimensional
  Extended {Hubbard} Model: a Functional Renormalization Group Study}},\
  }\href@noop {} {\bibfield  {journal} {\bibinfo  {journal} {J. Phys.: Condens.
  Matter}\ }\textbf {\bibinfo {volume} {31}},\ \bibinfo {pages} {125601}
  (\bibinfo {year} {2019})}\BibitemShut {NoStop}%
\bibitem [{\citenamefont {Shinjo}\ \emph {et~al.}(2019)\citenamefont {Shinjo},
  \citenamefont {Sasaki}, \citenamefont {Hase}, \citenamefont {Sota},
  \citenamefont {Ejima}, \citenamefont {Yunoki},\ and\ \citenamefont
  {Tohyama}}]{shinjo2019machine}%
  \BibitemOpen
  \bibfield  {author} {\bibinfo {author} {\bibfnamefont {K.}~\bibnamefont
  {Shinjo}}, \bibinfo {author} {\bibfnamefont {K.}~\bibnamefont {Sasaki}},
  \bibinfo {author} {\bibfnamefont {S.}~\bibnamefont {Hase}}, \bibinfo {author}
  {\bibfnamefont {S.}~\bibnamefont {Sota}}, \bibinfo {author} {\bibfnamefont
  {S.}~\bibnamefont {Ejima}}, \bibinfo {author} {\bibfnamefont
  {S.}~\bibnamefont {Yunoki}},\ and\ \bibinfo {author} {\bibfnamefont
  {T.}~\bibnamefont {Tohyama}},\ }\bibfield  {title} {\bibinfo {title}
  {\textit{Machine Learning Phase Diagram in the Half-Filled One-Dimensional
  Extended Hubbard Model}},\ }\href@noop {} {\bibfield  {journal} {\bibinfo
  {journal} {Journal of the Physical Society of Japan}\ }\textbf {\bibinfo
  {volume} {88}},\ \bibinfo {pages} {065001} (\bibinfo {year}
  {2019})}\BibitemShut {NoStop}%
\bibitem [{\citenamefont {Suresh}\ \emph {et~al.}(2022)\citenamefont {Suresh},
  \citenamefont {Soares}, \citenamefont {Mondal}, \citenamefont {Pires},
  \citenamefont {Lopes}, \citenamefont {Ferreira}, \citenamefont {Feiguin},
  \citenamefont {Plech{\'a}{\v{c}}},\ and\ \citenamefont
  {Nikoli{\'c}}}]{suresh2022electron}%
  \BibitemOpen
  \bibfield  {author} {\bibinfo {author} {\bibfnamefont {A.}~\bibnamefont
  {Suresh}}, \bibinfo {author} {\bibfnamefont {R.}~\bibnamefont {Soares}},
  \bibinfo {author} {\bibfnamefont {P.}~\bibnamefont {Mondal}}, \bibinfo
  {author} {\bibfnamefont {J.}~\bibnamefont {Pires}}, \bibinfo {author}
  {\bibfnamefont {J.}~\bibnamefont {Lopes}}, \bibinfo {author} {\bibfnamefont
  {A.}~\bibnamefont {Ferreira}}, \bibinfo {author} {\bibfnamefont
  {A.}~\bibnamefont {Feiguin}}, \bibinfo {author} {\bibfnamefont
  {P.}~\bibnamefont {Plech{\'a}{\v{c}}}},\ and\ \bibinfo {author}
  {\bibfnamefont {B.}~\bibnamefont {Nikoli{\'c}}},\ }\bibfield  {title}
  {\bibinfo {title} {\textit{Electron-mediated entanglement of two distant
  macroscopic ferromagnets within a nonequilibrium spintronic device}},\
  }\href@noop {} {\bibfield  {journal} {\bibinfo  {journal} {arXiv:2210.06634}\
  } (\bibinfo {year} {2022})}\BibitemShut {NoStop}%
\bibitem [{\citenamefont {Eckert}\ \emph {et~al.}(2002)\citenamefont {Eckert},
  \citenamefont {Schliemann}, \citenamefont {Bruß},\ and\ \citenamefont
  {Lewenstein}}]{eckert2002quantum}%
  \BibitemOpen
  \bibfield  {author} {\bibinfo {author} {\bibfnamefont {K.}~\bibnamefont
  {Eckert}}, \bibinfo {author} {\bibfnamefont {J.}~\bibnamefont {Schliemann}},
  \bibinfo {author} {\bibfnamefont {D.}~\bibnamefont {Bruß}},\ and\ \bibinfo
  {author} {\bibfnamefont {M.}~\bibnamefont {Lewenstein}},\ }\bibfield  {title}
  {\bibinfo {title} {\textit{Quantum Correlations in Systems of
  Indistinguishable Particles}},\ }\href@noop {} {\bibfield  {journal}
  {\bibinfo  {journal} {Ann. Phys.}\ }\textbf {\bibinfo {volume} {299}},\
  \bibinfo {pages} {88} (\bibinfo {year} {2002})}\BibitemShut {NoStop}%
\bibitem [{\citenamefont {Schliemann}\ \emph
  {et~al.}(2001{\natexlab{a}})\citenamefont {Schliemann}, \citenamefont
  {Cirac}, \citenamefont {Ku{\'s}}, \citenamefont {Lewenstein},\ and\
  \citenamefont {Loss}}]{schliemann2001quantum}%
  \BibitemOpen
  \bibfield  {author} {\bibinfo {author} {\bibfnamefont {J.}~\bibnamefont
  {Schliemann}}, \bibinfo {author} {\bibfnamefont {J.~I.}\ \bibnamefont
  {Cirac}}, \bibinfo {author} {\bibfnamefont {M.}~\bibnamefont {Ku{\'s}}},
  \bibinfo {author} {\bibfnamefont {M.}~\bibnamefont {Lewenstein}},\ and\
  \bibinfo {author} {\bibfnamefont {D.}~\bibnamefont {Loss}},\ }\bibfield
  {title} {\bibinfo {title} {\textit{Quantum Correlations in Two-fermion
  Systems}},\ }\href@noop {} {\bibfield  {journal} {\bibinfo  {journal} {Phys.
  Rev. A}\ }\textbf {\bibinfo {volume} {64}},\ \bibinfo {pages} {022303}
  (\bibinfo {year} {2001}{\natexlab{a}})}\BibitemShut {NoStop}%
\bibitem [{\citenamefont {Schliemann}\ \emph
  {et~al.}(2001{\natexlab{b}})\citenamefont {Schliemann}, \citenamefont
  {Loss},\ and\ \citenamefont {MacDonald}}]{schliemann2001double}%
  \BibitemOpen
  \bibfield  {author} {\bibinfo {author} {\bibfnamefont {J.}~\bibnamefont
  {Schliemann}}, \bibinfo {author} {\bibfnamefont {D.}~\bibnamefont {Loss}},\
  and\ \bibinfo {author} {\bibfnamefont {A.}~\bibnamefont {MacDonald}},\
  }\bibfield  {title} {\bibinfo {title} {\textit{Double-occupancy Errors,
  Adiabaticity, and Entanglement of Spin Qubits in Quantum Dots}},\ }\href@noop
  {} {\bibfield  {journal} {\bibinfo  {journal} {Phys. Rev. B}\ }\textbf
  {\bibinfo {volume} {63}},\ \bibinfo {pages} {085311} (\bibinfo {year}
  {2001}{\natexlab{b}})}\BibitemShut {NoStop}%
\bibitem [{\citenamefont {Kraus}\ \emph {et~al.}(2009)\citenamefont {Kraus},
  \citenamefont {Wolf}, \citenamefont {Cirac},\ and\ \citenamefont
  {Giedke}}]{kraus2009pairing}%
  \BibitemOpen
  \bibfield  {author} {\bibinfo {author} {\bibfnamefont {C.~V.}\ \bibnamefont
  {Kraus}}, \bibinfo {author} {\bibfnamefont {M.~M.}\ \bibnamefont {Wolf}},
  \bibinfo {author} {\bibfnamefont {J.~I.}\ \bibnamefont {Cirac}},\ and\
  \bibinfo {author} {\bibfnamefont {G.}~\bibnamefont {Giedke}},\ }\bibfield
  {title} {\bibinfo {title} {\textit{Pairing in Fermionic Systems: A
  Quantum-Information Perspective}},\ }\href@noop {} {\bibfield  {journal}
  {\bibinfo  {journal} {Phys. Rev. A}\ }\textbf {\bibinfo {volume} {79}},\
  \bibinfo {pages} {012306} (\bibinfo {year} {2009})}\BibitemShut {NoStop}%
\bibitem [{\citenamefont {Braicovich}\ \emph {et~al.}(2010)\citenamefont
  {Braicovich}, \citenamefont {van~den Brink}, \citenamefont {Bisogni},
  \citenamefont {Sala}, \citenamefont {Ament}, \citenamefont {Brookes},
  \citenamefont {De~Luca}, \citenamefont {Salluzzo}, \citenamefont {Schmitt},
  \citenamefont {Strocov},\ and\ \citenamefont
  {Ghiringhelli}}]{Braicovich2010magnetic}%
  \BibitemOpen
  \bibfield  {author} {\bibinfo {author} {\bibfnamefont {L.}~\bibnamefont
  {Braicovich}}, \bibinfo {author} {\bibfnamefont {J.}~\bibnamefont {van~den
  Brink}}, \bibinfo {author} {\bibfnamefont {V.}~\bibnamefont {Bisogni}},
  \bibinfo {author} {\bibfnamefont {M.~M.}\ \bibnamefont {Sala}}, \bibinfo
  {author} {\bibfnamefont {L.~J.~P.}\ \bibnamefont {Ament}}, \bibinfo {author}
  {\bibfnamefont {N.~B.}\ \bibnamefont {Brookes}}, \bibinfo {author}
  {\bibfnamefont {G.~M.}\ \bibnamefont {De~Luca}}, \bibinfo {author}
  {\bibfnamefont {M.}~\bibnamefont {Salluzzo}}, \bibinfo {author}
  {\bibfnamefont {T.}~\bibnamefont {Schmitt}}, \bibinfo {author} {\bibfnamefont
  {V.~N.}\ \bibnamefont {Strocov}},\ and\ \bibinfo {author} {\bibfnamefont
  {G.}~\bibnamefont {Ghiringhelli}},\ }\bibfield  {title} {\bibinfo {title}
  {Magnetic excitations and phase separation in the underdoped
  ${\mathrm{la}}_{2\ensuremath{-}x}{\mathrm{sr}}_{x}{\mathrm{cuo}}_{4}$
  superconductor measured by resonant inelastic x-ray scattering},\ }\href@noop
  {} {\bibfield  {journal} {\bibinfo  {journal} {Phys. Rev. Lett.}\ }\textbf
  {\bibinfo {volume} {104}},\ \bibinfo {pages} {077002} (\bibinfo {year}
  {2010})}\BibitemShut {NoStop}%
\bibitem [{\citenamefont {Haverkort}(2010)}]{Haverkort2010theory}%
  \BibitemOpen
  \bibfield  {author} {\bibinfo {author} {\bibfnamefont {M.~W.}\ \bibnamefont
  {Haverkort}},\ }\bibfield  {title} {\bibinfo {title} {Theory of resonant
  inelastic x-ray scattering by collective magnetic excitations},\ }\href@noop
  {} {\bibfield  {journal} {\bibinfo  {journal} {Phys. Rev. Lett.}\ }\textbf
  {\bibinfo {volume} {105}},\ \bibinfo {pages} {167404} (\bibinfo {year}
  {2010})}\BibitemShut {NoStop}%
\bibitem [{\citenamefont {Robarts}\ \emph {et~al.}(2021)\citenamefont
  {Robarts}, \citenamefont {Garc\'{\i}a-Fern\'andez}, \citenamefont {Li},
  \citenamefont {Nag}, \citenamefont {Walters}, \citenamefont {Headings},
  \citenamefont {Hayden},\ and\ \citenamefont {Zhou}}]{Robarts2021dynamical}%
  \BibitemOpen
  \bibfield  {author} {\bibinfo {author} {\bibfnamefont {H.~C.}\ \bibnamefont
  {Robarts}}, \bibinfo {author} {\bibfnamefont {M.}~\bibnamefont
  {Garc\'{\i}a-Fern\'andez}}, \bibinfo {author} {\bibfnamefont
  {J.}~\bibnamefont {Li}}, \bibinfo {author} {\bibfnamefont {A.}~\bibnamefont
  {Nag}}, \bibinfo {author} {\bibfnamefont {A.~C.}\ \bibnamefont {Walters}},
  \bibinfo {author} {\bibfnamefont {N.~E.}\ \bibnamefont {Headings}}, \bibinfo
  {author} {\bibfnamefont {S.~M.}\ \bibnamefont {Hayden}},\ and\ \bibinfo
  {author} {\bibfnamefont {K.-J.}\ \bibnamefont {Zhou}},\ }\bibfield  {title}
  {\bibinfo {title} {Dynamical spin susceptibility in
  ${\mathrm{la}}_{2}{\mathrm{cuo}}_{4}$ studied by resonant inelastic x-ray
  scattering},\ }\href@noop {} {\bibfield  {journal} {\bibinfo  {journal}
  {Phys. Rev. B}\ }\textbf {\bibinfo {volume} {103}},\ \bibinfo {pages}
  {224427} (\bibinfo {year} {2021})}\BibitemShut {NoStop}%
\bibitem [{\citenamefont {Tsutsui}\ \emph {et~al.}(2000)\citenamefont
  {Tsutsui}, \citenamefont {Kondo}, \citenamefont {Tohyama},\ and\
  \citenamefont {Maekawa}}]{tsutsui2000resonant}%
  \BibitemOpen
  \bibfield  {author} {\bibinfo {author} {\bibfnamefont {K.}~\bibnamefont
  {Tsutsui}}, \bibinfo {author} {\bibfnamefont {H.}~\bibnamefont {Kondo}},
  \bibinfo {author} {\bibfnamefont {T.}~\bibnamefont {Tohyama}},\ and\ \bibinfo
  {author} {\bibfnamefont {S.}~\bibnamefont {Maekawa}},\ }\bibfield  {title}
  {\bibinfo {title} {\textit{Resonant Inelastic X-ray Scattering Spectrum in
  High-Tc Cuprates}},\ }\href@noop {} {\bibfield  {journal} {\bibinfo
  {journal} {Physica B: Condensed Matter}\ }\textbf {\bibinfo {volume} {284}},\
  \bibinfo {pages} {457} (\bibinfo {year} {2000})}\BibitemShut {NoStop}%
\bibitem [{\citenamefont {Kourtis}\ \emph {et~al.}(2012)\citenamefont
  {Kourtis}, \citenamefont {van~den Brink},\ and\ \citenamefont
  {Daghofer}}]{kourtis2012exact}%
  \BibitemOpen
  \bibfield  {author} {\bibinfo {author} {\bibfnamefont {S.}~\bibnamefont
  {Kourtis}}, \bibinfo {author} {\bibfnamefont {J.}~\bibnamefont {van~den
  Brink}},\ and\ \bibinfo {author} {\bibfnamefont {M.}~\bibnamefont
  {Daghofer}},\ }\bibfield  {title} {\bibinfo {title} {\textit{Exact
  Diagonalization Results for Resonant Inelastic X-ray Scattering Spectra of
  One-dimensional Mott Insulators}},\ }\href@noop {} {\bibfield  {journal}
  {\bibinfo  {journal} {Phys. Rev. B}\ }\textbf {\bibinfo {volume} {85}},\
  \bibinfo {pages} {064423} (\bibinfo {year} {2012})}\BibitemShut {NoStop}%
\bibitem [{\citenamefont {Wiseman}\ and\ \citenamefont
  {Vaccaro}(2003)}]{wiseman2003entanglement}%
  \BibitemOpen
  \bibfield  {author} {\bibinfo {author} {\bibfnamefont {H.~M.}\ \bibnamefont
  {Wiseman}}\ and\ \bibinfo {author} {\bibfnamefont {J.~A.}\ \bibnamefont
  {Vaccaro}},\ }\bibfield  {title} {\bibinfo {title} {\textit{Entanglement of
  Indistinguishable Particles Shared Between Two Parties}},\ }\href@noop {}
  {\bibfield  {journal} {\bibinfo  {journal} {Physical review letters}\
  }\textbf {\bibinfo {volume} {91}},\ \bibinfo {pages} {097902} (\bibinfo
  {year} {2003})}\BibitemShut {NoStop}%
\bibitem [{\citenamefont {Ba\~nuls}\ \emph {et~al.}(2007)\citenamefont
  {Ba\~nuls}, \citenamefont {Cirac},\ and\ \citenamefont
  {Wolf}}]{banuls2007entanglement}%
  \BibitemOpen
  \bibfield  {author} {\bibinfo {author} {\bibfnamefont {M.-C.}\ \bibnamefont
  {Ba\~nuls}}, \bibinfo {author} {\bibfnamefont {J.~I.}\ \bibnamefont
  {Cirac}},\ and\ \bibinfo {author} {\bibfnamefont {M.~M.}\ \bibnamefont
  {Wolf}},\ }\bibfield  {title} {\bibinfo {title} {\textit{Entanglement in
  Fermionic Systems}},\ }\href@noop {} {\bibfield  {journal} {\bibinfo
  {journal} {Phys. Rev. A}\ }\textbf {\bibinfo {volume} {76}},\ \bibinfo
  {pages} {022311} (\bibinfo {year} {2007})}\BibitemShut {NoStop}%
\end{thebibliography}%

\end{document}